\documentclass{aastex6}

\usepackage{gensymb}

\newcommand{\feh}{\hbox{$ [\mathrm{Fe}/\mathrm{H}]$}}
\newcommand{\ea}{et al.}

\newcommand{\beq}{\begin{equation}}
\newcommand{\eeq}{\end{equation}}
\newcommand{\hst}{\textit{HST} }

\newcommand{\hip}{\textit{Hipparcos} }

\newcommand{\gaia}{\textit{Gaia} }
\newcommand{\hstns}{\textit{HST} }

\newcommand{\cht}{$\chi^2~$}

\begin{document}
\DeclareGraphicsExtensions{.pdf,.gif,.jpg}

\title{Testing Metal Poor Stellar Models and Isochrones with HST Parallaxes of Metal Poor Stars}

\shorttitle{Parallaxes of Metal-Poor Stars}
\shortauthors{Chaboyer et al.}

\author{B.\ Chaboyer\altaffilmark{1,2}, 
B.E.\ McArthur\altaffilmark{3},
E.\ O'Malley\altaffilmark{1},
G.F.\ Benedict\altaffilmark{3}, 
G.\ A.Feiden\altaffilmark{4}, 
T.E.\ Harrison\altaffilmark{5},
A.\ McWilliam\altaffilmark{6},
E.P.\ Nelan\altaffilmark{7},
R.J.\ Patterson\altaffilmark{8},
\and A.\ Sarajedini\altaffilmark{9}
}

\altaffiltext{1}{Dept.\ of Physics and Astronomy, Dartmouth College,
Hanover, NH 03755 USA \email{Brian.Chaboyer@Dartmouth.edu}}
\altaffiltext{2}{Max-Planck-Institut f\"{u}r Astrophysik, Karl-Schwarzschild-Str. 1, D-85748 Garching bei M\"{u}nchen, Germany }
\altaffiltext{3}{McDonald Observatory, University of Texas, Austin, TX 78712 USA  }
\altaffiltext{4}{Department of Physics and Astronomy, Uppsala University, Box 516, SE-751 20, Uppsala, Sweden}
\altaffiltext{5}{Dept.\ of Astronomy, New Mexico State University, Las Cruces, NM 88003 USA   }
\altaffiltext{6}{The Observatories of the Carnegie Institute of Washington, Pasadena, CA 91101 USA} 
\altaffiltext{7}{Space Telescope Science Institute, Baltimore, MD 21218 USA  }
\altaffiltext{8}{Dept.\  of Astronomy, University of Virginia, Charlottesville, VA 22904 USA} 
\altaffiltext{9}{Dept.\  of Astronomy, University of Florida, Gainesville, FL 32611 USA }

\begin{abstract}
Hubble Space Telescope (HST) fine guidance sensor observations were used to obtain parallaxes of eight metal-poor ($\feh < -1.4$) stars.  
 The parallaxes of these stars  determined by  the new $\hip$  reduction average 17\% accuracy, in contrast to  our new $\hst$ parallaxes which average 1\%  accuracy and have  errors on the individual parallaxes ranging from 85 to 144 {\it microarcseconds}. This parallax data has been combined with HST ACS photometry in the F606W and F814W filters to obtain the absolute magnitudes of the stars with an accuracy of 0.02 to 0.03 magnitudes. Six of these stars are on the main sequence (with $-2.7 < \feh < -1.8$), and suitable for testing metal-poor stellar evolution models and determining the distances to metal-poor globular clusters.  Using the abundances obtained by O'Malley \ea\ (2016) we find that standard stellar models using the \citet{VC2003} color transformation do a reasonable job of matching five of the main sequence stars, with HD 54639 ($\feh = -2.5$) being anomalous in its location in the color-magnitude diagram.  Stellar models and isochrones were generated using a Monte Carlo analysis to take into account uncertainties in the models.   Isochrones which fit the parallax stars were used to determine the distances and ages of nine globular clusters (with $ -2.4 \le \feh \le -1.9$).  Averaging together the age of all nine clusters, leads to an absolute age of the oldest, most metal-poor globular clusters of $12.7\pm 1.0\,$Gyr, where the quoted uncertainty takes into account the known uncertainties in the stellar models and isochrones, along with the uncertainty in the distance and reddening of the clusters.  

\end{abstract}


\section{Introduction}
Stellar evolution models and isochrones are widely used in astrophysics to determine the properties of stars and integrated stellar populations.  The predicted absolute magnitudes and colors of stars are frequently used, yet there are few tests of these basic properties of stellar models and isochrones for metal-poor stars.  Stars on the main sequence provide a stringent test of stellar evolution models, as their properties vary little with age.  In contrast, the properties of more evolved stars are sensitive to the age of the star, and stellar evolution models are typically assumed to be correct and used to determine the ages of stars which have evolved off the main sequence.  To test metal-poor stellar models and isochrones, it is important to have high quality absolute magnitudes and colors of single, main sequence stars.  

The $\hip$ catalog \citep{hip, hip2}  provided accurate parallaxes for a large number of stars and was used to investigate a large number of issues in stellar astrophysics, including the distances and ages of globular clusters \citep[e.g.][]{reid97,gratton97,pont98,chaboyer98,caretta00,grundahl02,gratton03}. One of the key limitations of the $\hip$ catalog for studying metal-poor stars is that it  only contained one single main sequence star with $\feh < -1.5$ whose parallax was known to a sufficient accuracy ($\sim 10\%$) to provide a reasonable test of stellar models.   To investigate the reliability of low metallicity stellar models prior to the \gaia data release, and to illustrate the types of studies enabled by high quality parallaxes, we have obtained HST fine guidance sensor (FGS) observations of nine metal-poor stars, and parallaxes of eight of these stars are presented in this paper.  The HST observations started on Oct. 28, 2008 and were completed on June 3, 2013.  In addition to illustrating the type of science possible with high quality parallaxes of metal-poor stars, this work provides accurate parallaxes which can be used to cross-check the Tycho-\gaia astrometric solution (TGAS)
\citep{michalik15, gaia16, lindegren16}, which was released after this paper was submitted for publication.  
The uncertainties in our parallaxes range from 85 to 144 $\mu$as, which are lower than the TGAS parallax uncertainties, which range from  $220$ to $860\,\mu$as for our target stars \citep{lindegren16}.  

The stars selected for HST observations were selected from various lists of metal-poor stars.  The key criteria used to select stars for the HST observations were  (a) estimated $[\mathrm{Fe/H}] < -1.5$, (b) the star was not a known member of a multiple star system, and  (c) that the star be on the main sequence.  To determine if a star was on the main sequence, we selected stars which were relatively cool ($V-I > 0.7$) and which had high surface gravities (based upon estimates from high-resolution spectroscopy, or  St\"{o}mgren photometry).  

Details of our observations and data reductions are presented in \S \ref{SEC:OBS}. The derived parallaxes are discussed in \S \ref{SEC:PAR}, where we determine that two of the stars are in the main-sequence turn-off or sub-giant branch phase of evolution and so are not useful for testing stellar models.  
These two  stars turned out to suffer from significant reddening and as a result were bluer than was assumed during our sample selection. The main sequence stars are compared to stellar isochrones in \S \ref{SEC:ISO}, and the isochrones which fit these calibrating stars are used to derive distances and ages of a number of metal-poor globular clusters in \S \ref{SEC:GC}.  Our key results are summarized in \S \ref{SEC:SUMM}.

\section{Observations  \label{SEC:OBS} }

\subsection{$\hst$ Photometry of Program Stars}

In order to compare the magnitudes of our program stars with HST color-magnitude diagrams of globular clusters, each program star was observed with ACS/WFC in the F606W and F814W filters. 
The CTE-corrected ACS/WFC images for the program stars were retrieved from MAST. Each
of them was multiplied by the geometric correction image, and the image quality file was
applied in order to mask-out pixels of lower quality. Photometry with a 0.5 arcsec 
aperture was performed on the program stars in the resultant images. These instrumental
magnitudes were corrected for exposure time, matched to form colors, and calibrated to
the VEGAMag and ground-based VI systems using the \citet{sirianni05} photometric 
transformations.  Ground based photometry for all of our program stars were obtained using the 
NMSU 1 m telescope, the MDM 1.3m telescope and the SMARTS 0.9 m telescope.  Further details are provided in \S \ref{groundphot}.

A summary of HST and ground based photometry of the program stars is presented in Table \ref{photometry}.  For each star, we list our own ground based measurements,  along with  photoelectric photometer observations from the literature.   For several stars, there is a considerable range in the $V$ and $V - I$ measurements. We recommend  averaging the $V$ and $V-I$ measurements  to obtain the best estimate of the apparent magnitude and color of each star.  Given the scatter in the data, an uncertainty of 0.01 mag in the observed V magnitudes and $V-I$ colors of the stars seems appropriate.  

\begin{deluxetable}{llllrc}
\tabletypesize{\footnotesize}
\tablecaption{Photometry of Program Stars  \label{photometry} }
\tablehead{
\colhead{ID}&
\colhead{Telescope} &
\colhead{$V$ } &
\colhead{$V - I$ } & 
\colhead{F606W} &
\colhead{F606W --- F814W} 
}
\startdata
HIP 46120Ê&    HST &      10.097ÊÊ&   0.752 Ê& Ê$9.938\pm 0.0015$ÊÊ& $0.566\pm 0.0023$\\
& SSO\tablenotemark{a} & 10.15 & 0.725  &   \nodata   & \nodata\\ 
HIP 54639Ê&    HST &      11.354ÊÊ &  0.917ÊÊ& $11.149\pm 0.0017$ÊÊ&  $0.688\pm 0.0044$\\
 & MDM &      11.377  & ÊÊ0.911  &   \nodata   & \nodata\\ 
 & KPNO\tablenotemark{b} & 11.40 & \nodata   & \nodata & \nodata \\
HIP 56291\tablenotemark{c}ÊÊ & HST & 11.519Ê& 0.865Ê&Ê$11.328\pm 0.0024$ Ê&Ê $0.649\pm 0.0069$\\ÊÊ
 & MDM  &   11.575 & 0.845ÊÊ&  \nodata   & \nodata\\ 
 & KPNO\tablenotemark{b} & 11.53 & \nodata   & \nodata & \nodata \\
HIP 87062Ê& HST   &  10.565Ê&Ê0.848Ê&Ê $10.379\pm 0.0003$Ê&Ê $0.636\pm 0.0021$\\
 & NMSU  &  10.57Ê&  0.83  & \nodata   & \nodata\\ 
 & MDM   &   10.56Ê&  0.82ÊÊ& \nodata   & \nodata\\ 
 & KPNO\tablenotemark{d} & 10.60 & \nodata   & \nodata & \nodata \\
HIP 87788Ê& HST &    11.298Ê & 0.856Ê& $11.109\pm 0.0031$Ê&Ê $0.642\pm 0.0034$\\Ê 
 & NMSU  &  11.33Ê&  0.87 & \nodata   & \nodata\\ 
 & MDM   &  11.29Ê&  0.83Ê&Ê \nodata   & \nodata\\ 
  & KPNO\tablenotemark{d} & 11.29 & \nodata   & \nodata & \nodata \\
HIP 98492ÊÊ& HST &    11.559Ê&Ê0.832ÊÊ& $11.377\pm 0.0051$ÊÊ& $0.625\pm 0.0055$\\ 
 & NMSU   & 11.58ÊÊ&  0.80Ê&Ê \nodata   & \nodata\\ 
 & MDM    & 11.57Ê&  0.80  &  \nodata   & \nodata\\ 
  & KPNO\tablenotemark{b} & 11.59 & \nodata   & \nodata & \nodata \\ 
  & OAN-SPM\tablenotemark{e}  & 11.568 & \nodata & \nodata & \nodata \\
HIP 103269Ê&  HST &    10.249Ê&Ê0.772Ê&Ê $10.084\pm 0.0034$ &  $0.581\pm 0.0043$\\ÊÊ
 & NMSU   & 10.28Ê&  0.79ÊÊ& \nodata   & \nodata\\ 
 & MDM    & 10.29Ê&  0.75 & \nodata   & \nodata\\ 
  & KPNO\tablenotemark{b} &   10.27 & \nodata   & \nodata & \nodata \\
HIP 106924Ê&  HST &    10.328Ê&Ê0.799Ê&Ê Ê$10.156\pm 0.0024$Ê&Ê $0.601\pm 0.0054$\\ÊÊ 
 & NMSU  &  10.42Ê&ÊÊ0.87ÊÊ& \nodata   & \nodata\\ 
 & MDM    & 10.36ÊÊ& 0.74 & \nodata   & \nodata\\
 & KPNO\tablenotemark{b} &   10.34 & \nodata   & \nodata & \nodata \\
HIP 108200Ê&  HST &    10.974Ê&Ê0.856Ê& $10.785\pm 0.0031$Ê&Ê $0.643\pm 0.0063$\\Ê 
 & NMSU  &  10.97Ê&ÊÊ0.83ÊÊ& \nodata   & \nodata\\Ê
 & MDM    & 11.00Ê &  0.82ÊÊ& \nodata   & \nodata\\
  & KPNO\tablenotemark{d} &   11.03 & \nodata   & \nodata & \nodata \\ 
\enddata 
\tablenotetext{a}{Bessel, M.S.\ 1990 (Cousins (Kron-Cape) system)}
\tablenotetext{b}{Carney \& Latham 1987}
\tablenotetext{c}{Missing FGS parallax}
\tablenotetext{d}{Carney \textit{et al.}\ 1994}
\tablenotetext{e}{Silva, Schuster \& Contreras 2012}
\end{deluxetable}

\subsection{ $\hst$ Astrometry Observations}

We used  $\hst$ FGS-1r,  a two-axis interferometer, in position (POS) ``fringe-tracking"  mode\footnote{A detailed 
\href{http://www.stsci.edu/hst/fgs/documents/instrumenthandbook/}{Instrument Handbook} can be found on the Space Telescope Science Institute website.  } to make the astrometric observations.     \cite{Nelan07} describes the FGS instrument.    The reduction and calibration of the data is described in   \cite{Ben07}. A new improved Optical Field Angle Distortion (OFAD) derived by McArthur, was used to reduce and calibrate the data which is  available with the reduction pipeline.   This astrometric data  are available from the \href{http://www.stsci.edu/hst/scheduling/program_information}{HST Program Schedule and Information 
website}, in proposal numbers 11704 and 12320.  Eighty-nine orbits of  $\hst$  astrometric observations  were made between December 2008 and June 2013.  
Every orbit contains several observations of  the target and surrounding reference stars. The latest calibrations parameters and the   two part pipeline used to reduce the raw data to the values used in this modeling is available  from the Space Telescope Science Institute in IRAF STSDAS and in a standalone version available from one of the coauthors, the $\hstns$ FGS Instrument Scientist at STSCI Ed Nelan.  

With respect to its reference frame, an {\hst } target parallax is relative, not absolute.   For a conversion from a relative to and absolute frame, we can use a statistically-
derived correction  \citep[c.f.][]{Wva95}, or  we can use the derived spectroscopic parallaxes of the reference frame stars as input to the model in a Bayesian approach.  For finer accuracy, we use the Bayesian approach, in which
we estimate the absolute parallaxes of the reference frame stars  using spectra to classify the temperature and 
luminosity class of each star, and then combine these with $UBVRIJHK$ 
photometry to determine their visual extinctions, as described in \citet{Harr13}. We require the 
absolute magnitude, $M_V$, and $V$-band absorption, $A_V$ for the equation
\begin{equation}
\pi_{\rm abs} = 10^{-(V-M_V+5-A_V)/5}
\end{equation}
Our modeling then produces not a relative parallax, but an absolute parallax.


Full details of our astrometric reductions are provided in the Appendix.  Below, we summarize the key details.  
Each reference frame contained from 5 to 9 reference stars plus the target.    The positions  ($x',y'$) of the target and reference stars change with each observation set because  $\hst$ rolls with the observations shown in Table~\ref{tab:atmlog}.  We use parameters  for scale, rotation, and offset in an overlapping plate model.   We relate each plate's parameters to a master  constrained plate, usually one of the central observations.    The astrometric model also includes the time-dependent movements of each star,  given by the absolute parallax  $\pi_{abs}$ and the proper motion components, $\mu_{\alpha}$ and $\mu_{\delta}$, which are transformed into the roll of the constraint plate.  We also include instrumentally caused position shift parameters  for  lateral color.   The standard coordinate catalog positions $\xi$ and $\eta$ are the result of  modeling  these equations of condition:
\begin{eqnarray}
x^\prime & = & x + lc_x(B-V) - \Delta XFx \label{eq:cflcx}\\
y^\prime & = & y + lc_x(B-V) - \Delta XFy \label{eq:cflcy}
\end{eqnarray}
\begin{eqnarray}
\xi & =  & Ax^\prime + By^\prime + C - \mu_\xi \Delta t  - P_\xi\pi\label{eq:astmodxi6} \\
\eta &  = &  Dx^\prime + Ey^\prime+ F -  \mu_\eta \Delta t  - P_\eta\pi \label{eq:astmodeta6}
\end{eqnarray}
where $\it x$ and $\it y$  are the measured coordinates from $\hst$; $\it lc_x$ and $\it lc_y$ are the lateral color corrections, and  $B-V$ are the $B - V$ colors of each star.  $A$, $B$, $D$,  and $E$, are scale and rotation plate constants, and $C$ and $F$ are offsets.   The constraint plate defines $A$ and $E$ equal to 1, $B$ and $D$ equal to 0, and $C$ and $F$ equal to 0.   $\mu_\xi$ and $\mu_\eta$ are proper motions, $\Delta t$ is the epoch difference from the mean epoch, $P_\xi$ and $P_\eta$ are parallax factors,  and $\it \pi$ is  the parallax.   We get the parallax factors from a JPL Earth orbit predictor
 \citep[][upgraded to version DE405]{Sta90}. 
We  use a model in the GaussFit language  \citep{Jefferys88} utilizing robust estimation to derive a simultaneous solution for all parameters.  The resulting astrometric catalogs
from the combined  modeling are shown in Table~\ref{tab:cat}.

Condition equations relate an initial   and final parameter value.  All input data supplied to the equations of condition (such as reference star proper motions from catalogs and spectrophotometric parallaxes) are input to the model with errors.      The lateral color term of all stars and the  $\hst$ roll reported by the spacecraft of  the constraint plate are also input with their errors in additional  equations of condition.  These additional equations of condition  allow the \cht minimization process to adjust parameter values by amounts constrained by their input errors.   In this quasi-Bayesian approach,  prior knowledge is  not  input as a hardwired quantity known to infinite precision, but  input as an observation with associated error.   For all metal-poor target  (not reference) stars  no priors were used for parallax or proper motion.

\section{  $\hst$ Parallax and Proper Motions  of the target stars \label{SEC:PAR} }

The parallaxes   of the metal-poor target stars are shown in  Table~\ref{tab:pcomp}, including independently  determined values from the updated $\hip$ catalogue \citep{hip2} for comparison.    Our determination yielded uncertainties between 85 and 144 micro-arseconds, with an average uncertainty of 1\%   on the parallax, a very high accuracy.  In contrast, the new $\hip$ reduction yielded uncertainties  of 17\%, {while the recently released TGAS parallaxes typically have uncertainties a factor of two larger than the HST uncertainties.     The proper motions  shown in  Table~\ref{tab:pmcomp} agree with {\hip} and URAT1 values, with  $\hst$ motions having smaller uncertainties.   $\hst$ proper motions are relative to the reference frame.    No a priori values for proper motion and parallax were used as input for the target stars, only for the reference frame.  

\begin{deluxetable}{lrrrr}
\tablewidth{0in}
\tablecaption{Metal-Poor Target Star Parallaxes    \label{tab:pcomp}}
\tablehead{\colhead{HIP ID}&
\colhead{HST (mas)} &
\colhead{ HIP07 (mas)}  &
\colhead{TGAS17 (mas)}
}
\startdata
46120	&	15.011	$\pm$	0.119	&	15.2	$\pm$	0.98   & $14.938\pm 0.211$ \\
54639	&	11.116	$\pm$	0.113	&	15.69$\pm$	2.79   & $12.258\pm  0.232$\\
87062	&	8.205	$\pm$	0.110	&	9.59	$\pm$	2.21   & $8.383\pm 0.860$\\
87788	&	10.830	$\pm$	0.127	&	10.01$\pm$	2.79   & $10.972\pm 0.258$\\
98492	&	3.487	$\pm$	0.144	&	9.78	$\pm$	2.77   & $2.484\pm 0.371$\\
103269	&	14.118	$\pm$	0.099	&	14.86$\pm$	1.31   & $13.760\pm 0.220$\\
106924	&	14.474	$\pm$	0.100	&	15.11$\pm$	1.26 & \nodata \\
108200	&	12.397	$\pm$	0.085	&	12.33$\pm$	1.76 & \nodata \\
\enddata
\end{deluxetable}

\begin{deluxetable}{lcccccc}
\tabletypesize{\scriptsize}
\tablewidth{0in}
\tablecaption{Metal-Poor Target Star Proper Motions   \label{tab:pmcomp}}
\tablehead{\colhead{ID}&
\colhead{HST $\mu_{RA}$} &
\colhead{HST$\mu_{DEC}$} &
\colhead{HIP $\mu_{RA}$} &
\colhead{HIP$\mu_{DEC}$} &
\colhead{PPMXL cat $\mu_{RA}$} &
\colhead{PPMXL cat$\mu_{DEC}$}
}
\startdata
46120&203.410$\pm$0.116	&	1236.325$\pm$0.123	&	202.17$\pm$1.13	&	1236.93$\pm$1		&202.1$\pm$1	&	1237.2$\pm$1	\\
54639&	-567.057$\pm$	0.111	&	-517.165$\pm$	0.101	&	-567.16$\pm$3.62	&	-509.01$\pm$2.06		&568.5$\pm$2.2	&	517.7$\pm$2.1	\\
87062&244.342	$\pm$0.079	&	-366.504$\pm$0.088	&	242.57$\pm$2.29	&	-364.62$\pm$1.4		&244.8$\pm$1.6	&	-365.3$\pm$1.2	\\
87788&	5.88$\pm$0.078	&	-625.569$\pm$	0.088	&	6.67	$\pm$	2.85	&	-626.66$\pm$1.88	 &-0.6$\pm$	8.7	&	-629.1$\pm$5.6	\\
98492&			-186.256$\pm$	0.111	&	-186.256$\pm$0.115	&	-195.17$\pm$3.27	&	-194.57$\pm$2.61		&-186.6$\pm$6.2	&	-185.9$\pm$	6.2	\\
103269&			57.749$\pm$0.111	&	-389.392$\pm$0.117	&	57.62$\pm$1.24	&	-390.72$\pm$1.24	&	55.9	$\pm$1.1	&	-391$\pm$1	\\
106924&			-382.527$\pm$	0.092	&	232.585$\pm$0.106	&	-381.67$\pm$1.14	&	232.74$\pm$1.13	&	-381.6$\pm$1.1	&	232.7$\pm$1	\\
108200&		760.661$\pm$0.092	&	127.455$\pm$0.099	&	762.57$\pm$1.19	&	132.16$\pm$1.96		&761.6$\pm$1.3	&	128.5$\pm$2	\\
\enddata
\end{deluxetable}

 The key properties of our target stars are summarized in Table \ref{tabstars}.  High resolution spectroscopic abundances and reddening estimates are from \citet{OMalPress}. The reddening estimates are based upon the strength of the interstellar Na I line in the high resolution spectra of the target stars.  If no interstellar Na I lines were detected, then the reddening was estimated to be less than 0.001 mag, and was assumed to be zero for that star. For stars with non-zero reddening, an uncertainty of 10\% in the reddening was added in quadrature with the uncertainty in the parallax to determine the uncertainty in the absolute magnitudes.
  No \citet{lk73} corrections have been applied to Table \ref{tabstars} as the high accuracy of the parallaxes make these corrections  negligible
 (less than 0.0008 mag).  
  The stars are plotted in an color-absolute magnitude diagram in Figure \ref{fig:cmdhst}.  Below we briefly comment on each of the target stars.

\begin{deluxetable}{lrrccrcrr}
\tablecaption{Summary of Parallax Star Properties  \label{tabstars} }
\tablehead{
\colhead{ID} &
\colhead{V} &
\colhead{V -- I} &
\colhead{E(B -- V)} &
\colhead{(V-I)$_o$} &
\colhead{M$_\mathrm{V}$} &  
\colhead{M$_\mathrm{F606W}\tablenotemark{a}$} & 
\colhead{[Fe/H]}  & 
\colhead{[$\alpha$/Fe]\tablenotemark{b} } 
}
\startdata
46120   & 10.12 & 0.752 & 0.00 & 0.738 &   $6.00\pm0.02$ & 5.81 & $-2.22$ & 0.29\\
54639   & 11.38 & 0.914 & 0.00 & 0.914 &  $6.61\pm 0.02$& 6.37 & $-2.50$ & 0.19  \\
87062   & 10.57 & 0.833 & 0.06 & 0.750 &  $ 4.95\pm 0.03$& 4.77 &$-1.56$ &   0.1 \\
87788   & 11.30 & 0.852 & 0.00 & 0.852 & $6.47\pm0.03$  & 6.27 & $-2.66$ & 0.61\\
98492   & 11.57 & 0.811 & 0.11  & 0.659 &  $3.94\pm 0.09$  &3.78 &  $-1.40$ & 0.4\\
103269 & 10.27 & 0.771 & 0.00 & 0.771 &  $6.02\pm0.02$ & 5.83 &  $-1.83$ & 0.06\\
106924 & 10.36 & 0.803 & 0.00 & 0.803 & $6.16\pm0.02$ & 5.95 &  $-2.23$ & 0.23\\
108200 & 10.99 & 0.835 & 0.02 & 0.807 & $6.40\pm0.01$ & 6.19 &   $-1.83$ & 0.01\\
\enddata 
\tablenotetext{a}{Uncertainty the same as the M$_\mathrm{V}$ values. }
\tablenotetext{b}{Based upon the abundance of calcium, an $\alpha$-capture element. }
\end{deluxetable}

\begin{figure}
\centering
\includegraphics[scale=0.4]{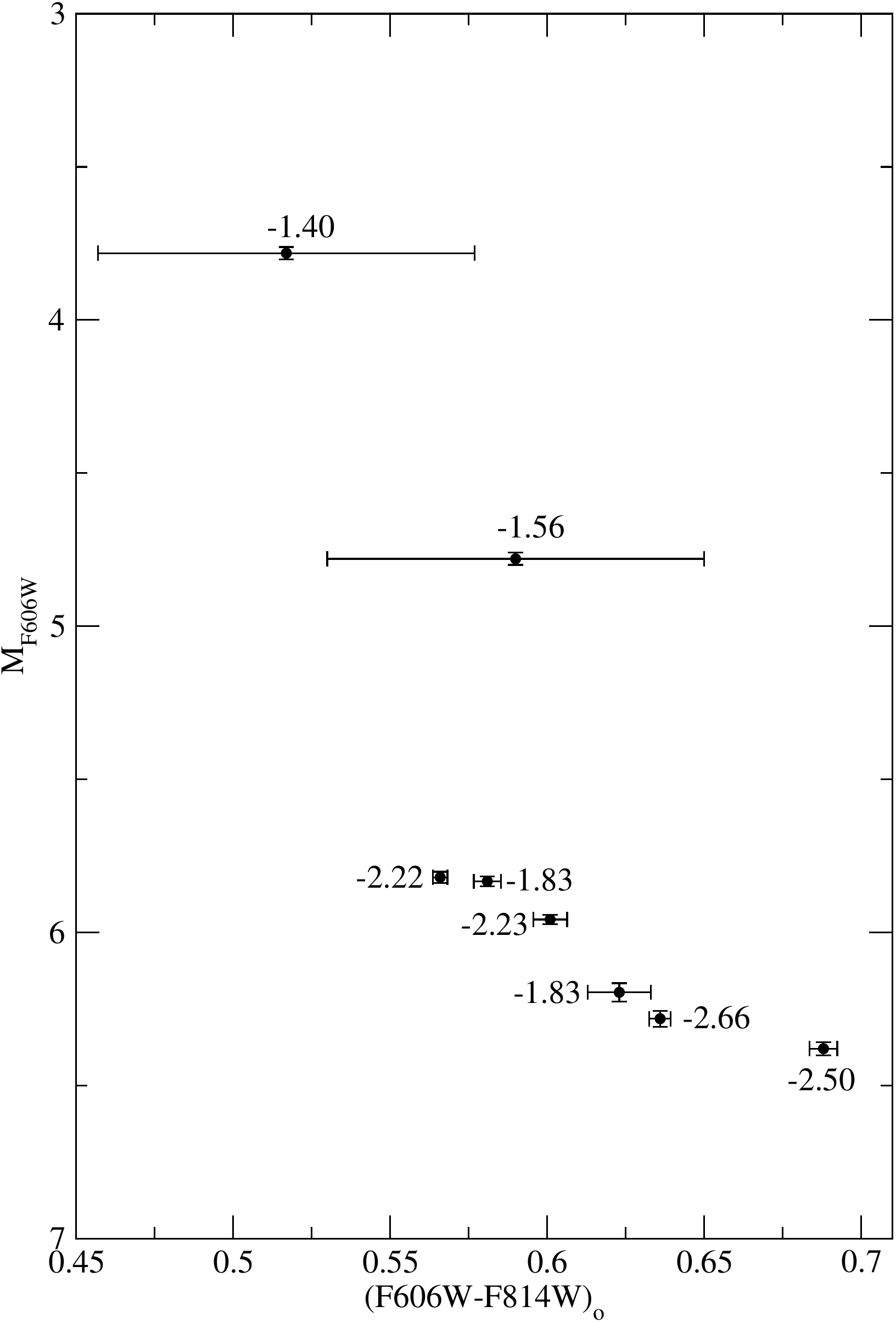}
\caption{The location of the stars in the color-absolute magnitude diagram. Each star is labeled with its $\feh$ value (see Table \ref{tabstars}). \label{fig:cmdhst} }
\end{figure}

HIP 46120 is typical main sequence star, with $\feh = -2.24$ and will serve as an excellent calibrator of the main sequence for the most metal-poor globular clusters.  The $\hst$,  $\hip$ and TGAS  parallaxes agree with each other for this star. 

HIP 54639 is the reddest (and intrinsically faintest) star in the sample.  This star has [Fe/H] = -2.5, and   isochrones would prefer that this star be fainter by about 0.2 mag.  The $\hst$ and TGAS parallaxes agree with each other, while the $\hip$ parallaxe agrees at the $1.6\,\sigma$ level for this star. The  $\hip$ parallax  of 15.679 mas (this is a faint star for $\hip$ with high error) would make this star fainter by 0.7 mag, the $\hst$  parallax  of 11.116 mas  has the star closer to the theoretical isochrone than the $\hip$ value. 


Initial observations of HIP 56291 had  $\hst$ target acquisition failures.   The first was because of incorrect target position and that was corrected, but then there were  star acquisition failures, most likely because the reference stars  in this field were all fainter than the catalogs indicated.  New reference stars were chosen,  but some proved to be double stars, which locked on to different components.   It may be possible to recover enough data to get a parallax, but it would be of significantly lower quality and as a result, a parallax was not determined for this star.  
  
 The $\hst$, TGAS and $\hip$ parallaxes for HIP 87062  are in agreement. 
High resolution spectra show the presence of interstellar Na I lines \citep{OMalPress} indicating that this star has $E(B-V) = 0.06$ magnitudes. As a result, the star is bluer than indicated by its colors, and the $\hst$ parallax indicate that the star is in  the main sequence turn-off, region (with $M_v = 4.95$) and hence not  suitable to use for main sequence fitting, or to test the stellar evolution isochrones.  The location of this star in a color-magnitude diagram differs considerably from the other stars, and from theoretical isochrones.  This suggests that we have may be using  an incorrect reddening for this star.  The Str\"{o}mgren photometry from \citet{schuster06} implies $E(B-V) = 0.12$ suggesting that our reddening value could be in error.  Using this reddening value puts the  star's location in the color-magnitude diagram in better agreement with other stars.

HIP  87788 is the most metal-poor star in our sample, with $\feh = -2.69$.  The $\hst$, TGAS and  $\hip$ parallaxes agree.  This star is on the main sequence, and with  no known reddening and an accurate parallax HIP 87788  is an excellent calibrator for the lowest metallicity stellar stellar models.

The parallax of the faintest star in our sample, HIP 98492 is discrepant at the $2.2\,\sigma$ level between  $\hst$  (3.49 mas) and  $\hip$  (9.78 mas).  This star is very faint for $\hip$.  The TGAS parallax (2.48 mas) is much closer to the HST value, but formally, the TGAS and HST parallaxes disagree $2.5\,\sigma$ level.  We note that \cite{stassun16} have found that the TGAS parallaxes are too small by 0.25 mas (when compared to eclipsing binaries), and offsetting the TGAS parallax by this amount would reduce the discrepancy with the HST parallax to $1.9\,\sigma$.  High resolution spectra 
 show the  presence of interstellar Na I lines \citep{OMalPress} indicating that this star has $E(B-V) = 0.11$ magnitudes. \citet{silva12} obtained Str\"{o}mgren photometry of this star and also estimate  $E(B-V) = 0.11$.  This star is in the main sequence turn-off or sub-giant branch region. As a result, it is not suitable for testing stellar evolution models.  Its location in the color-magnitude diagram is somewhat anomalous, which could be due to an incorrect reddening estimate.  
   
 HIP  103269 is a relatively metal-rich main sequence star in our sample, with $\feh = -1.85$. The $\hst$ and $\hip$ parallaxes agree at the $0.6\,\sigma$ level for this star and place it on the main sequence.  The TGAS parallax agrees at the $1.5\,\sigma$ level with the HST parallax.   Offsetting the TGAS parallax by 0.25 mas, brings the TGAS and HST parallax into agreement at the $0.4\,\sigma$ level.  With no reddening, and a well determined absolute magnitude, HIP 103269 is another excellent calibrator for low metallicity stellar models.
 
HIP  106924   is a typical main sequence star, with $\feh = -2.22$ and will serve as an excellent calibrator of the main sequence for the most metal-poor globular clusters.  The $\hst$ and $\hip$ parallaxes agree with each other at the $0.5\,\sigma$ level for this star. This star is not in the TGAS catalogue.

HIP  108200 is another  relatively metal-rich main sequence star in our sample (with $\feh = -1.83$). It has a small reddening of $E(B-V) = 0.02\,$mag based upon  the interstellar Na I lines \citep{OMalPress}.The $\hst$ and $\hip$ parallaxes agree at the $0.6\,\sigma$ level for this star and place it on the main sequence.  With a small reddening, and a well determined absolute magnitude, HIP 108200 is another good calibrator for low metallicity stellar models.  This star is not in the TGAS catalogue.

\section{Comparison to Theoretical Isochrones \label{SEC:ISO} }
The reliability of theoretical stellar models and isochrones is tested by comparison of these stars in color and magnitude to the Dartmouth  isochrones \citep{Dott2008}.  The theoretical uncertainty in stellar evolution models is estimated by constructing 2000 independent isochrones via a Monte Carlo analysis similar to \citet{BC2006}. This Monte Carlo analysis uses probability distributions for the various input parameters which have intrinsic uncertainties associated with their value.  For example, a comparison between different opacity calculations suggests that current high temperature ($T \ge 10^7\,$K) opacities are uncertain at the 3\% level and so when constructing the stellar models, the tabulated opacities are multiplied by a number randomly drawn from a Gaussian distribution with a mean of 1, and $\sigma = 0.03$.
The probability density distributions of the stellar evolution parameters varied in the Monte Carlo simulation are provided in Table~\ref{table:MCParams}.  
Starting with the probability density distributions used by  \citet{BC2006}, we updated the nuclear reaction rates, helium abundance,  and choice of model atmosphere to reflect more recent measurements/calculations. 
References for our choice of parameter distributions  are given in Table \ref{table:MCParams} and our choice of the mixing length distribution merits further discussion.

\begin{deluxetable}{l c c c}
\tablecaption{Monte Carlo Stellar Evolution Parameter Density Distributions \label{table:MCParams}}
\tablehead{
\colhead{Parameter} & 
\colhead{Distribution} & 
\colhead{Standard} & 
\colhead{Type}
}
\startdata
He mass fraction ($Y$)\dotfill & 0.24725 - 0.24757 & \citet{planck14} & Uniform\\
Mixing length\dotfill & 1.00 - 1.70 & N/A & Uniform\\
Convective overshoot\dotfill & $0.0H_p$ - $0.2H_p$ & N/A & Uniform\\
Atmospheric $T(\tau)$\dotfill & 33.3/33.3/33.3& \cite{edd26} or & Triinary\\
& & \cite{ks66}, or \\
 & & \cite{Haus1999}\\
Low-$T$ opacities\dotfill & 0.7 - 1.3 & \cite{ferg05} & Uniform\\
High-$T$ opacities\dotfill & 1.00\% $\pm$ 3\% ($T \geq 10^7$ K) & \cite{Iglesias96}  & Gaussian\\
Diffusion coefficients\dotfill & 0.5 - 1.3 & \cite{thoul94} & Uniform\\
$p+p \rightarrow {}^2\mathrm{H} + e^+ +\nu_e$& 1\% $\pm$ 1\% & \cite{adelberger11} & Gaussian\\
$^{3}\mathrm{He} + ^{3}\mathrm{He} \rightarrow ^{4} \mathrm{He} + 2p$\dotfill & 1\% $\pm$ 5\% &  \cite{adelberger11} & Gaussian\\
$^{3}\mathrm{He} + ^{4} \mathrm{He} \rightarrow ^{7} \mathrm{Be} + \gamma$\dotfill &1\% $\pm$ 2\% & \cite{deboer14} & Gaussian\\
$^{12}\mathrm{C} + p \rightarrow ^{13}\mathrm{N} + \gamma$\dotfill & 1\% $\pm$ 36\% & 
\cite{nacre13} & Gaussian\\
$^{13}\mathrm{C} + p \rightarrow ^{14}\mathrm{N} + \gamma$\dotfill & 1\% $\pm$ 15\% & 
\cite{chak15}& Gaussian\\
$^{14}\mathrm{N} + p \rightarrow ^{15}\mathrm{O} + \gamma$\dotfill & 1\% $\pm$ 7\% & 
 \cite{adelberger11}& Gaussian\\
$^{16}\mathrm{O} + p \rightarrow ^{17}\mathrm{F} + \gamma$\dotfill & 1\% $\pm$ 16\% & 
 \cite{adelberger98}& Gaussian\\
Triple-$\alpha$ reaction rate\dotfill & 1\% $\pm$ 15\% & \cite{nacre99} & Gaussian\\
Neutrino cooling rate\dotfill & 1\% $\pm$ 5\% & \cite{haft94} & Gaussian\\
Conductive opacities\dotfill & 1\% $\pm$ 20\% & \cite{hubbard69} plus  & Gaussian\\
& & \cite{canuto70}\\
\enddata

\tablecomments{As in Bjork \& Chaboyer (2006), parameters below Atmospheric $T(\tau)$ are treated as multiplicative factors applied to standard tables and formulas.}
\end{deluxetable}

The stellar models use a mixing length prescription to describe convection.  This theory has two free parameters (the mixing length itself, $\alpha$,  and the degree of overshoot past the formal edge of the convective boundary), which are varied in the Monte Carlo simulation.  Typically, the mixing length is determined by calibrating a solar model to match the observed properties of the Sun, and this solar calibrated mixing length is used for calculating metal-poor stellar models.  However, recent interferometric observations of the radius of the metal poor ($\feh  = -2.2$) star HD 140283 by \citet{creevey15} require that stellar models for this star use a mixing length which is substantially below the solar value.  Astroseismic studies using Kepler data have found a systematic metallicity dependance of the mixing length, with lower metallicity stars requiring the use of a lower mixing length \citep{bonaca12,tanner14}.    To take this into account, the  probability density distribution for the mixing length is taken to be a  uniform distribution with $ 1.00 \le \alpha \le 1.70$, which is substantially below the solar calibrated mixing length ($\alpha \simeq 1.9$) for our stellar models. 

The Monte Carlo simulation allows for the variation of both $[\alpha/\mathrm{Fe}]$ and [Fe/H].  Table \ref{table:Abund} provides mean [Fe/H] and [$\alpha$/Fe] density distribution used in the MC simulations for each metal-poor subdwarf with the [Fe/H] standard deviation given from \citet{OMalPress} as $\pm$0.08 dex. Two different pairs of stars have very similar abundances, and a single MC simulation was run for a given pair of stars.  In total, four different composition probability distributions were used, with 2000 MC isochrones being produced for each composition distribution and compared with the observations.

\begin{deluxetable}{l c l}
\tablecaption{Subdwarf Abundances \label{table:Abund}}
\tablehead{ 
\colhead{Subdwarf} & 
\colhead{[Fe/H]} & 
\colhead{[$\alpha$/Fe] distributions }
}
\startdata
HIP 46120 & $-2.22$ & 0.20 (50\%), 0.40 (50\%)\\
HIP 54639 & $-2.50$ & 0.20 (80\%), 0.40 (20\%)\\
HIP 87788 & $-2.66$ & 0.40 (25\%), 0.60 (50\%), 0.80 (25\%)\\
HIP 103269 & $-1.83$ & 0.20 (50\%), 0.40 (50\%)\\
HIP 106924 & $-2.23$ & 0.20 (50\%), 0.40 (50\%)\\
HIP 108200 & $-1.83$ & 0.20 (50\%), 0.40 (50\%)\\
\enddata
\end{deluxetable}

The stellar evolution tracks cover a stellar mass range of 0.30 to 1.00~M$_{\odot}$ and provide the physical parameters for the simulated star from the zero age main-sequence through its evolution along the red giant branch.  The stellar evolution tracks are used to produce isochrones in a standard fashion.  
The conversion from luminosities and effective temperatures to absolute magnitudes and colors used 
 both synthetic color-temperature transformations from the PHOENIX model atmosphere grid \citep{Haus1999} and the semi-empirical color-temperature relations of \citet{VC2003}.  
 The PHOENIX transformation isochrones  will be referred to as ISO-P, and the \citet{VC2003} transformed isochrones will be referred to as ISO-VC in the remainder of this paper.   

The location of the stars in the color-absolute magnitude plane were compared  to the isochrones for the stellar [Fe/H] using a reduced $\chi^2$ minimization analysis.  The $\chi^2$ for each MC isochrone is defined as:
\begin{equation}
\chi^2 = \bigg{(}\frac{d}{\sigma}\bigg{)}^2
\end{equation}
where $d$ is the perpendicular distance between the target star and the main sequence (MS) of a 12~Gyr isochrone in a CMD and $\sigma$ is the error associated with both the color and magnitude of our data.  The reduced $\chi^2$ for each model is 
\begin{equation}
\chi^2_{red} = \frac{\sum_{n=1}^5\chi^2}{K}
\end{equation}
where $K$ is the number of degrees of freedom, which in this case is  $K=5$  as HIP 54639 was discarded from the fit (see below) and so leaving five stars to fit to our models with zero fit parameters.

A comparison of ISO-P versus ISO-VC fits to the field stars is shown for HIP 103269 in Figure~\ref{fig:StarCMD}. The color of each field star was overestimated by the median ISO-P isochrone while underestimated by the ISO-VC distribution. This was  a general trend found for all the stars, except for HIP 54639.  The ISO-VC $\feh =-2.50$ models showed strong disagreement in their comparisons to HIP 54639, with the models being too red, while the ISO-P models gave an acceptable fit to this stars.  This is not too surprising, as  HIP 54639 is much redder compared the other main sequence stars in our sample.  HIP 54639 is also the faintest star in our sample, however there is no indication from deep globular cluster color-magnitude diagrams \citep{Sara2007} that the main sequence becomes markable redder at these magnitudes. As a result, it is most likely that the position of HIP 54639 is anomalous, perhaps due to it being an unresolved binary.  As a result, HIP 54639 was removed from subsequent analysis for both color-temperature transformations.

\begin{figure}
\centering
\includegraphics[scale=0.48]{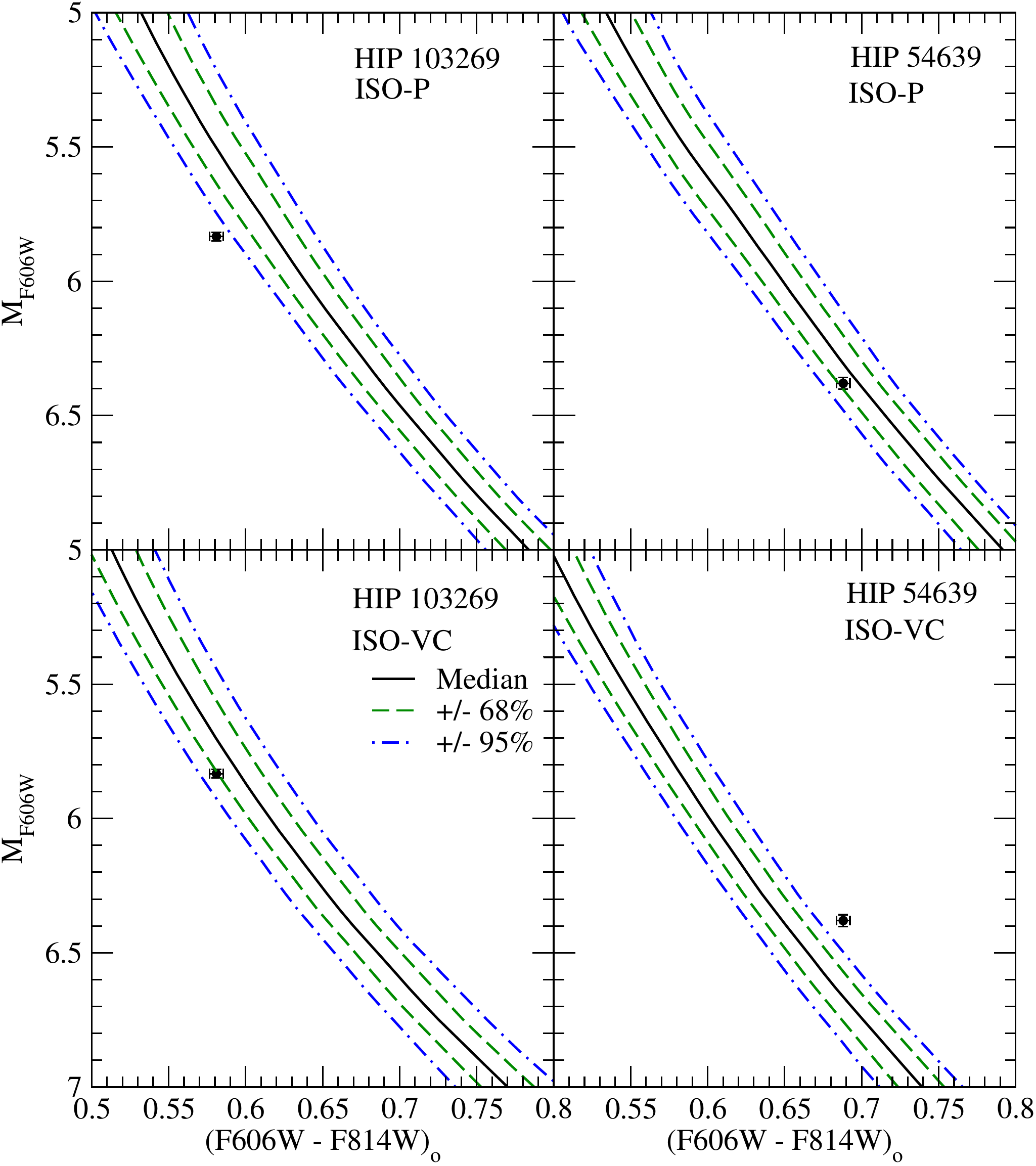}
\caption{Comparison of  HIP 103269 (left, with $\feh = -1.83$) and HIP 54639 (right, with $\feh = -2.50$) to Dartmouth stellar evolution isochrones. Shown here are the median isochrone (black solid), $1\,\sigma$ (green dashed) and $2\, \sigma$ (blue dot-dash) deviations for ISO-P (upper) and ISO-VC (lower).  The color of HIP 103269  was over predicted by more than 95\% of the ISO-P isochrones, while the color matches within $1\,\sigma$ the ISO-VC isochrones.  This is true for the four other main sequence stars, with the only exception being HIP 54639 whose color matches within $1\,\sigma$ the ISO-P isochrones.  
\label{fig:StarCMD} }
\end{figure}

The median deviation of the ISO-P models from the stars was $8.5\,\sigma$,  with 99.8\% falling outside of $2\,\sigma$ and 98\% falling outside $3\,\sigma$.  For the ISO-VC models, the median deviation from the stars was $2.1\sigma$ ,with 56\% of the isochrones falling outside of $2\sigma$ and only 21\% outside of $3\sigma$.   The distribution of $\chi^2$ values is shown in Figure~\ref{fig:HistHST} for  two isochrones sets.  It is clear from this analysis that the \cite{VC2003} color-transformation does a much better job of matching the observed properties of metal-poor main sequence stars, and is the preferred color-transformation.  These isochrones and their construction parameters will be used in determining the distances and ages of several GCs.

\begin{figure}
\centering
\includegraphics[scale=0.6]{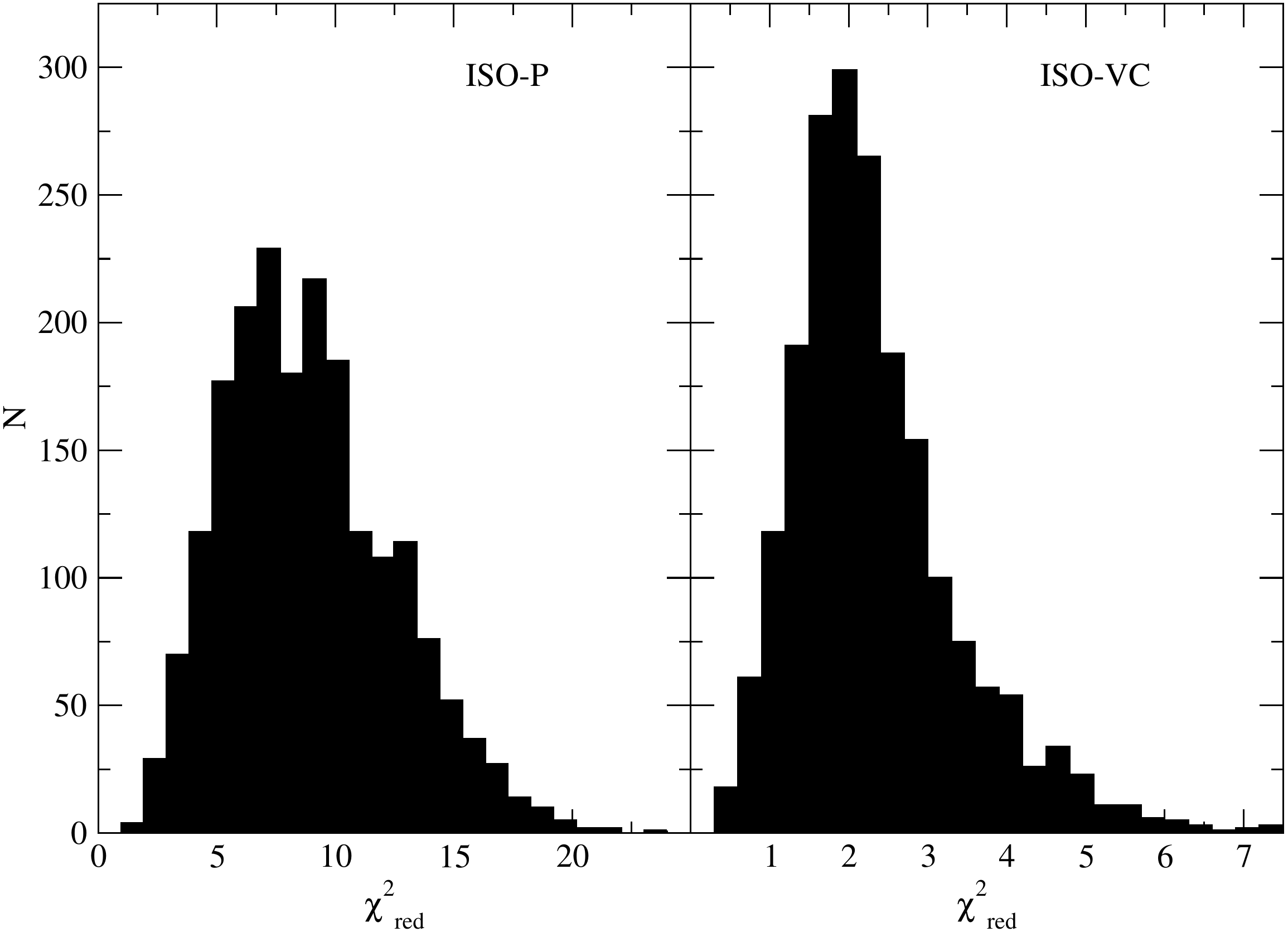}
\caption{\small{Distribution of $\chi^2$ values shown for ISO-P (left) and ISO-V (right) using HST magnitudes.} \label{fig:HistHST}}
\end{figure}


\subsection{Stellar Parameters}
To understand the relationship between the input parameters and the goodness of fit to the parallax stars, it is instructive to perform an analysis of covariance (ANCOVA) on our results.  
An ANCOVA  was conducted with the statistical software R \citep{R2005} to test the effects and interactions of the input parameters on the fit of the models to the very metal-poor field stars.  Table~\ref{table:ancova} provides the results of this analysis in the form of the minimal adequate model, both for the entire isochrone set, and just the ICO-VC isochornes.  As was obvious in our initial investigation into the model fits (see Figure~\ref{fig:HistHST}), the choice of color temperature relation produces very different distributions of fits for the same model parameters.  The ANCOVA confirms this result, but also finds the underlying effects due to the input parameters themselves.  Of the 20 input parameters used in constructing these models, six  individual parameters produce highly significant effects on the resulting reduced $\chi^2$ fit.   As an example, the mixing length has a very strong effect on the model fit ($t=-30.5$).  The effect is clearly evident when we look at the reduced $\chi^2$ as a function of mixing length as shown in Figure~\ref{fig:cmix}, highlighting the advantages of the ANCOVA method of extracting this information from the complex dataset.

\begin{deluxetable}{lrrrr}
\tablecaption{ANCOVA of Model Input Parameters \label{table:ancova}}
\tablehead{
\colhead{Coefficients} & 
\colhead{Estimates} & 
\colhead{Std.\ Error} & 
\colhead{$t$-Value} & 
\colhead{Prob($<|t|$)} \\ 
}
\startdata
\multicolumn{5}{c}{All Models}\\
Mixing length & $-7.47$ & 0.211 & $-35.46$ & $< 2\times 10^{-16}$ \\
$p+p \rightarrow {}^2\mathrm{H} + e^+ +\nu_e$  & 6.46 & 0.270 & 23.9 & $< 2\times 10^{-16}$  \\
Diffusion Coefficients & 2.41 & 0.293 & 8.25 & $< 2\times 10^{-16}$  \\
Atmospheric $T(\tau)$ & 0.111& 0.0196 & 5.69 & $1.4\times 10^{-8}$  \\
$^{3}\mathrm{He} + ^{4} \mathrm{He} \rightarrow ^{7} \mathrm{Be} + \gamma$ 
 & 11.90 & 2.29 & 5.21 & $2.0\times 10^{-7}$  \\
High-$T$ opacities & 2.73 & 1.02 & 2.67 & 0.0076  \\
\hline
\multicolumn{5}{c}{Only ISO-VC Models}\\
Mixing Length &  $-1.37$ & 0.097 & $-14.18$ &  $< 2\times 10^{-16}$ \\
$p+p \rightarrow {}^2\mathrm{H} + e^+ +\nu_e$   & 1.18 & 0.124& 9.47 & $< 2\times 10^{-16}$\\
$[\mathrm{Fe/H}]$ & 1.10 & 0.245 & 4.50 & $7.1\times 10^{-6}$ \\ 
$[\alpha/\mathrm{Fe}]$ & 0.502 & 0.195 & $ 2.57 $ & 0.010 \\
$^{3}\mathrm{He} + ^{4} \mathrm{He} \rightarrow ^{7} \mathrm{Be} + \gamma$  & $-2.61$ & 1.05 & 
$-2.48$ & 0.013\\
High-$T$ opacities & 1.63 & 0.667 & 2.45 & 0.014  \\
\enddata
\end{deluxetable}

\begin{figure}
\centering
\includegraphics[scale=0.5]{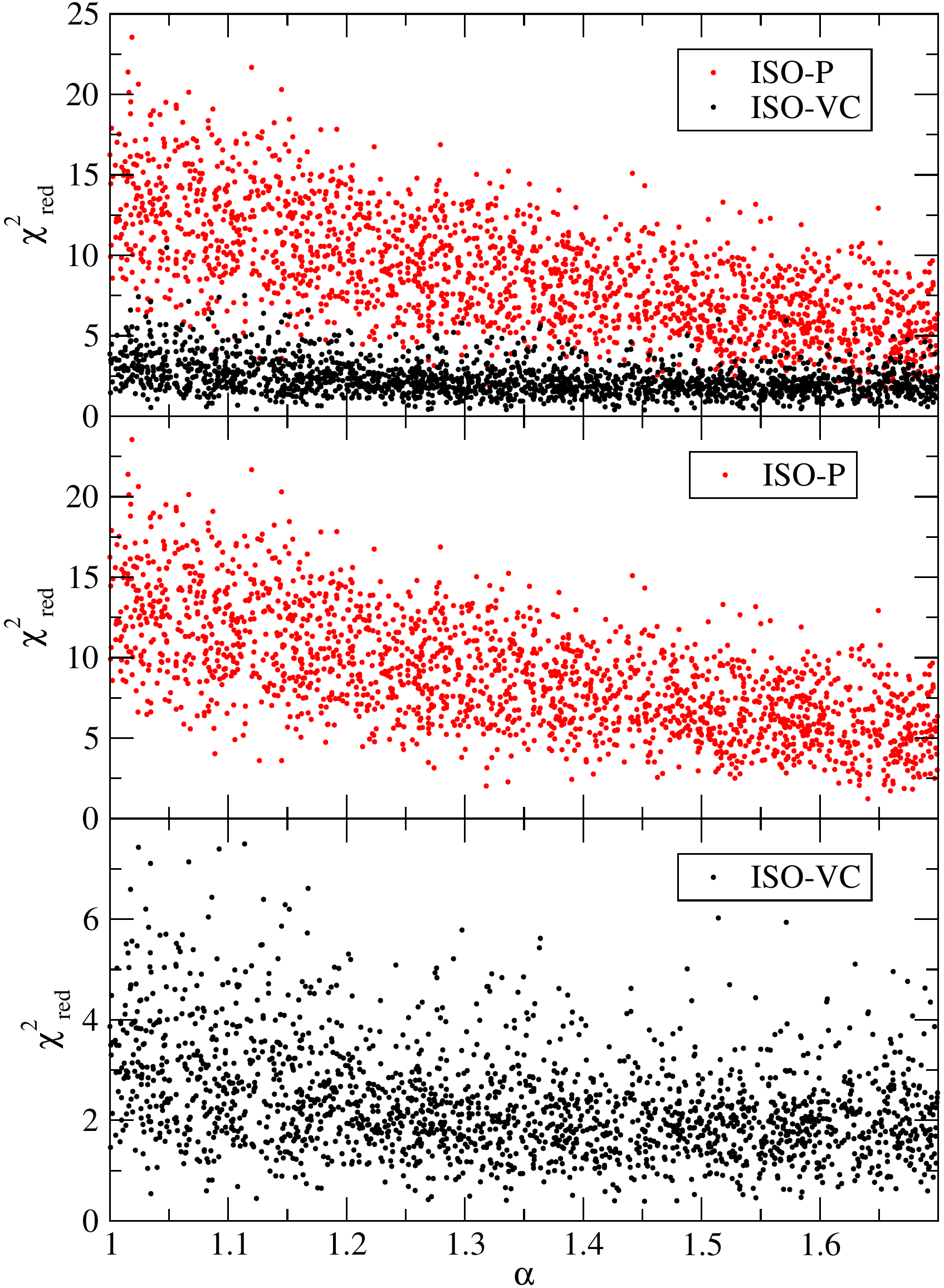}
\caption{\small{Reduced $\chi^2$ as a function of mixing length, $\alpha$.  It is clear the color-temperature relation plays a large role in determining the goodness of fit and this effect must be accounted for in determining the best values for each parameter.} \label{fig:cmix}}
\end{figure}

Regardless of which isochrone set we adopt, the most important physical parameters in determining the goodness of fit to the parallax stars are the mixing length and the $p + p$ reaction rate.
These have been identified in previous studies as important uncertainties in lower mass, metal-poor stellar models \citep[e.g.][]{Chab02}.  The only surprise result is that we find the uncertainty in the ${}^3\mathrm{He} +   
{}^{4}\mathrm{He} \rightarrow ^{7}\mathrm{Be} + \gamma$ reaction has an important impact on our fits to metal-poor stars.  This is the slowest reaction in the PP-II chain, which becomes important at higher temperatures.  The significance of this result is much lower with the ICO-VC models, and it would be worth investigating this issue in more detail with dedicated models. 


\section{Distances and Ages of GCs \label{SEC:GC} }
Main sequence fitting, whereby the absolute magnitude of stars with well determined parallexes are compared to the apparent magnitude  of the main sequence in distance star clusters, is a commonly used technique to determine the distances to star clusters. As mentioned in the introduction, a number of groups used $\hip$ parallaxes of metal-poor stars to determine the distances to globular clusters  \citep[e.g.][]{reid97,gratton97,pont98,chaboyer98,caretta00,grundahl02,gratton03}.  Since these works, it has become clear than many, if not all globular clusters contain multiple stellar populations \citep[e.g.\ see reviews by][]{piotto09,gratton12}.  These different stellar populations are characterized by different chemical abundances, and it is likely that self-enrichment has occurred in many, if not all globular clusters.   These multiple stellar populations have been identified using both spectroscopy \citep{gratton12} and photometrically \citep{piotto09}. It is unlikely that field stars experienced the same formation scenario, and the chemical composition of field stars is most similar to the primordial population of stars found in globular clusters.  In order to use main sequence fitting to determine distances to GCs, it is important that one uses GC stars which have a primordial composition.

Photometric studies have found that when a blue filter is chosen (such as F225W or F336W) then the apparently tight principle sequences observed in ground based color-magnitude diagrams (CMDs) are often split into multiple sequences indicative of different stellar populations.   For example, some globular clusters contain populations with enhanced helium abundances 
\citep[e.g.][]{milone12a,milone15}, and  the different populations trace slightly different main sequences, with the helium enriched stars being redder on the main sequence than the stars with a primordial helium abundance.  However, these distinct main sequences are often just seen when using blue filters, and when using the F606W and F814W filters, all the stars within the globular clusters 
NGC 6397 \citep{milone12a} and NGC 6752 \citep{milone13} fall along the same main sequence.   This is not always the case, and in 47 Tuc \citep{milone12b}, NGC 2808 \citep{milone15a} and NGC 7089 \citep{milone15b} one sees a small offset on the main sequence between the helium enriched population and the primordial population when looking at the CMDs in the F606W and F814W filters.  

The  \emph{HST} GC Treasury Project \citep{Sara2007}, which obtained deep CMDs of a 65 globular clusters, was conducted in the F606W and F814W filters.  The photometry from this project was downloaded from MAST, and used to determine the median ridge-line along the main sequence and subgiant branches for a number of GCs. In order to determine if this ridge-line traces the primordial stars, one needs to examine UV photometry to identify the different stellar populations.  \cite{piotto15}  have obtained UV photometry for most of GCs observed by \cite{Sara2007}.  Careful analysis of this photometry has been used to identify multiple populations in most GCs studied \citep{milone16}.  However, 
this photometry is not publicly  available at this time, despite claims to the contrary in \cite{piotto15}.   Without having access to the UV photometry, which can be used to determine which stars are primordial, we need to adopt another approach to determine the location of the primordial main sequence in the F606W--F814W CMD. 

UV CMDs are available for the clusters at Piotto's website, \url{http://groups.dfa.unipd.it/ESPG/treasury.php} in graphical form. In looking at these diagrams, clusters which show an offset between the primordial and helium enhanced stars on the main sequence in their F606W--F814W CMDs (47 Tuc, NGC 2808 and 7089), clearly show multiple main sequence populations in their UV CMDs, while for those clusters whose main sequence multiple populations are all on the same ridge-line in their F606W--F814W CMDs (NGC 6397 and 6752)  one does not see evidence for multiple main sequence populations in the graphical UV CMDs.  To ensure that the median ridge-lines in the F606W--F814W CMDs are tracing the primordial population, UV CMDs  of a number of metal-poor GCs were examined, and those which showed no obvious evidence for multiple main sequence populations in their graphical UV CMDs were used in our main sequence fitting study.  The location of the principle sequences (main sequence and  turn-off region) in these clusters F606W--F814W CMDs is not affected by the presence of multiple populations, and the  observed location of the principal sequences is indicative of the primordial population.  

The location of the main sequence (in all filters) is sensitive to the abundance of iron and the $\alpha-$capture elements. To take this into account, distances will only be determined to GCs which have similar abundances to the HST parallax stars. The calibrating parallax  stars have similar, although not identical, abundances to the primordial population of stars in our chosen GCs.  To take into account the small differences in [Fe/H] and [$\alpha$/Fe] between the HST parallax stars and the primordial GC stars, we do not directly compare the HST parallax  stars to the main sequence stars in globular clusters. Instead, as described in the previous section, the HST parallax stars were used to determine what parameters yield theoretical isochrones which correctly predict the main sequence location of metal-poor stars.  These same set of parameters, but with [Fe/H] and [$\alpha$/Fe] values appropriate to a given GC are then used to generate isochrones and these isochrones are used to determine the distance and age of that GC. 


\subsection{Sample Clusters}
A sample of  nine metal-poor ($\feh < -1.9$) and relatively un-reddened  ($E(B-V)\leq0.10\,$mag)  GCs  with photometry from  the \emph{HST} GC Treasury Project \citep{Sara2007} were selected from the Harris (1996 version 2010) GC catalog for the  main sequence fitting analysis.  The photometry for these clusters was obtained with the same filters on ACS/WFC  as our target metal-poor field stars.  Hence, the photometry of both the GCs and our field stars are on the exact same system, which removes one possible source of systematic error from our analysis.  

In studying the photometry of similar metallicity GCs, the \citet{Harris1996} reddening values give inconsistent colors of the RGB.  This is demonstrated in Figure~\ref{fig:RGBcolor} where the red giant branch (RGB)  color of NGC~5024 (black, $\feh =-2.10$ is compared to  to NGC~6101 (red, 
$\feh = -1.98$).  The top panel of Figure~\ref{fig:RGBcolor} uses the Harris (1996, version 2010) reddening and a relative distance modulus of 0.75 mag such that the MSs overlap. GCs of similar metallicity are expected to have comparable RGB colors and HB magnitudes when adjusted to the same relative distance; the fact that this is not seen suggests the reddening of one or both of these clusters may be incorrect.  Due to this issue, the \citet{Dutra2000} far-infrared (FIR) reddening values (based on the reddening maps of \citet{Schlegel1998} with reddening errors of $\pm0.01$ mag) were adopted for the MS fitting.  The comparison of RGB colors for NGC~5024 and NGC~6101 using the FIR reddening values shows much better agreement in the bottom panel of Figure~\ref{fig:RGBcolor} using a relative distance modulus of 0.55 mag to match the MSs.

\begin{figure}
\centering
\includegraphics[scale=0.5]{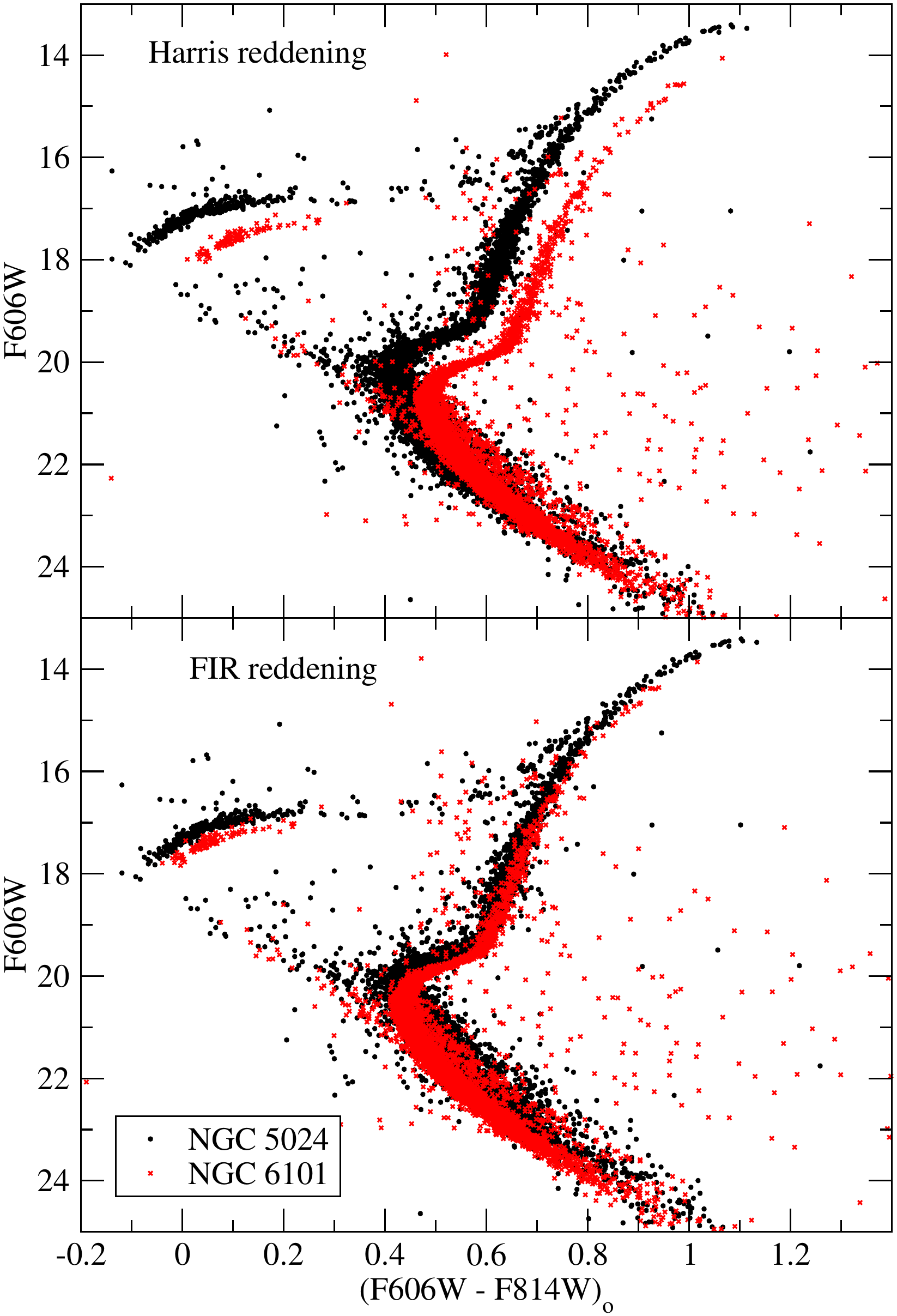}
\caption{\small{GCs of similar metallicity expected to have comparable RGB colors and HB magnitudes when adjusted to same relative distance.  \emph{Upper} - Dereddened CMDs of NGC~5024 (black) and NGC~6101 (red) using Harris (1996, version 2010) adjusted to same relative distance.  We do not see overlapping RGBs nor HBs, suggesting reddening of one or both may be incorrect.  \emph{Bottom} - Dedreddened CMDs with Dutra et al. (2000) FIR reddening show much better agreement.} \label{fig:RGBcolor}}
\end{figure}

Some may question the use of FIR reddening for objects within the Milky Way due to the fact that FIR reddening values should represent the integrated dust column density throughout the whole Galaxy in any direction; therefore, FIR reddening values may over-estimate the reddening for nearby objects.  However, the GCs used for this study are sufficiently far away and mostly out of the plane of the Galaxy that the reddening should not be over-estimated by the FIR reddening values. The FIR reddening values agree fairly well with the reddenings found by \citet{Dott2010}  and \citet{Van2013}  who studied the cluster CMDs in detail to determine the cluster ages.

Table~\ref{table:Clusters} provides information on the clusters' locations and reddening from \citet{Dutra2000} along with metallicity from \citet{Harris1996} and the metallicity bin used in this study.
The \citet{sirianni05} reddening relation for ACS/WFC  of $E(F606W - F814W) = 0.984\, E(B - V)$ was used to convert the \citet{Dutra2000} reddening to the HST photometric system. 
 
 
 \begin{deluxetable}{l c c c c c c c c}
\tablecaption{MS-fitting Cluster Data \label{table:Clusters}}
\tablehead{
\colhead{Cluster} & 
\colhead{Messier} & 
\colhead{l($^\circ$)} & 
\colhead{b($^\circ$)} & 
\colhead{d$_{\mathrm{Sun}}$(kpc)} & 
\colhead{$E(B-V)$} & 
\colhead{$E(B-V)_{FIR}$ }& 
\colhead{Harris [Fe/H]} & 
\colhead{[Fe/H] bin}
}
\startdata
NGC 4590 & M 68 & 299.63 & 36.05 & 10.2 & 0.05 & 0.06 & -2.23 & -2.25\\ 
NGC 5024 & M 53 &332.96 & 79.76 & 18.3 & 0.02 & 0.03 & -2.10 & -2.10\\
NGC 5053 & & 335.69 & 78.94 & 16.4 & 0.04 & 0.02 & -2.27 & -2.25\\
NGC 5466 & & 42.15 & 73.59 & 17.0 & 0.00 & 0.02 & -2.31\tablenotemark{a} & -2.36\\
NGC 6101 & & 317.75 & -15.82 & 15.3 & 0.05 & 0.10 & -1.98 & -1.96\\
NGC 6341 & M 92 & 68.34 & 34.86 & 8.2 & 0.02 & 0.02 & -2.34 & -2.36\\
NGC 6809 & M 55 & 8.80 & -23.27 & 5.4 & 0.07 & 0.14 & -1.94 & -1.96\\
NGC 7078 & M 15 & 65.01 & -27.31 & 10.3 & 0.10 & 0.11 & -2.37 & -2.36\\
NGC 7099 & M30 & 27.18 & -46.83 & 8.0 & 0.03 & 0.05 & -2.27 & -2.25\\
\enddata
\tablenotetext{a}{[Fe/H] of NGC 5466 taken from \citep{caretta09}.}
\end{deluxetable}

 
\subsection{MS-Fitting Distances}
MS-fitting distances were determined for each cluster, taking into account observational uncertainty (including reddening errors). The best-fitting distance modulus was based on a fit of the median observed MS ridge-line shifted in both color and magnitude to the isochrone MS. For a given cluster, by visual inspection of the de-reddened CMD, the MS was defined in color and magnitude. The median MS ridge-line (within the color range of the field stars with parallaxes) was then calculated by determining the median color in 0.2 magnitude overlapping bins, therefore ensuring the median color found in each bin is not isolated.  The median ridge-line was selected to ensure that the results are not affected by the red binary sequence found in GCs.

This median MS ridge-line is used to determine the distance modulus for each cluster using the 12 Gyr isochrones in our suite of models.  The shape of the MS in the color range of field stars did not constrain the reddening. Shifting the median MS ridge-line in both color and magnitude led to equally well-fitting distance modulus for any color in the $E(F606W-F814W)$ range for that cluster. Therefore, we are able to calculate the distance modulus for the cluster based on the median $E(F606W-F814W)$ and propagate the uncertainty in the reddening to the uncertainty in the distance modulus using standard techniques. Since $\mathrm{A}_v=3.1\,E(B-V)$, the $\pm0.01$ uncertainty in reddening corresponds to a $\pm0.03$ uncertainty in the distance modulus for a given isochrone. The distance modulus determined with a given isochrone spans a range of $\sim 0.7$ mag, in the case of M92 the distance modulus ranges from 14.90 to 15.61 mag using ISO-P or from 14.60 to 15.21 mag using ISO-VC.

Not all isochrones have the same goodness of fit to the calibrating field stars with parallaxes, and the resultant distance modulus  need to be weighted to take into account their probability of correctly representing actual stars.  
Specifically, a $p$-value is calculated for each isochrone based on its $\chi^2$ vaule using $K=5$ degrees of freedom.  The weight applied to each isochrone was obtained by normalizing the $p$-value distribution.  Table~\ref{table:HSTresults} gives the weighted mean distance modulus for each cluster.  Since the ISO-P isochrones gave very poor fits to the calibrating field stars, they have very little weight when  combined with the ISO-V isochrones, and the combined result, which represents our best estimate for the distance modulus to each cluster, is heavily weighted in favor of the ISO-VC isochrones.  The average uncertainty in distance modulus is $\sigma_{DM}=0.10$ mag which incorporates the photometric uncertainty and the uncertainty in stellar models.  We see an offset of $\sim$0.15 mag between ISO-P and ISO-VC distance modulus determinations in the sense the ISO-P distance moduli are greater than the ISO-VC distance moduli.
To illustrate the technique, Figure~\ref{fig:M92_comp} compares data for M92 (NGC 6341) to the median ISO-VC isochrone and the calibrating field stars assuming the best fitting distance modulus shown in Table~\ref{table:HSTresults}.


\begin{deluxetable}{lcccccccc}
\tabletypesize{\small}
\tablecolumns{9}
\tablewidth{0pc}
\tablecaption{Cluster Distance Modulus and Age Weighted by HST Fit \label{table:HSTresults}}
\tablehead{
\colhead{}& \multicolumn{2}{c}{ISO-P} & & \multicolumn{2}{c}{ISO-V} & & \multicolumn{2}{c}{Combined}\\
\cline{2-3}\cline{5-6}\cline{8-9}
\colhead{Cluster} & 
\colhead{$(m-M)_{F606W}$} & 
\colhead{Age (Gyr)} && 
\colhead{$(m-M)_{F606W}$} & 
\colhead{Age (Gyr)} && 
\colhead{$(m-M)_{F606W}$} & 
\colhead{Age (Gyr)}}
\startdata
NGC 4590 & $15.63\pm0.13$ & $10.2\pm1.2$ && $15.52\pm0.11$ & $12.4\pm1.2$ && $15.52\pm0.11$ & $12.4\pm1.2$\\
NGC 5024 & $16.84\pm0.13$ & $10.3\pm1.3$ && $16.67\pm0.10$ & $13.2\pm1.2$ && $16.68\pm0.10$ & $13.2\pm1.2$\\
NGC 5053 & $16.58\pm0.13$ & $10.2\pm1.2$ && $16.47\pm0.10$ & $12.3\pm1.2$ && $16.47\pm0.10$ & $12.3\pm1.2$\\
NGC 5466 & $16.42\pm0.13$ & $11.0\pm1.3$ && $16.24\pm0.10$ & $13.9\pm1.1$ && $16.24\pm0.10$ & $13.9\pm1.1$\\
NGC 6101 & $16.35\pm0.14$ & $10.7\pm1.5$ && $16.21\pm0.11$ & $13.4\pm1.4$ && $16.21\pm0.11$ & $13.4\pm1.4$\\
NGC 6341 & $15.07\pm0.13$ & $10.3\pm1.0$ && $14.88\pm0.10$ & $13.2\pm1.1$ && $14.88\pm0.10$ & $13.2\pm1.1$\\
NGC 6809 & $14.50\pm0.14$ & $9.3\pm1.5$   && $14.35\pm0.11$ & $11.9\pm1.4$ && $14.35\pm0.11$ & $11.9\pm1.4$\\
NGC 7078 & $15.93\pm0.12$ & $9.2\pm1.1$   && $15.74\pm0.10$ & $12.1\pm1.1$ && $15.74\pm0.10$ & $12.1\pm1.1$\\
NGC 7099 & $15.12\pm0.13$ & $10.1\pm1.2$  && $15.02\pm0.10$ & $12.2\pm1.2$ && $15.02\pm0.10$ & $12.2\pm1.2$\\
\enddata
\end{deluxetable}

\begin{figure}
\centering
\includegraphics[scale=0.6]{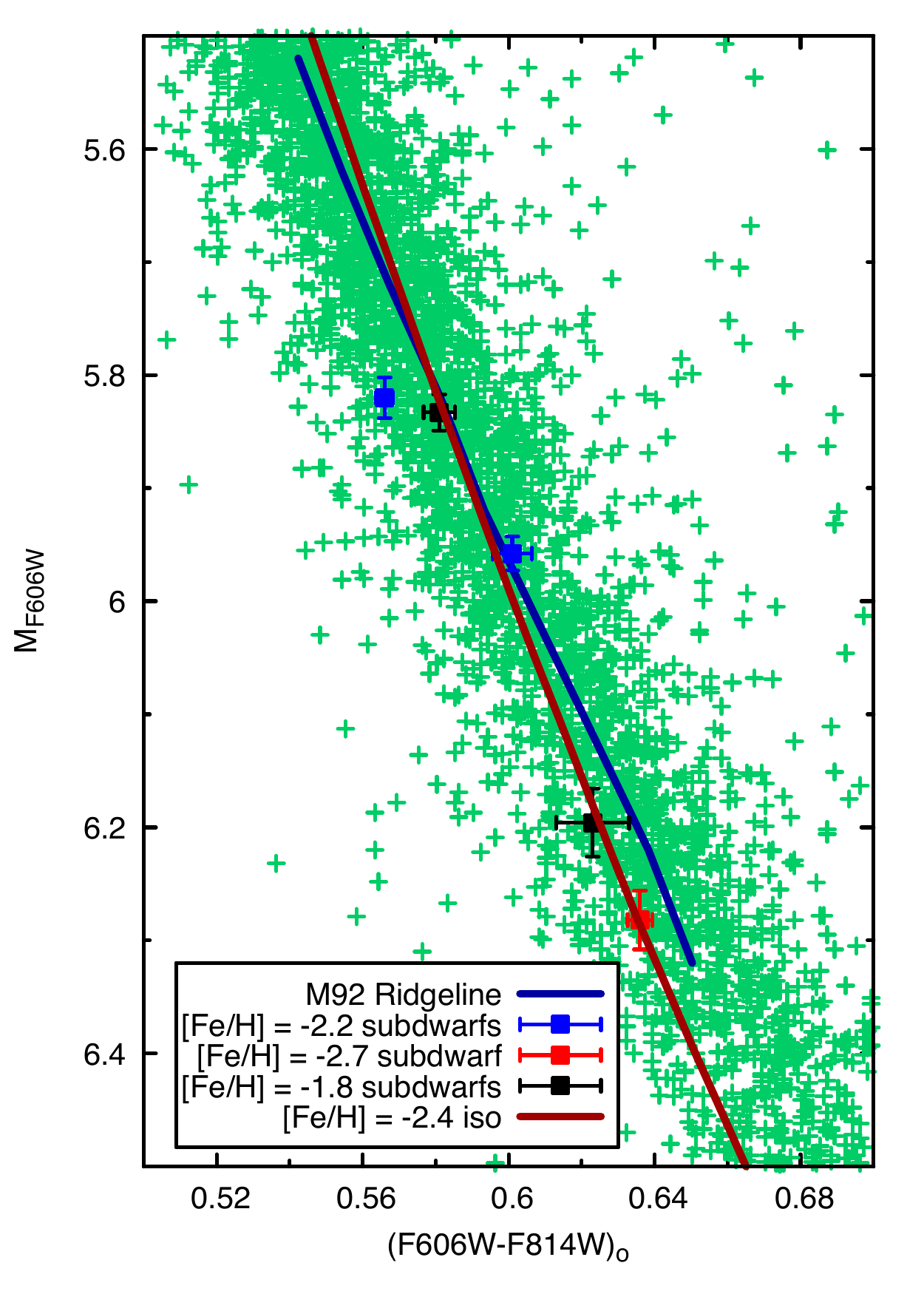}
\caption{\small{Comparison between the M92 data (green points, median ridgeline of the data blue solid line) from \citet{Sara2007} with the calibrating field stars and the median ISO-VC isochrone from our Monte Carlo model with the M92 composition. 
\label{fig:M92_comp} } }
\end{figure}

\subsection{Cluster Ages}
The cluster distance modulus given by each individual isochrone is used to determine a distribution of ages for the cluster. The location of the sub-giant branch (SGB) in the clusters is compared to the SGB of the isochrone at ages ranging from 8 Gyr to 15 Gyr.  The magnitude of the SGB is defined in \citet{Chab1996} as the point brighter than the main-sequence turn-off (MSTO) and 0.05 mag redder.  \citet{Chab1996} showed the SGB provides the same level of theoretical uncertainty as the MSTO but is easier to measure on observational CMDs which leads to lower observational uncertainty.  We visually inspect the CMDs of the clusters to determine the SGB and use a linear regression of isochrone age vs. SGB magnitude to find the cluster age.  The same weighting scheme is applied to this distribution of ages to derive our final ages based on the goodness of fit of each isochrone to the field star data.  The ages for each GC are provided in Table~\ref{table:HSTresults}, with average uncertainties of 
$\sim 1.2\,$Gyr.

The offset of $\sim$0.15 mag in distance modulus between ISO-P and ISO-VC plays directly into the 2.6 Gyr offset we see in the ages between these two isochrone sets. Note that the inverse relationship between the distance modulus and age determined for each cluster is inherent in this analysis and is shown very clearly for M92 in Figure~\ref{fig:DistAge} which plots the ISO-VC  age and distance muduli distributions (on a relative scale).  
\begin{figure}
\centering
\includegraphics[scale=0.5]{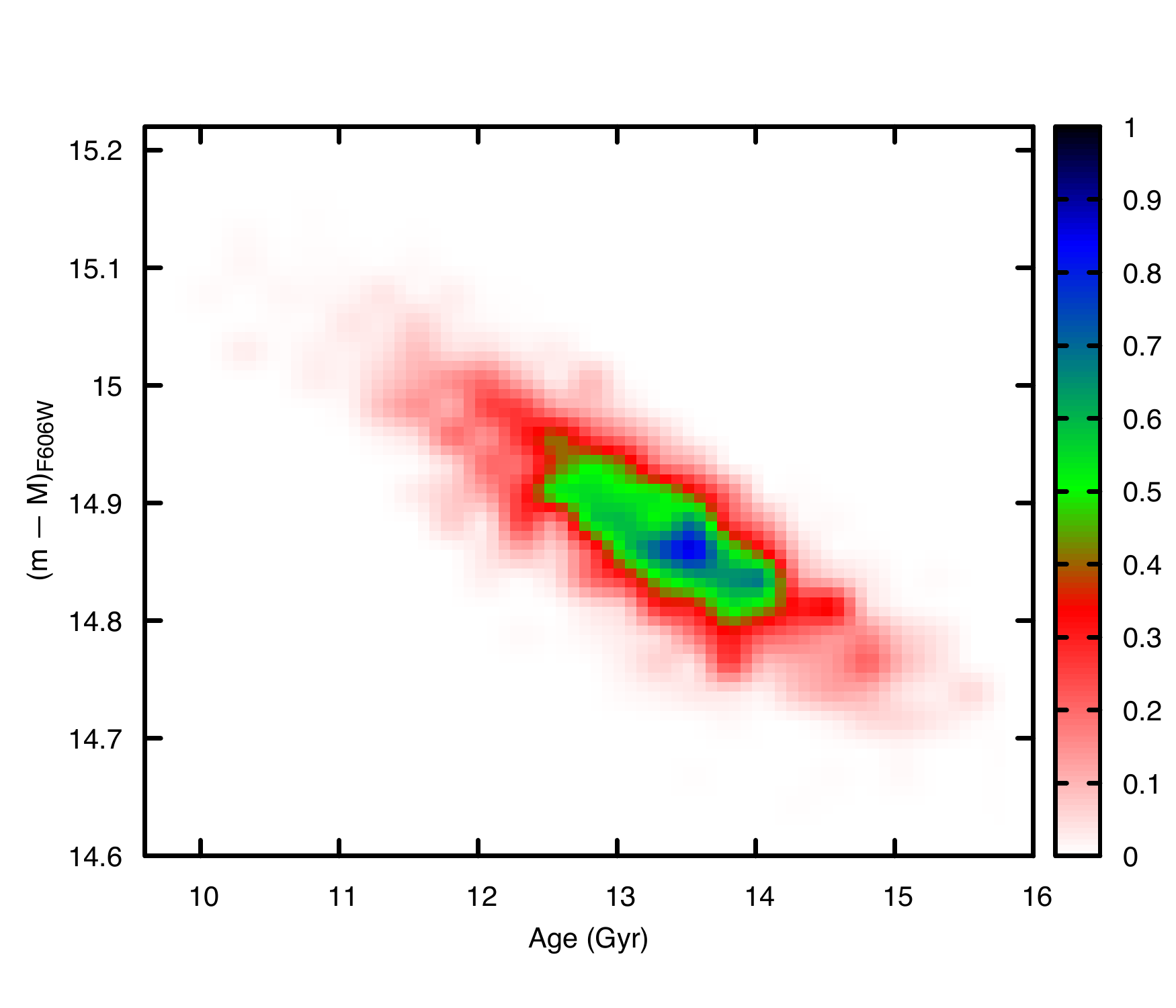}
\caption{\small{Probability distribution of the estimated age and distance modulus of M92 (on a relative scale) using  ISO-VC  isochrones.  The inverse relationship between distance modulus and age is inherent in this analysis and is very clear.} \label{fig:DistAge}}
\end{figure}

\subsection{Discussion \label{sec:discus} }
Several sources provide estimates of GC distances and ages using varying methods of analysis.  The \citet{Harris1996} GC catalog, which was used to determine initial estimates of [Fe/H] and $E(B-V)$ in this study, also provides distance moduli based on the horizontal branch (HB) magnitude observed in the CMD.  On average, our best estimates for the distance moduli (from the combined models) are 0.30 mag larger than those listed in \citet{Harris1996}. This difference is too large to be explained by simply the use of FIR reddening values. As shown earlier, a 0.01 mag increase in $E(B-V)$ corresponds to 0.03 mag increase in the distance modulus. The majority of our clusters used reddenings which are only 0.02 mag larger than listed by \citet{Harris1996}, with a few exceptions increasing to 0.05 mag, suggesting that the difference in reddening estimates only accounts for  $\sim 0.08\,$mag of the difference in distance moduli.  

\citet{Dott2010} determine distance moduli by fitting isochrones to  same \emph{HST} photometry from \citet{Sara2007} used in our study. We are using the same stellar evolution code as \citet{Dott2010}.  On average, our distance moduli are 0.16 mag larger than those found by \citet{Dott2010}.  The key difference\footnote{In addition, \citet{Dott2010} allowed both [Fe/H] and $E(B-V)$ to vary to find the best fit.} with the current study is that \citet{Dott2010} did not have field star parallaxes to constrain their isochrones, and hence just used isochrones constructed with the best estimates for the various input parameters.  This would correspond to our `median' Monte Carlo isochrones and since the median isochrones do not provide the best fit to the parallax stars, the isochrones we are effectively using differ from those used by \citet{Dott2010} even though we are using the same stellar evolution code. 

\citet{fritz11} determined used HST FGS parallaxes for six RR Lyr stars and used RR Lyr stars as standard candles to determine distances to three GCs (NGC 4590, 6341 and 7078) in our sample.  Using the same reddening values, the  \citet{fritz11}  distance moduli are, on  average, smaller by 0.21 mag than the values determined in this paper. \citet{Van2013} (who adopted similar reddenings to those used in this study)  fit theoretical zero-age horizontal branch models to determine the distance to six of the clusters in our sample and our distance moduli are on average 0.24 mag larger than the \citet{Van2013} values.  

Most recently, \citet{watkins15} have combined dynamical modeling with measurements of proper motion dispersions and line of slight velocity dispersions to determine dynamical distances to 15 GCs.  To compare to our distance determinations, we converted the \citet{watkins15} distances to distance moduli using our preferred reddening value for each cluster.  Only two of the cluster in our sample are in the \citet{watkins15} sample.  For NGC 6341, \citet{watkins15} have $(m-M)_{F606W} = 14.81\pm 0.07\,$mag, which is similar to our distance moduli of $(m-M)_{F606W} = 14.88\pm 0.10\,$mag.  In contrast, our distance modulus for NGC 7078 of $(m-M)_{F606W} = 15.74\pm 0.10\,$mag disagrees strongly with that found by \citet{watkins15} $(m-M)_{F606W} = 15.39\pm 0.03\,$mag.  

In contrast to our distance results, our age determinations largely agree with previous work.  For example, 
\citet{Dott2010}  determined the age of GCs by isochrone fitting from the TO through the SGB and their  ages are  
 0.1 to 1.2 Gyr older than the combined ages in this work for seven out of the nine clusters.  The ages derived for NGC 5466 and 6101 in this work are both older that that found in Dotter et al. (2010) by 0.9 and 0.4 Gyr, respectively.  On average, our ages are $0.4\,$Gyr younger than those found by \citet{Dott2010}.  

\citet{Van2013} also use the photometry provided in \citet{Sara2007} to derive ages. In this case they allow for larger observational errorbars, but utilize the same \citet{Schlegel1998} reddening maps.  The authors  use the stellar evolution models from \citet{Van2012} and find the distance moduli using the zero-age horizontal branch magnitude.  They determine ages   by fitting isochrones from the TO through the SGB. 
The  difference between the \cite{Van2012} ages and our ages span a range of 1.4 Gyr younger to 1.1 Gyr older, with a mean age difference of 0.2 Gyr.  

We note that both \citet{Dott2010} and \cite{Van2013} used a solar calibrated mixing length in their stellar models, while we use a mixing length which is well below the solar value in our isochrones.  Changing the mixing length will impact the shape of the isochrones, and so it is likely that the ages derived by \citet{Dott2010} and \citet{Van2013} would change if they adopted a non-solar mixing length.  
  The luminosity of the  SGB is only minimally affected by the choice of the mixing length \citep{Chab1996} and so the ages which we derive are less sensitive to the choice of the mixing length.  The fact that our ages agree with previous work is fortuitous, as it appears that the younger ages one would expect from adopting the larger distance moduli find in this study, is offset by using a smaller mixing length, which increases our derived stellar ages.

\section{Summary \label{SEC:SUMM}}
Accurate HST FGS1r parallaxes have been derived for 8 metal-poor stars.  Six of these stars, with 
$-2.7 < \feh < -1.8$, are on the main sequence, and suitable for testing metal-poor stellar models and isochrones.  These stars were also observed with ACS/WFC in the F606W and F814W filters, with uncertainties in their apparent magnitudes being less than $0.005\,$mag. Typical parallax uncertainties are of order 1\%, leading to uncertainties of order $\pm 0.02\,$ mag in the absolute magnitude of the stars.  
Using a Monte Carlo approach to take into account the uncertainties in the stellar models and isochrones, we found that  isochrones constructed with the \citet{VC2003} color calibration provide 
a much better fit to the location of the stars in a color-absolute magnitude diagram than those isochrones constructed using the the color calibration based upon the PHOENIX model atmospheres \citep{Haus1999}.  

Monte Carlo isochrones which provided an acceptable fit to the location of parallax stars  were used to determine the distance (via main sequence fitting) and age (via the luminosity of the SGB) of nine metal-poor globular clusters (with $-2.4 <  \feh < -1.9$) which have excellent ACS/WFC photometry from \citet{Sara2007}.  Our distance moduli are of order $0.2\,$mag larger than previous distance determinations to metal-poor clusters.  Thus,  just as main sequence fitting results using \hip parallaxes of more metal-rich stars than in our sample led to the conclusion that globular clusters  were more distant than found by previous work  \citep[e.g.][]{reid97,gratton97,pont98,chaboyer98,caretta00,grundahl02,gratton03}, we find that our \hst parallaxes lead of metal-poor stars lead to main sequence fitting distances which are larger than previous work.  The reason for this discrepancy is unclear, and warrants further study.  In determining these main distances, we examined graphical UV CMDs in an attempt to only use GCs whose main sequences do not appear to show evidence for substantial helium enhancement (which would bias our results). Once the UV photometric data is publicly  available, it would be useful to check that these results to ensure that location of the main sequence median ridge-line is reflective of the primordial stellar population.  

Our absolute ages from these clusters range from $11.9\pm 1.4\,$Gyr to $13.9\pm 1.1\,$Gyr, in agreement with previous work.  There is no convincing evidence that an intrinsic age difference exists between the different clusters.  Averaging together the age of all nine clusters, leads to an absolute age of the oldest, most metal-poor globular clusters of $12.7\pm 1.0\,$Gyr, where the quoted uncertainty takes into account the known uncertainties in the stellar models and isochrones, along with the uncertainty in the distance and reddening of the clusters.

\acknowledgments
Support for this work was provided by NASA through grants GO-11704 and GO-12320  from the Space Telescope Science Institute, which is operated by the Association of Universities for Research in Astronomy (AURA), Inc., under NASA contract NAS5-26555. 
This material is based upon work supported by the National Science Foundation Graduate Research Fellowship under Grant No. DGE-1313911. Any opinion, findings, and conclusions or recommendations expressed in this material are those of the authors(s) and do not necessarily reflect the views of the National Science Foundation.
 This publication makes use of data products from the Two Micron All Sky Survey, which is a joint project of the University of Massachusetts and the Infrared Processing and Analysis Center/California Institute of Technology, funded by NASA and the NSF. This research has made use of the SIMBAD database, operated at CDS, Strasbourg, France, 
and the NASA Astrophysics Data System Abstract Service.  This work has made use of data from the European Space Agency (ESA)
mission {\it Gaia} (\url{http://www.cosmos.esa.int/gaia}), processed by
the {\it Gaia} Data Processing and Analysis Consortium (DPAC,
\url{http://www.cosmos.esa.int/web/gaia/dpac/consortium}). Funding
for the DPAC has been provided by national institutions, in particular
the institutions participating in the {\it Gaia} Multilateral Agreement.

  \facility{HST} \facility{McGraw-Hill}\facility{NMSU:1m}


\appendix

\section{FGS Data Reduction \label{sec:FGS} }
Astrometric data  from the FGS are retrieved by download from the $\hstns$ online archival retrieval system  and then processed through the two-part FGS  pipeline system. The low-level calibration pipeline  extracts the astrometry measurements (usually from  1 to 2 minutes of fringe position information that was acquired at a 40 Hz rate,  yielding several thousand discrete measurements)  and after outlier removal (from cosmic ray hits etc.) calculates the median and performs a per-observation error estimation.    The high-level calibration pipeline corrects the observations with  the  time-variant OFAD,  compensates the velocity aberration, processes the time tags, and calculates the parallax factors with  the JPL Earth orbit predictor \citep{Sta90}.   The OFAD calibration is presented in several calibration papers \citep{McA97,McA02,McA06} and ongoing stability tests (LTSTABs) are used to maintain this calibration.   FGS1R Instrumental systematics (such as intra-orbit drift and color and filter effects) are also corrected for.   We have not found after 24 years of calibration additional systematics in our data at a level that is higher than our detection limit.  Regression analysis between $\hip$  and $\hst$ parallax measurements has shown (with the exception of the Pleiades, \citet{Sod05} ) not only good agreement between the  parallaxes determined by the two instruments, but also an {\it overestimation} of error in $\hst$ astrometric measurements \citep{Ben07,McA10}.  

 The distribution of  the reference stars in the  field is shown in the Digital Sky Survey images in Figure~ \ref{fig:dss}.   
The  position of each star is measured by the FGS sequentially. Each epoch contains multiple visits, of the metal-poor target  and  reference stars, providing $x(t)$ and $y(t)$ positions.  These positions are measured  in the $\hst$ reference frame in seconds of arc. During an orbit, positional drift occurs and was corrected   for with an adaptable polynomial fitting routine amplified to model the high intra-orbit drift seen in some of these observations.  The F583W filter was used for all observations.  The observation dates, the number of measurements per epoch of the target metal-poor stars,  the HST orientation angles and the number of plate parameters are listed on Table~\ref{tab:atmlog}.  
 Available only on-line as a machine-readable table  (Table~ \ref{tab:hstdata}) is the  $\hst$ astrometric data for the targets and their reference stars.   The most current  calibration the data should be retrieved from the  $\hstns$ online archival retrieval system  and processed through the low and high level  pipeline system.
 
 \clearpage
 
 \begin{figure}
\gridline {\fig{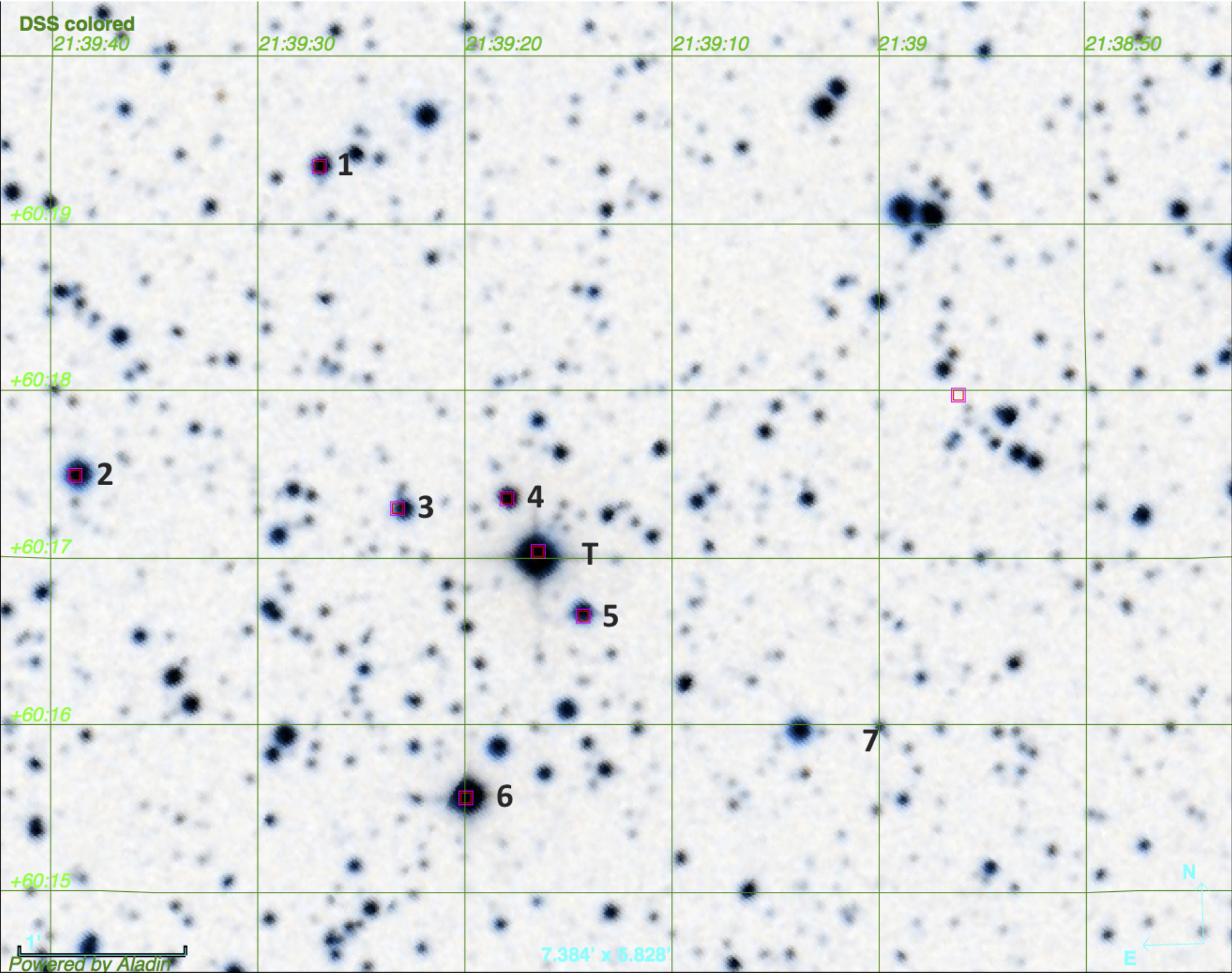}{0.33\textwidth}{HIP106924}
          \fig{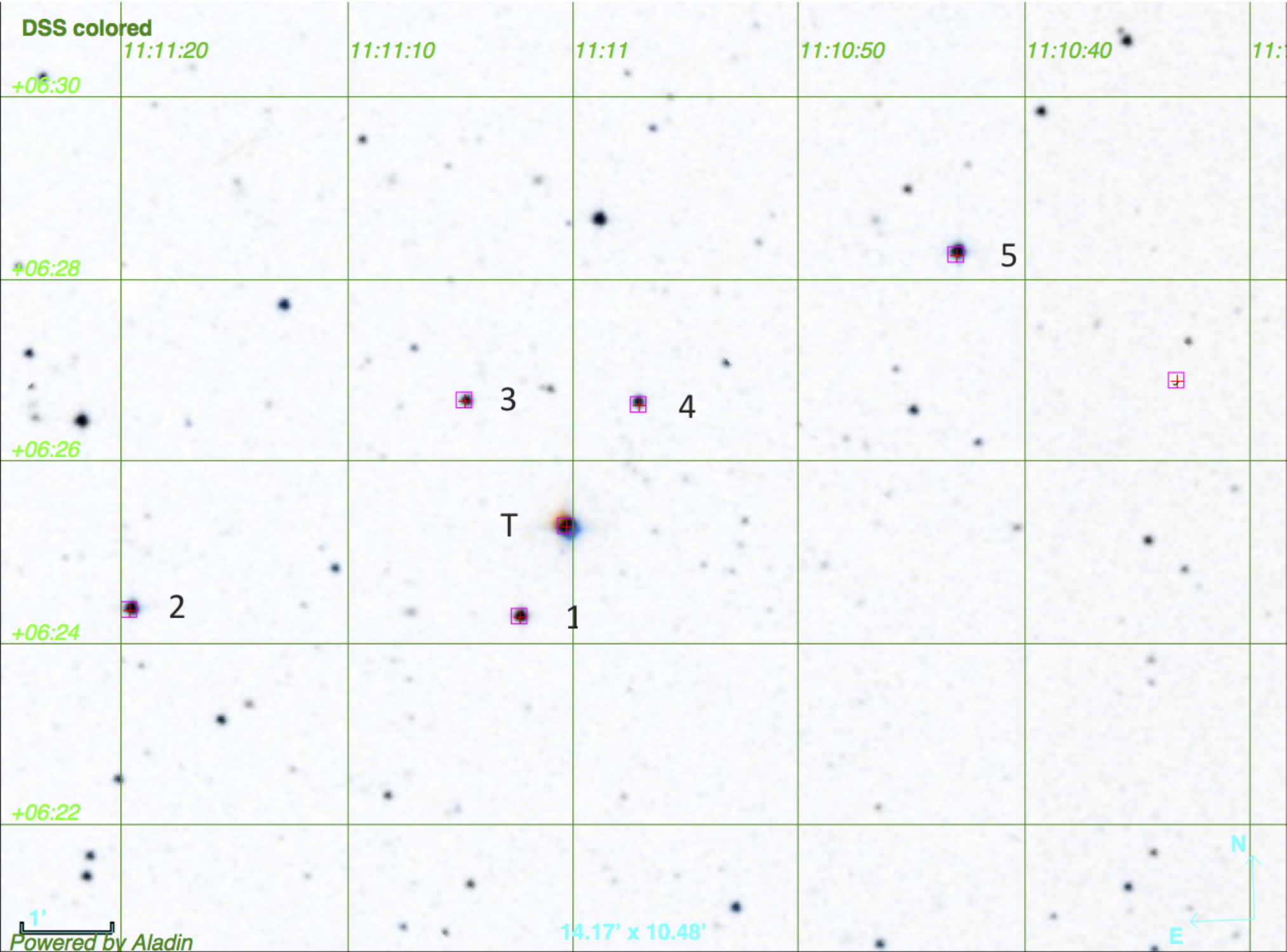}{0.33\textwidth}{HIP54639}
          \fig{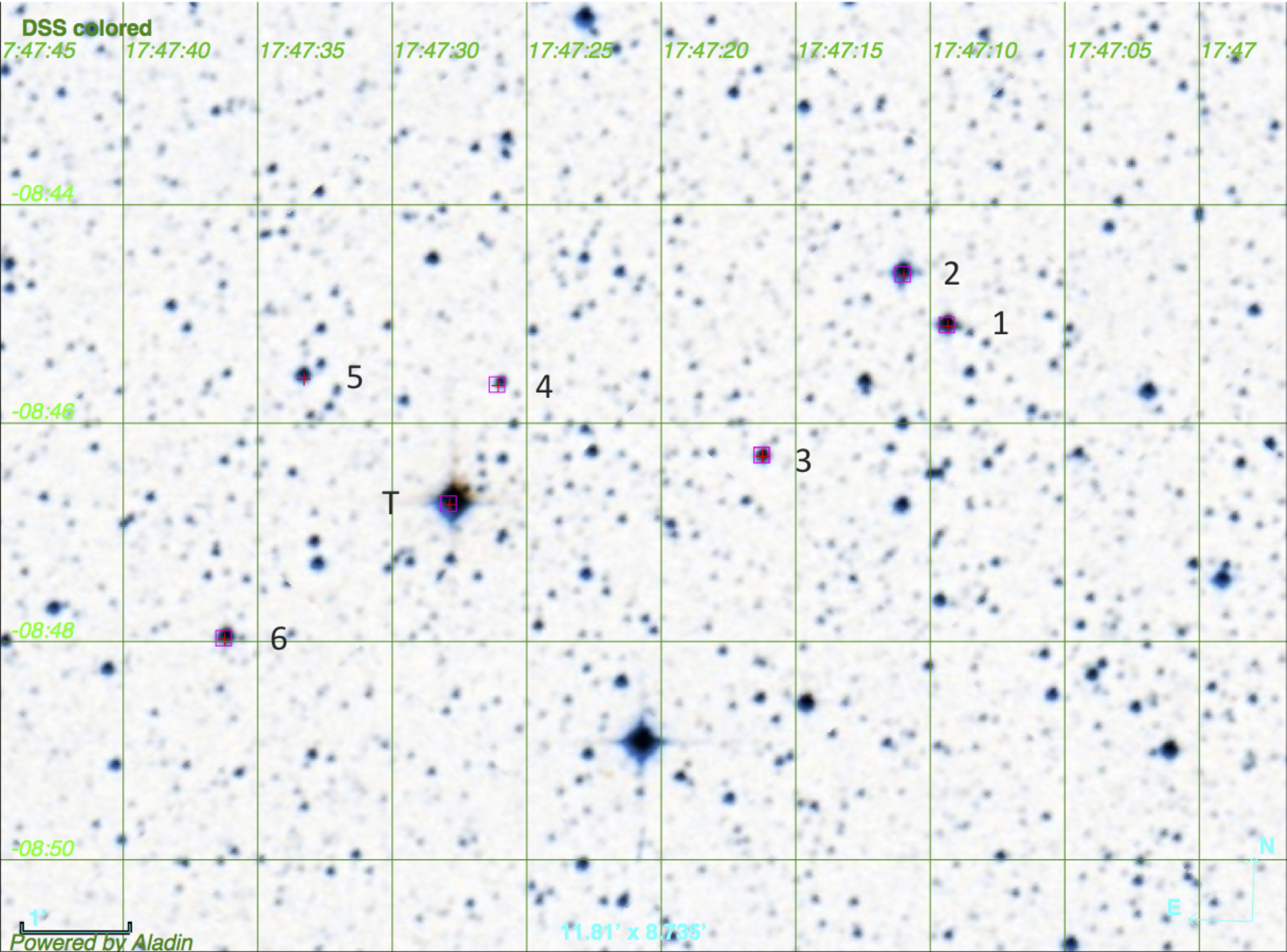}{0.33\textwidth}{HIP87062}
                    }
 \gridline{\fig{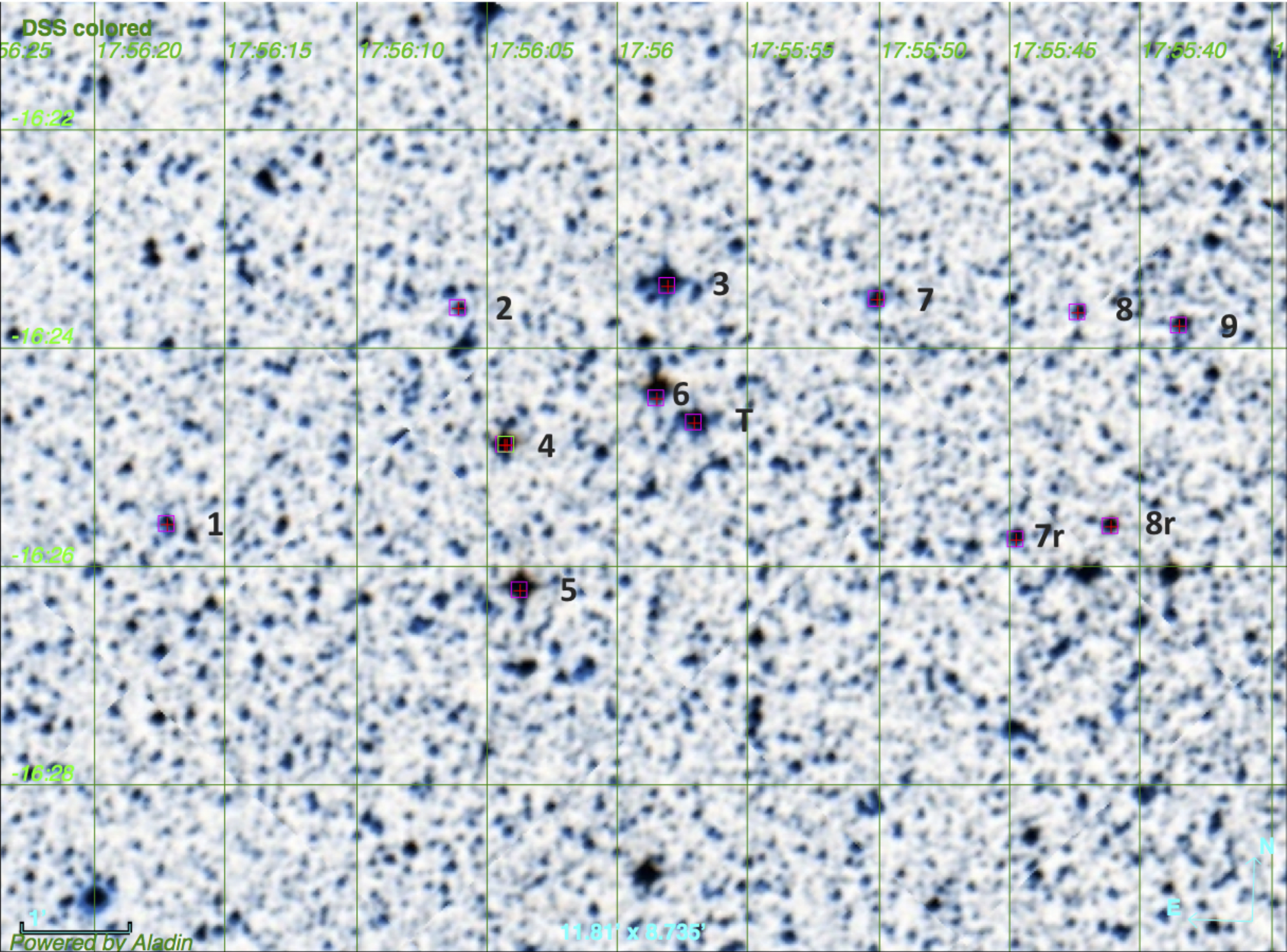}{0.33\textwidth}{HIP87788}
   \fig{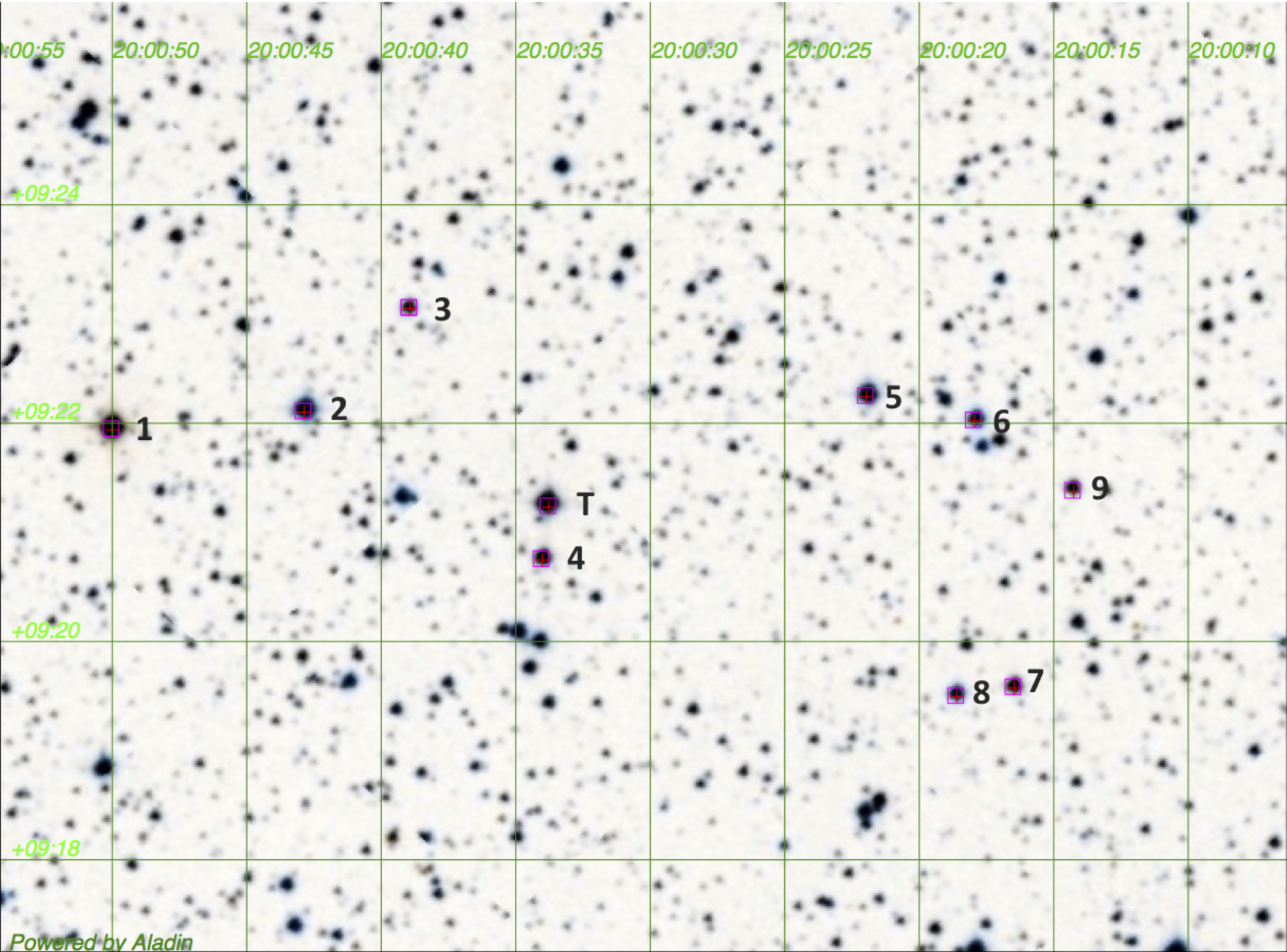}{0.33\textwidth}{HIP98492}
    \fig{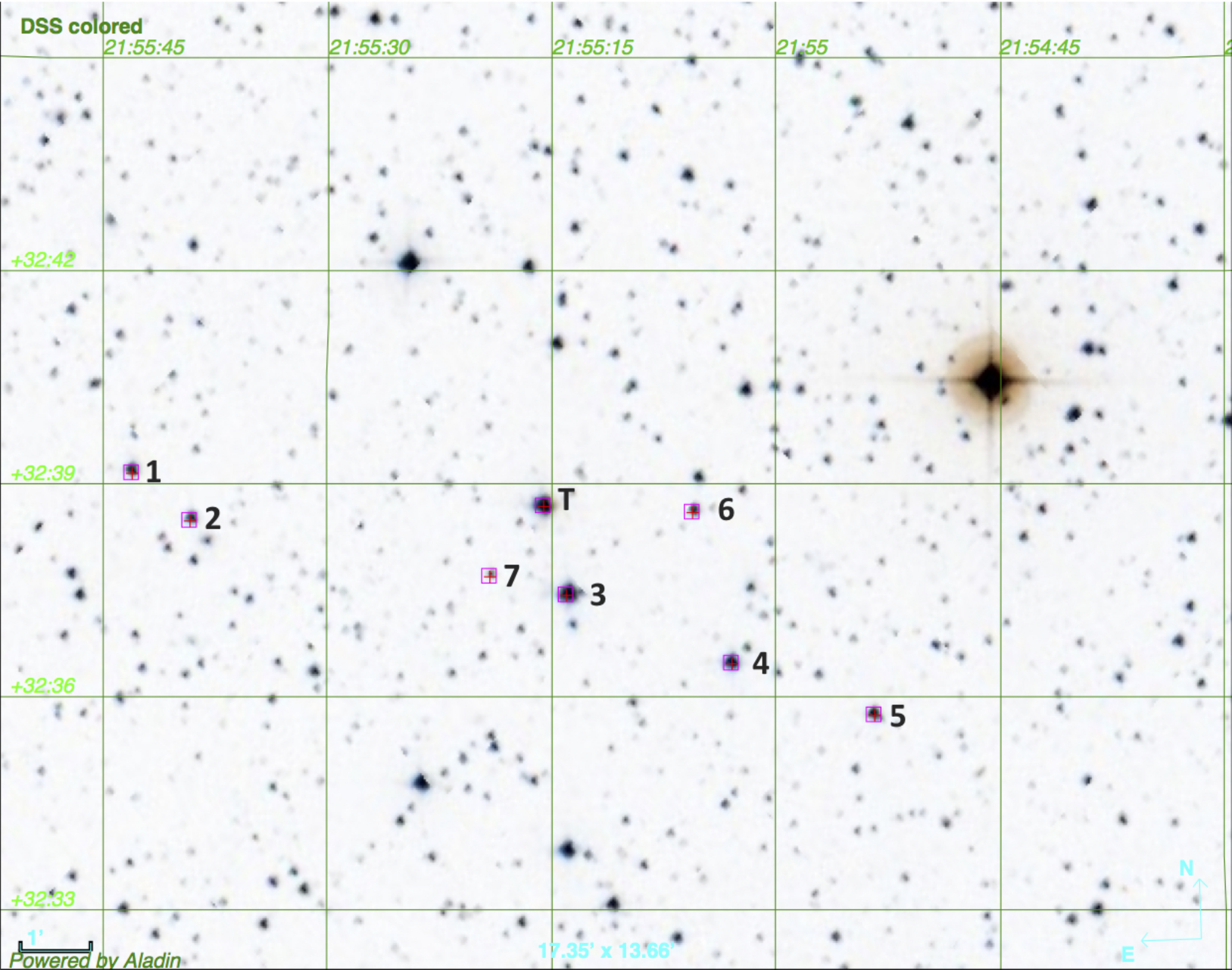}{0.33\textwidth}{HIP108200}
          }

 \gridline{\fig{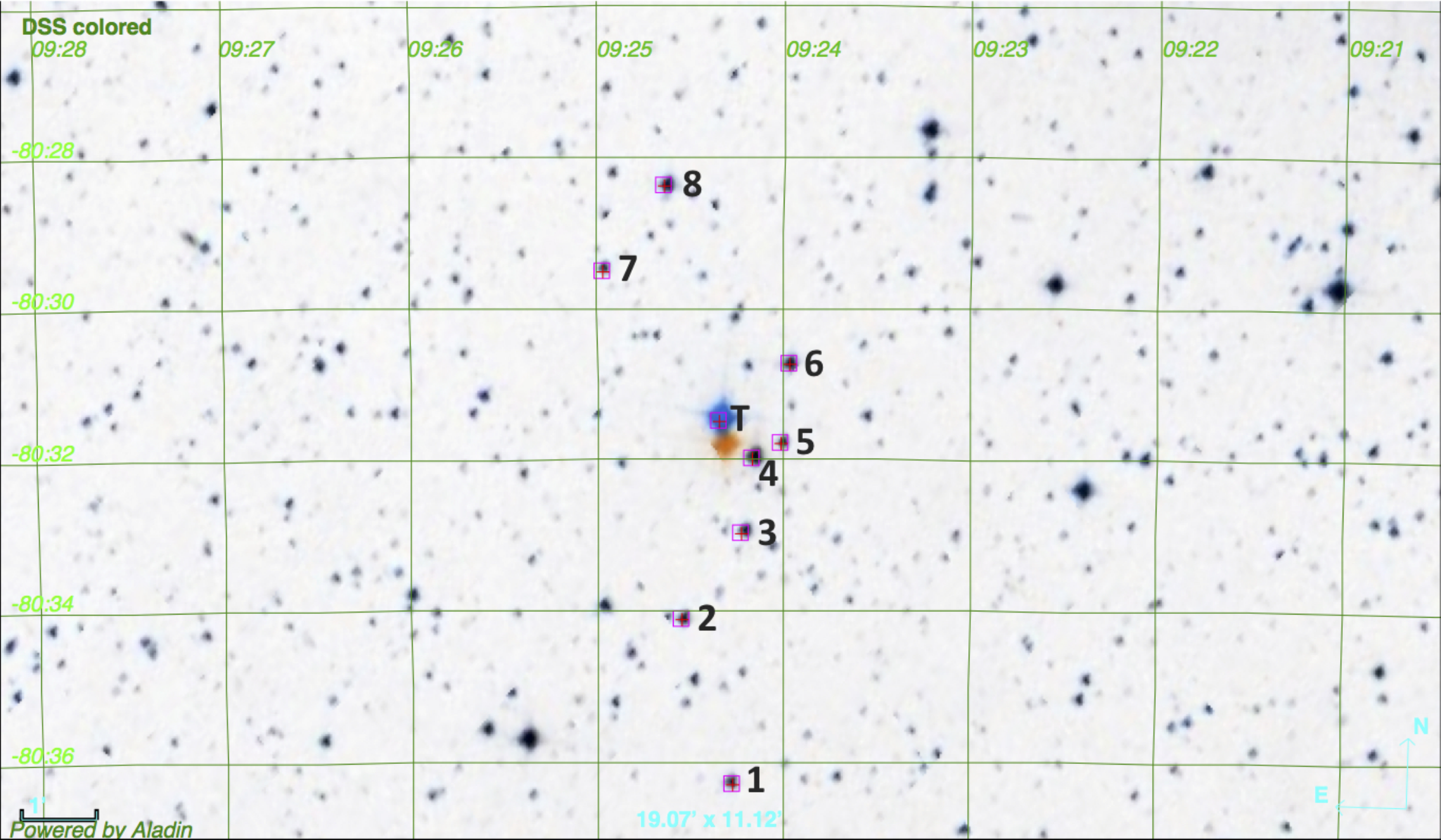}{0.45\textwidth}{HIP46120}
  \fig{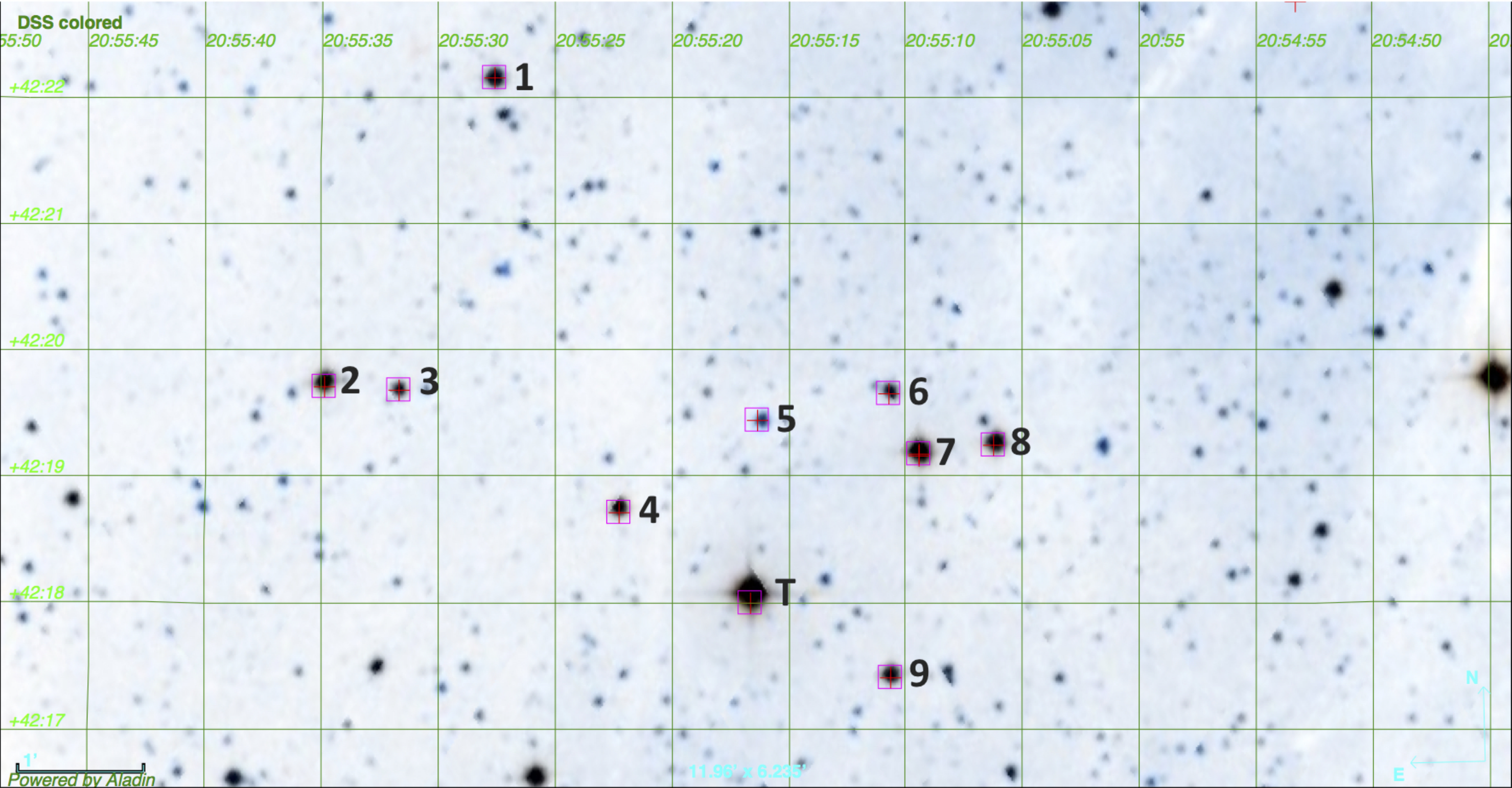}{0.5\textwidth}{HIP103269}
                  }

\caption{Metal-poor star fields with astrometric reference stars and target (T) labelled.  The stars are listed in Table~\ref{tab:cat}.  The DSS images  were  made using {\it Aladin}. }
\label{fig:dss}
\end{figure}

\clearpage

\begin{deluxetable}{cccc}
\tablewidth{0in}
\tablecaption{$\hst$Astrometric Observation Targets Log\label{tab:atmlog}}
\tablehead{\colhead{Orbit}&\colhead{Date}&\colhead{N$_{obs}$}&\colhead{$\hst$ Roll}}
\startdata
\hline
\multicolumn{4}{c}{HIP 46120} \\
\hline 
1	&	27-Feb-2009		&	5	&	159.1263	\\
2	&	5-Mar-2009		&	6	&	156.1014	\\
3	&	15-Mar-2009		&	6	&	161.1411	\\
4	&	7-Aug-2009		&	6	&	353.805	\\
5	&	20-Aug-2009		&	6	&	353.805	\\
6	&	28-Aug-2009		&	6	&	353.805	\\
7	&	21-Aug-2010		&	6	&	353.8054	\\
8	&	21-Feb-2011		&	5	&	168.5837	\\
9	&	25-Feb-2011		&	6	&	169.1866	\\
10	&	7-Aug-2011		&	6	&	353.8055	\\
11	&	26-Aug-2011		&	6	&	353.8055	\\ 
\hline
\multicolumn{4}{c}{HIP 54639} \\
\hline
1	&	9-Dec-2008		&	7	&	256.9946	\\
2	&	15-Dec-2008		&	5	&	256.9946	\\
3\tablenotemark{a}	&	21-Dec-2008		&	7	&	256.9946	\\
4	&	23-Mar-2010		&	7	&	67.50748	\\
5	&	3-Jun-2010		&	7	&	67.50748	\\
6	&	17-Jun-2010		&	7	&	67.50748	\\
7	&	9-Dec-2010		&	7	&	256.9946	\\
8	&	4-Jun-2011		&	7	&	67.78232	\\
9	&	17-Jun-2011		&	7	&	67.78232	\\
10	&	9-Dec-2011		&	7	&	256.9946	\\
11	&	14-Dec-2011		&	7	&	256.9946	\\ 
\hline
\multicolumn{4}{c}{HIP 87062} \\
\hline
1	&	9-Mar-2009		&	5	&	270.9999	\\
2	&	13-Mar-2009		&	5	&	270.9999	\\
3	&	10-May-2009		&	5	&	271.9994	\\
4	&	24-Jul-2009		&	5	&	88.99068	\\
5	&	11-Sep-2009		&	5	&	88.99068	\\
6	&	18-Sep-2009		&	5	&	88.99068	\\
7	&	19-Sep-2010		&	5	&	88.99069	\\
8	&	3-Mar-2011		&	5	&	263.3456	\\
9	&	18-Mar-2011		&	5	&	271.5355	\\
10	&	2-Sep-2011		&	5	&	88.9907	\\
11	&	16-Sep-2011		&	5	&	88.9907	\\ 
\hline
\multicolumn{4}{c}{HIP 87788} \\
\hline
1	&	20-Mar-2009		&	4	&	270.0006	\\
2	&	2-Apr-2009		&	6	&	270.0006	\\
3	&	18-Jul-2009		&	6	&	84.99417	\\
4	&	11-Sep-2009		&	6	&	84.99417	\\
5\tablenotemark{b}	&	22-Sep-2009		&	6	&	84.99417	\\
6	&	12-Oct-2009		&	6	&	87.49677	\\
7	&	1-Jun-2010		&	6	&	270.0006	\\
8	&	22-Sep-2010		&	6	&	84.99417	\\
9	&	20-Mar-2011		&	4	&	270.0006	\\
10	&	30-Mar-2011		&	6	&	270.0006	\\
11	&	10-Sep-2011		&	6	&	84.99417	\\
12	&	22-Sep-2011		&	6	&	84.99417 \\ 
\hline
\multicolumn{4}{c}{HIP 98492} \\
\hline
1	&	25-Apr-2009		&	5	&	287.0096	\\
2	&	4-May-2009		&	6	&	287.0096	\\
3	&	19-Sep-2009		&	6	&	99.99505	\\
4	&	16-Oct-2009		&	6	&	99.99505	\\
5	&	26-Oct-2009		&	6	&	99.99505	\\
6	&	1-Jun-2010		&	6	&	289.0107	\\
7	&	24-Oct-2010		&	6	&	99.99506	\\
8	&	24-Apr-2011		&	5	&	287.0096	\\
9	&	13-May-2011		&	6	&	287.0096	\\
10	&	22-Sep-2011		&	6	&	99.99506	\\
11	&	26-Oct-2011		&	6	&	99.99506	\\ 
\hline
\multicolumn{4}{c}{HIP 103269} \\
\hline
1	&	16-Jun-2009		&	7	&	302.1005	\\
2	&	15-Nov-2009		&	7	&	116.8969	\\
3	&	25-Nov-2009		&	7	&	116.8969	\\
4	&	30-Nov-2009		&	7	&	116.8969	\\
5	&	26-May-2010		&	7	&	302.1005	\\
6	&	6-Jun-2010		&	7	&	302.1005	\\
7	&	31-Oct-2010		&	7	&	116.8969	\\
8	&	26-May-2011		&	7	&	302.1005	\\
9	&	16-Jun-2011		&	7	&	302.1005	\\
10	&	15-Nov-2011		&	7	&	116.8969	\\
11	&	30-Nov-2011		&	7	&	116.8969	\\ 
\hline
\multicolumn{4}{c}{HIP 106924} \\
\hline
1	&	19-Dec-2008		&	5	&	142.711	\\
2	&	2-Jan-2009		&	5	&	142.711	\\
3\tablenotemark{a}	&	5-Jan-2009		&	6	&	142.711	\\
4	&	2-Jul-2009		&	6	&	323.2631	\\
5	&	12-Jul-2009		&	6	&	323.2631	\\
6	&	20-Jul-2009		&	6	&	323.2631	\\
7	&	2-Aug-2010		&	6	&	336.3016	\\
8	&	19-Dec-2010		&	5	&	142.7101	\\
9	&	7-Jan-2011		&	6	&	142.7101	\\
10	&	4-Jul-2011		&	6	&	323.2635	\\
11	&	21-Jul-2011		&	6	&	323.2635\\ 
\hline
\multicolumn{4}{c}{HIP 108200} \\
\hline
1	&	19-Jun-2009		&	6	&	290.0325	\\
2	&	13-Nov-2009		&	5	&	109.956	\\
3	&	18-Nov-2009		&	6	&	109.956	\\
4	&	4-Dec-2009		&	6	&	109.956	\\
5	&	21-May-2010		&	6	&	290.0323	\\
6	&	31-May-2010		&	6	&	290.0323	\\
7	&	3-Dec-2010		&	5	&	109.9559	\\
8	&	21-May-2011		&	6	&	290.0322	\\
9	&	18-Jun-2011		&	6	&	290.0322	\\
10	&	11-Nov-2011		&	6	&	109.9557	\\
11	&	4-Dec-2011		&	6	&	109.9557	\\
\enddata
\tablenotetext{a}{ FGS1R AMA-Adjustment after this observation}
\tablenotetext{b}{ Observation  set failed}
\end{deluxetable}

\begin{deluxetable}{lll}
\tablewidth{0in}
\tablecaption{Table Format for Astrometric Data for Target and Reference Stars \tablenotemark{a}\label{tab:hstdata}}
\tablehead{
\colhead{Column}&
\colhead{Format}&
\colhead{Desription}
}
\startdata
1&	A9	&HST Observation ID\\ 
2&	A8	&HST observing mode; POSITION or TRANSFER\\
3&	A10	&Raw target ID from proposal\\
4&	I5	&Proposal number of observations\\
5&	I4	&tar number of observation from proposal\\
6&	F5.2	&Predicted V band magnitude\\
7&	F5.2	&Actual V band magnitude\\
8&	F8.5	&Predicted Right Ascension; decimal degrees (J2000)\\
9&	F8.5	&Predicted Declination; decimal degrees (J2000)\\
10&	F8.5	&Right Ascension of V1 telescope axis; degrees (1)\\
11&	F8.5	&Declination of V1 telescope axis; degrees (1)\\
12&	F8.4	&Roll about V3 of telescope\\
13&	F12.6	&X position of telescope\\
14&	F12.6	&Y position of telescope\\
15&	F12.6	&Z position of telescope\\
16&	F9.6	&X velocity of telescope\\
17&	F9.6	&Y velocity of telescope\\
18&	F9.6	&Z velocity of telescope\\
19&	F16.8	&Observation Julian Date; corrected\\
20&	F14.8	&Modified Julian observation time\\
21&	I4	&Year of observation\\
22&	I3	&Day of observation\\
23&	A8	&The hour:min:sec of observation\\
24&	I1	&FGS used as the astrometer\\
25&	A5	&Filter used for observation\\
26&	I3	&Set number of observation group\\
27&	F13.9	&Earth plus HST X velocity\\
28&	F13.9	&Earth plus HST Y velocity\\
29&	F13.9	&Earth plus HST Z velocity\\
30&	F10.5	&Uncorrected X position calculated by average\\
31&	F9.5	&Uncorrected Y position calculated by average\\
32&	F7.5	&Standard deviation in Xave\\
33&	F7.5	&Standard deviation in Yave\\
34&	F8.5	&XY correlation\\
35&	F13.10	&XY covariance\\
36&	F14.9	&Uncorrected X position calculated by median\\
37&	F11.7	&Uncorrected Y position calculated by median\\
38&	F7.5	&X median average deviation\\
39&	F7.5	&Y median average deviation\\
40&	F6.2	&Seconds of FINELOCK data\\
41&	I4	&Number of samples from FINELOCK interval\\
42&	A5	&Instrument velocity aberration measured from\\
43&	I1	&S  FGS velocity aberration measured from\\
44&	F13.9	&X position of velocity aberration\\
45&	F11.7	&Y position of velocity aberration\\
46&	I1	&Dominant FGS guider\\
47&	F10.4	&Sum of pmts\\
48&	F9.4	&pmt from channel xa\\
49&	F9.4	&pmt from channel xb\\
50&	F9.4	&pmt from channel ya\\
51&	F9.4	&pmt from channel yb\\
52&	I3	&Number of samples for background\\
53&	I4	&Number of samples for walkdown pmt diff/sum in X axis\\
54&	F7.4	&X pmt diff/sum from fine lock dv interval\\
55&	F7.4	&X pmt diff/sum from walkdown interval\\
56&	F6.4	&Reference X axis S\_curve inverse slope of Upgren69\\
57&	F9.6	&X centroid shift in arcseconds; add to Xmed\\
58&	I4	&Number of samples for walkdown pmt diff/sum in Y axis\\
59&	F7.4	&Y pmt diff/sum from fine lock dv interval\\
60&	F7.4	&Y pmt diff/sum from walkdown interval\\
61&	F6.4	&Reference Y axis S\_curve inverse slope of Upgren69\\
62&	F9.6	&Y centroid shift in arcseconds; add to Ymed\\
63&	F8.3	&Course track X position\\
64&	F7.3	&Course track Y position\\
65&	I6	&Star selector position; theta A\\
66&	I6	&Star selector position; theta B\\
67&	E11.5	&Variance of X\\
68&	E11.5	&Variance of Y\\
69&	E11.5	&Median average deviation variance of X\\
70&	E11.5	&Median average deviation variance of Y\\
71&	F11.4	&Predicted star selector position; theta A\\
72&	F11.4	&Predicted star selector position; theta B\\
73&	F11.9	&Interpolated rhoA lever arm\\
74&	F11.9	&Interpolated kA lever arm\\
75&	F11.9	&Spacecraft quaternion\\
76&	F12.9	&Spacecraft quaternion\\
77&	F12.9	&Spacecraft quaternion\\
78&	F12.9	&Spacecraft quaternion\\
79&	F12.9	&Parallax factor {alpha}\\
80&	F12.9	&Parallax factor {delta}\\
81&	E14.7	&Sinfit position correction for X\\
82&	F14.9	&Final X corrected position with polynomial drift correction\\
83&	F13.9	&Final Y corrected position with polynomial drift correction\\
84&	A9	&Target name\\
\enddata
\tablenotetext{a}{Table \ref{tab:hstdata} is published in its entirety in the electronic edition.  The column descriptions and format are  shown here for guidance regarding its form and content.}
\end{deluxetable}

\subsection{Reference Star Spectroscopy and Photometry \label{groundphot}}
We obtained spectra of the reference frame stars for the northern targets
using the Dual Imaging
Spectrograph\footnotemark[2]\footnotetext[2]{http://www.apo.nmsu.edu/arc35m/Instruments/DIS/} (``DIS'') on the 3.5 m telescope at the Apache Point 
Observatory.  DIS simultaneously obtains spectra covering blue and red 
spectral regions, providing dispersions of 0.62 \AA/pix in the blue, and 
0.58 \AA/pix in the red with the high resolution gratings (1,200 line/mm). 
For HIP46120 and HIP54639, we obtained spectra of the reference frame stars 
using the R$-$C Spectrograph\footnotemark[3]\footnotetext[3]{http://www.ctio.noao.edu/spectrographs/4m\_R-C/4m\_R-C.html} on the Blanco 4 m telescope at 
Cerro Tololo Interamerican Observatory (program 2009A-0009). The KPGL1 
grating was used, and with the ``Loral 3K'' detector, provided a dispersion 
of 1.01 \AA/pix.

Optical photometry for the fields all of the program stars, except
HIP46120, was procured using the robotic New Mexico State University (NMSU) 
1 m telescope  \citep{Holtz10} at Apache Point Observatory and the MDM 1.3 m telescope. 
The NMSU 
1 m is equipped with an E2V 2048 sq. CCD camera, and the standard Bessell 
$UBVRI$ filter set. THE MDM 1.3 m  data was obtained with a STA-0500 4062 sq. CCD camera and a  standard BVRI filter set. Photometry of the field of HIP46120 was obtained using 
the Tek2K CCD imager\footnotemark[4]\footnotetext[4]{http://www.ctio.noao.edu/noao/content/tek2k} on the SMARTS 0.9 m telescope at CTIO (program 
2009A-0009). The images of the program object fields, along with the 
appropriate calibration data, were obtained in the usual fashion, reduced 
using IRAF, and flux calibrated with observations of Landolt standards. 

We use the transformations provided in 
\citet{Car01} to  convert the  2MASS $JHK$ values   to the 
Bessell \& Brett (1988)\nocite{Bes88} system. Table~\ref{tab:IR} lists $VJHK$ photometry for the target 
and reference stars indicated in Figures~\ref{fig:dss}. 
Figure~\ref{fig:JmKvsVmKrefs} show the  $(J-K)$ vs. $(V-K)$ 
color-color diagrams with  reference stars and 
targets labeled. 


\begin{deluxetable}{ccccccccc}
\tablewidth{0in}
\tablecaption{V and Near-IR Photometry of Target and Reference Stars \label{tab:IR}}
\tablehead{\colhead{ID}&
\colhead{$V$} &
\colhead{$U-B$} &
\colhead{$B-V$} &
\colhead{$V-R$} &
\colhead{$V-I$} &
\colhead{$J$} &
\colhead{$H$} &
\colhead{$K$} 
}
\startdata
\hline								
\multicolumn{9}{c}{HIP 46120} \\													
\hline				
Ref-1	&	14.34	$\pm$	0.02	&	0.13	$\pm$	0.04	&	0.74	$\pm$	0.04	&	0.42	$\pm$	0.03	&	0.84	$\pm$	0.03	&	12.91	&	12.52	&	12.49	\\
Ref-2	&	13.96	$\pm$	0.02	&	0.22	$\pm$	0.04	&	0.81	$\pm$	0.04	&	0.44	$\pm$	0.03	&	0.9	$\pm$	0.03	&	12.44	&	12.05	&	11.99	\\
Ref-3	&	14.02	$\pm$	0.02	&	0.59	$\pm$	0.04	&	1.00	$\pm$	0.04	&	0.53	$\pm$	0.03	&	1.05	$\pm$	0.03	&	12.23	&	11.73	&	11.65	\\
Ref-4	&	12.58	$\pm$	0.02	&	0.23	$\pm$	0.04	&	0.75	$\pm$	0.04	&	0.39	$\pm$	0.03	&	0.82	$\pm$	0.03	&	11.19	&	10.88	&	10.81	\\
Ref-5	&	14.76	$\pm$	0.02	&	0.35	$\pm$	0.05	&	0.82	$\pm$	0.04	&	0.43	$\pm$	0.03	&	0.86	$\pm$	0.03	&	13.3	&	12.97	&	12.85	\\
Ref-6	&	13.48	$\pm$	0.02	&	0.48	$\pm$	0.04	&	0.89	$\pm$	0.04	&	0.47	$\pm$	0.03	&	0.96	$\pm$	0.03	&	11.82	&	11.45	&	11.33	\\
Ref-7	&	14.91	$\pm$	0.02	&	0.81	$\pm$	0.06	&	1.01	$\pm$	0.05	&	0.6	$\pm$	0.03	&	1.13	$\pm$	0.03	&	12.99	&	12.48	&	12.39	\\
Ref-8	&	13.21	$\pm$	0.02	&	0.59	$\pm$	0.04	&	1.03	$\pm$	0.04	&	0.56	$\pm$	0.03	&	1.1	$\pm$	0.03	&	11.32	&	10.8	&	10.72	\\
HIP 46120\tablenotemark{a}	&		&		&		&		&	&		&		&		\\
\multicolumn{9}{c}{HIP 54639} \\													
\hline								
Ref-1	&	13.87	$\pm$	0.02	&	0	$\pm$	0.1	&	0.47	$\pm$	0.04	&	0.28	$\pm$	0.03	&	0.59	$\pm$	0.03	&	12.89	&	12.57	&	12.52	\\
Ref-2	&	13.45	$\pm$	0.02	&	0.9	$\pm$	0.13	&	0.97	$\pm$	0.07	&	0.53	$\pm$	0.03	&	1.01	$\pm$	0.03	&	11.71	&	11.2	&	11.07	\\
Ref-3	&	14.72	$\pm$	0.02	&	1.05	$\pm$	0.2	&	1.09	$\pm$	0.1	&	0.56	$\pm$	0.03	&	1.07	$\pm$	0.03	&	12.81	&	12.18	&	12.03	\\
Ref-4	&	14.72	$\pm$	0.02	&	1.28	$\pm$	0.2	&	1.06	$\pm$	0.1	&	0.66	$\pm$	0.03	&	1.24	$\pm$	0.03	&	12.66	&	11.99	&	11.91	\\
Ref-5	&	13.23	$\pm$	0.02	&	1.41	$\pm$	0.14	&	1.24	$\pm$	0.05	&	0.78	$\pm$	0.03	&	1.46	$\pm$	0.03	&	10.81	&	10.17	&	10.05	\\
HIP 54639	&		&		&		&		&	&		&		&		\\
\hline
\multicolumn{9}{c}{HIP 87062} \\													
\hline
Ref-1	&	13.36	$\pm$	0.02	&	0.51	$\pm$	0.1	&	1.07	$\pm$	0.04	&	0.67	$\pm$	0.03	&	1.36	$\pm$	0.03	&	10.99	&	10.58	&	10.25	\\
Ref-2	&	13.32	$\pm$	0.02	&	2.06	$\pm$	0.1	&	1.88	$\pm$	0.04	&	1.13	$\pm$	0.03	&	2.13	$\pm$	0.03	&	9.55	&	8.66	&	8.32	\\
Ref-3	&	14.58	$\pm$	0.02	&	0.77	$\pm$	0.1	&	1.24	$\pm$	0.05	&	0.73	$\pm$	0.03	&	1.44	$\pm$	0.03	&	12.05	&	11.55	&	11.42	\\
Ref-4	&	15.67	$\pm$	0.04	&	0.77	$\pm$	0.1	&	1.27	$\pm$	0.09	&	0.81	$\pm$	0.04	&	1.63	$\pm$	0.03	&	12.82	&	12.34	&	12.12	\\
Ref-5	&	14.65	$\pm$	0.02	&	2.02	$\pm$	0.1	&	2.02	$\pm$	0.07	&	1.27	$\pm$	0.03	&	2.47	$\pm$	0.03	&	10.32	&	9.36	&	9.05	\\
Ref-6	&	13.77	$\pm$	0.02	&	0.67	$\pm$	0.1	&	1.05	$\pm$	0.04	&	0.62	$\pm$	0.03	&	1.31	$\pm$	0.03	&	11.5	&	11.16	&	11.04	\\
HIP 87062	&	10.57	$\pm$	0.02	&	-0.06	$\pm$	0.1	&	0.59	$\pm$	0.04	&	0.37	$\pm$	0.03	&	0.83	$\pm$	0.03	&	9.17	&	8.86	&	8.77	\\
\hline	
\multicolumn{9}{c}{HIP 87788} \\						
\hline				
Ref-1	&	14.45	$\pm$	0.02	&	1.32	$\pm$	0.22	&	2.35	$\pm$	0.05	&	1.54	$\pm$	0.03	&	2.97	$\pm$	0.03	&	9.24	&	7.96	&	7.35	\\
Ref-2	&	14.57	$\pm$	0.02	&	1.64	$\pm$	0.24	&	2.13	$\pm$	0.05	&	1.23	$\pm$	0.03	&	2.34	$\pm$	0.03	&	10.38	&	9.36	&	9.16	\\
Ref-3	&	10.55	$\pm$	0.02	&	1.85	$\pm$	0.1	&	1.63	$\pm$	0.04	&	0.96	$\pm$	0.03	&	1.81	$\pm$	0.03	&	7.33	&	6.47	&	6.22	\\
Ref-4	&	11.98	$\pm$	0.02	&	-0.06	$\pm$	0.1	&	0.38	$\pm$	0.04	&	0.18	$\pm$	0.03	&	0.44	$\pm$	0.03	&	11.12	&	11.02	&	10.96	\\
Ref-5	&	11.04	$\pm$	0.02	&	0.21	$\pm$	0.1	&	0.54	$\pm$	0.04	&	0.26	$\pm$	0.03	&	0.62	$\pm$	0.03	&	9.87	&	9.73	&	9.63	\\
Ref-6	&	11.55	$\pm$	0.02	&	1.36	$\pm$	0.1	&	1.39	$\pm$	0.04	&	0.75	$\pm$	0.03	&	1.47	$\pm$	0.03	&	8.91	&	8.24	&	8.06	\\
Ref-7	&	13.49	$\pm$	0.02	&	1.82	$\pm$	0.15	&	2.06	$\pm$	0.04	&	1.39	$\pm$	0.03	&	2.93	$\pm$	0.03	&	8.47	&	7.42	&	7.04	\\
Ref-8	&	13.75	$\pm$	0.02	&	1.96	$\pm$	0.15	&	1.85	$\pm$	0.04	&	1.04	$\pm$	0.03	&	1.97	$\pm$	0.03	&	10.18	&	9.39	&	9.01	\\
Ref-9	&	13.47	$\pm$	0.02	&	1.05	$\pm$	0.13	&	1.3	$\pm$	0.04	&	0.71	$\pm$	0.03	&	1.42	$\pm$	0.03	&	10.85	&	10.25	&	10.05\\
Ref-7R	&	14.41	$\pm$	0.02	&		$\pm$		&	2.16	$\pm$	0.04	&		$\pm$		&		$\pm$		&		&		&	8.61	\\
Ref-8R	&	13.52	$\pm$	0.02	&		$\pm$		&	0.88	$\pm$	0.04	&		$\pm$		&		$\pm$	         &		&		&	11.33\\
HIP 87788	&	11.33	$\pm$	0.02	&	-0.16	$\pm$	0.07	&	0.74	$\pm$	0.04	&	0.39	$\pm$	0.03	&	0.87	$\pm$	0.03	&	9.94	&	9.54	&	9.46	\\
\hline	
\multicolumn{9}{c}{HIP 98492} \\				
\hline			
Ref-1	&	11.18	$\pm$	0.02	&	-0.06	$\pm$	0.06	&	0.08	$\pm$	0.03	&	0.03	$\pm$	0.02	&	0.12	$\pm$	0.02	&	10.93	&	10.96	&	10.94	\\
Ref-2	&	11.89	$\pm$	0.02	&	1.53	$\pm$	0.06	&	1.31	$\pm$	0.03	&	0.7	$\pm$	0.02	&	1.3	$\pm$	0.02	&	9.58	&	8.97	&	8.79	\\
Ref-3	&	14.19	$\pm$	0.02	&	0.75	$\pm$	0.1	&	1.01	$\pm$	0.03	&	0.56	$\pm$	0.02	&	1.07	$\pm$	0.02	&	12.26	&	11.97	&	11.69	\\
Ref-4	&	13.24	$\pm$	0.02	&	0.28	$\pm$	0.06	&	0.74	$\pm$	0.03	&	0.4	$\pm$	0.02	&	0.79	$\pm$	0.02	&	11.84	&	11.55	&	11.45	\\
Ref-5	&	12.35	$\pm$	0.02	&	0.99	$\pm$	0.06	&	1.04	$\pm$	0.03	&	0.55	$\pm$	0.02	&	1.05	$\pm$	0.02	&	10.48	&	10.02	&	9.91	\\
Ref-6	&	13.32	$\pm$	0.02	&	2.01	$\pm$	0.1	&	1.64	$\pm$	0.04	&	1.02	$\pm$	0.02	&	2.13	$\pm$	0.02	&	9.74	&	8.87	&	8.61	\\
Ref-7	&	13.61	$\pm$	0.02	&	0.12	$\pm$	0.06	&	0.57	$\pm$	0.03	&	0.32	$\pm$	0.02	&	0.66	$\pm$	0.02	&	12.5	&	12.27	&	12.2	\\
Ref-8	&	13.71	$\pm$	0.02	&	1.7	$\pm$	0.1	&	1.44	$\pm$	0.03	&	0.82	$\pm$	0.02	&	1.56	$\pm$	0.02	&	10.98	&	10.22	&	10.06	\\
Ref-9	&	14	$\pm$	0.02	&	0.28	$\pm$	0.08	&	0.7	$\pm$	0.03	&	0.38	$\pm$	0.02	&	0.77	$\pm$	0.02	&	12.64	&	12.4	&	12.31	\\
HIP 98492	&	11.58	$\pm$	0.02	&	-0.03	$\pm$	0.06	&	0.65	$\pm$	0.03	&	0.39	$\pm$	0.02	&	0.8	$\pm$	0.02	&	10.15	&	9.8	&	9.72	\\
\hline
\multicolumn{9}{c}{HIP 103269} \\		
\hline
Ref-1	&	13.04	$\pm$	0.02	&	0.85	$\pm$	0.1	&	0.99	$\pm$	0.04	&	0.59	$\pm$	0.03	&	1.13	$\pm$	0.03	&	11.13	&	10.61	&	10.54	\\
Ref-2	&	12.37	$\pm$	0.02	&	0.1	$\pm$	0.1	&	0.65	$\pm$	0.04	&	0.39	$\pm$	0.03	&	0.78	$\pm$	0.03	&	11.11	&	10.79	&	10.75	\\
Ref-3	&	14.64	$\pm$	0.02	&	0.34	$\pm$	0.11	&	0.8	$\pm$	0.04	&	0.46	$\pm$	0.03	&	0.93	$\pm$	0.03	&	13.09	&	12.7	&	12.7	\\
Ref-4	&	13.36	$\pm$	0.02	&	0.18	$\pm$	0.1	&	0.67	$\pm$	0.04	&	0.37	$\pm$	0.03	&	0.77	$\pm$	0.03	&	12.11	&	11.83	&	11.78	\\
Ref-5	&	15.46	$\pm$	0.02	&	-.-	$\pm$	2.61	&	0.07	$\pm$	1.85	&	0.03	$\pm$	3.58	&	0.03	$\pm$	9.42	&	8.13	&	7.69	&		\\
Ref-6	&	14.14	$\pm$	0.02	&	0.47	$\pm$	0.11	&	0.82	$\pm$	0.04	&	0.46	$\pm$	0.03	&	0.93	$\pm$	0.03	&	12.68	&	12.31	&	12.29	\\
Ref-7	&	11.59	$\pm$	0.02	&	0.03	$\pm$	0.1	&	0.55	$\pm$	0.04	&	0.3	$\pm$	0.03	&	0.65	$\pm$	0.03	&	10.56	&	10.29	&	10.26	\\
Ref-8	&	12.77	$\pm$	0.02	&	0.24	$\pm$	0.1	&	0.72	$\pm$	0.04	&	0.35	$\pm$	0.03	&	0.8	$\pm$	0.03	&	11.43	&	11.11	&	11.05	\\
Ref-9	&	13.49	$\pm$	0.02	&	0.26	$\pm$	0.1	&	0.77	$\pm$	0.04	&	0.39	$\pm$	0.03	&	0.83	$\pm$	0.03	&	12.13	&	11.73	&	11.71	\\
HIP 103269	&	10.28	$\pm$	0.02	&	-0.45	$\pm$	0.1	&	0.64	$\pm$	0.04	&	0.38	$\pm$	0.03	&	0.79	$\pm$	0.03	&	9.03	&	8.7	&	8.61	\\
\hline
\multicolumn{9}{c}{HIP 106924} \\	
\hline
Ref-1	&	14.64	$\pm$	0.02	&	0.5	$\pm$	0.1	&	1.06	$\pm$	0.04	&	0.7	$\pm$	0.03	&	1.28	$\pm$	0.03	&	12.47	&	12.05	&	11.89	\\
Ref-2	&	13.25	$\pm$	0.02	&	1.71	$\pm$	0.1	&	1.64	$\pm$	0.04	&	0.97	$\pm$	0.03	&	1.86	$\pm$	0.03	&	10.09	&	9.28	&	9.07	\\
Ref-3	&	14.73	$\pm$	0.02	&	1.5	$\pm$	0.11	&	1.54	$\pm$	0.05	&	0.92	$\pm$	0.03	&	1.76	$\pm$	0.03	&	11.66	&	11	&	10.78	\\
Ref-4	&	14.3	$\pm$	0.02	&	0.35	$\pm$	0.1	&	0.76	$\pm$	0.04	&	0.47	$\pm$	0.03	&	0.97	$\pm$	0.03	&	12.65	&	12.48	&	12.33	\\
Ref-5	&	14.63	$\pm$	0.02	&	1.68	$\pm$	0.12	&	1.61	$\pm$	0.04	&	0.93	$\pm$	0.03	&	1.75	$\pm$	0.03	&	11.64	&	10.89	&	10.68	\\
Ref-6	&	11.44	$\pm$	0.02	&	-0.02	$\pm$	0.1	&	0.56	$\pm$	0.04	&	0.28	$\pm$	0.03	&	0.62	$\pm$	0.03	&	10.41	&	10.17	&	10.14	\\
Ref-7	&	14.1	$\pm$	0.02	&	1.74	$\pm$	0.11	&	1.8	$\pm$	0.04	&	1.05	$\pm$	0.03	&	2.02	$\pm$	0.03	&	10.59	&	9.71	&	9.47	\\
HIP 106924	&	10.42	$\pm$	0.02	&	-0.36	$\pm$	0.1	&	0.57	$\pm$	0.04	&	0.43	$\pm$	0.03	&	0.87	$\pm$	0.03	&	9.04	&	8.86	&	8.57	\\
\hline
\multicolumn{9}{c}{HIP 108200} \\		
\hline	
Ref-1	&	13.06	$\pm$	0.02	&	0.25	$\pm$	0.1	&	0.69	$\pm$	0.04	&	0.37	$\pm$	0.03	&	0.76	$\pm$	0.03	&	11.78	&	11.51	&	11.4	\\
Ref-2	&	13.65	$\pm$	0.02	&	1.17	$\pm$	0.1	&	1.15	$\pm$	0.04	&	0.6	$\pm$	0.03	&	1.18	$\pm$	0.03	&	11.57	&	10.97	&	10.88	\\
Ref-3	&	11.15	$\pm$	0.02	&	1.32	$\pm$	0.1	&	1.17	$\pm$	0.04	&	0.65	$\pm$	0.03	&	1.22	$\pm$	0.03	&	9.06	&	8.49	&	8.34	\\
Ref-4	&	12.18	$\pm$	0.02	&	1.41	$\pm$	0.1	&	1.3	$\pm$	0.04	&	0.71	$\pm$	0.03	&	1.36	$\pm$	0.03	&	9.79	&	9.07	&	8.95	\\
Ref-5	&	13.18	$\pm$	0.02	&	0.16	$\pm$	0.1	&	0.65	$\pm$	0.04	&	0.35	$\pm$	0.03	&	0.72	$\pm$	0.03	&	11.96	&	11.65	&	11.6	\\
Ref-6	&	13.89	$\pm$	0.02	&	1.42	$\pm$	0.1	&	1.25	$\pm$	0.04	&	0.66	$\pm$	0.03	&	1.24	$\pm$	0.03	&	11.7	&	11.07	&	10.93	\\
Ref-7	&	15.29	$\pm$	0.02	&	0.16	$\pm$	0.1	&	0.67	$\pm$	0.04	&	0.41	$\pm$	0.03	&	0.84	$\pm$	0.03	&	13.88	&	13.61	&	13.43	\\
HIP 108200	&	10.97	$\pm$	0.02	&	-0.09	$\pm$	0.1	&	0.67	$\pm$	0.04	&	0.4	$\pm$	0.03	&	0.83	$\pm$	0.03	&	9.58	&	9.21	&	9.1	\\
\hline																
\enddata
\tablenotetext{a}{ Target saturated on all images}
\end{deluxetable}

\begin{figure}
\gridline{\fig{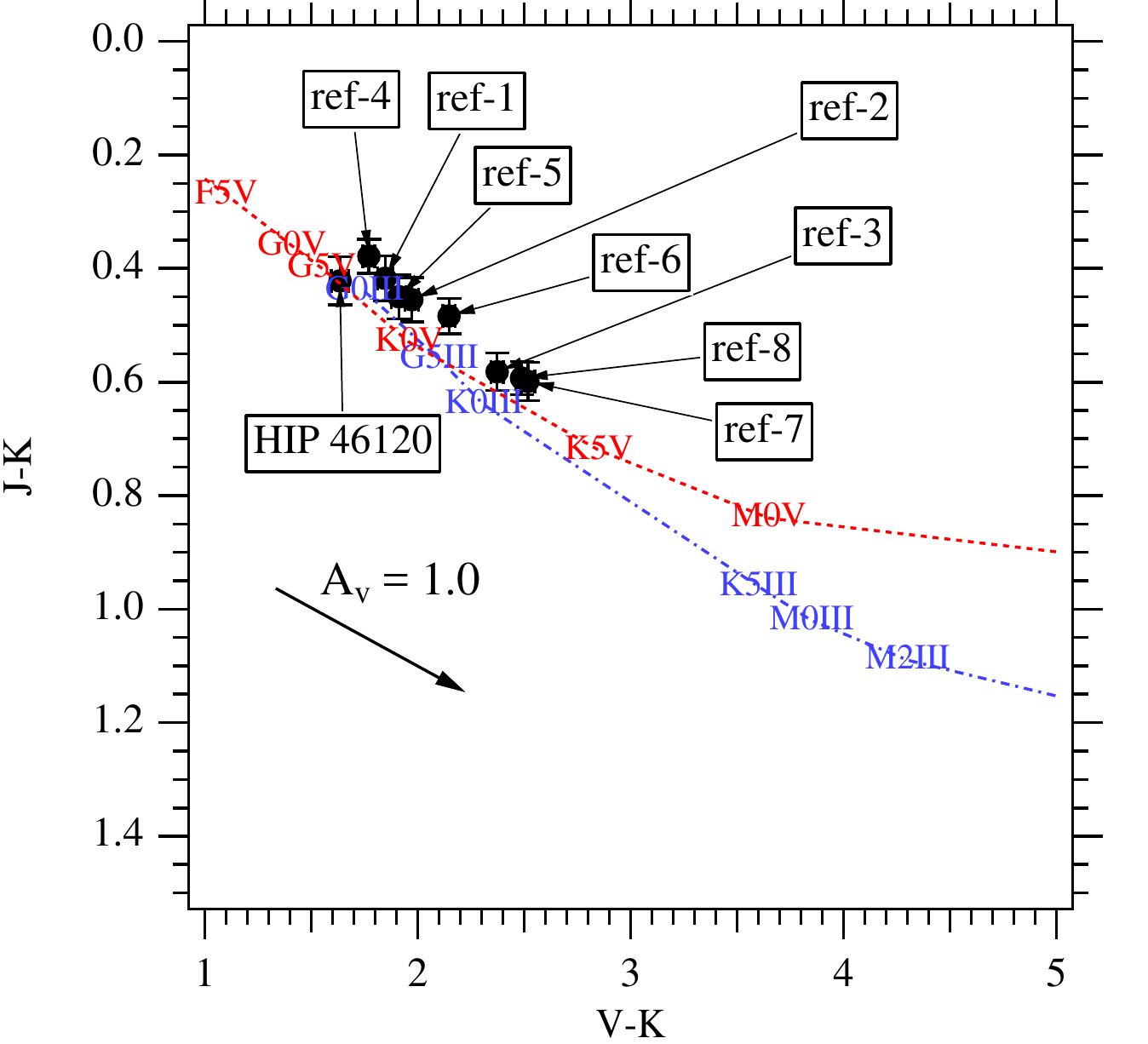}{0.33\textwidth}{}
          \fig{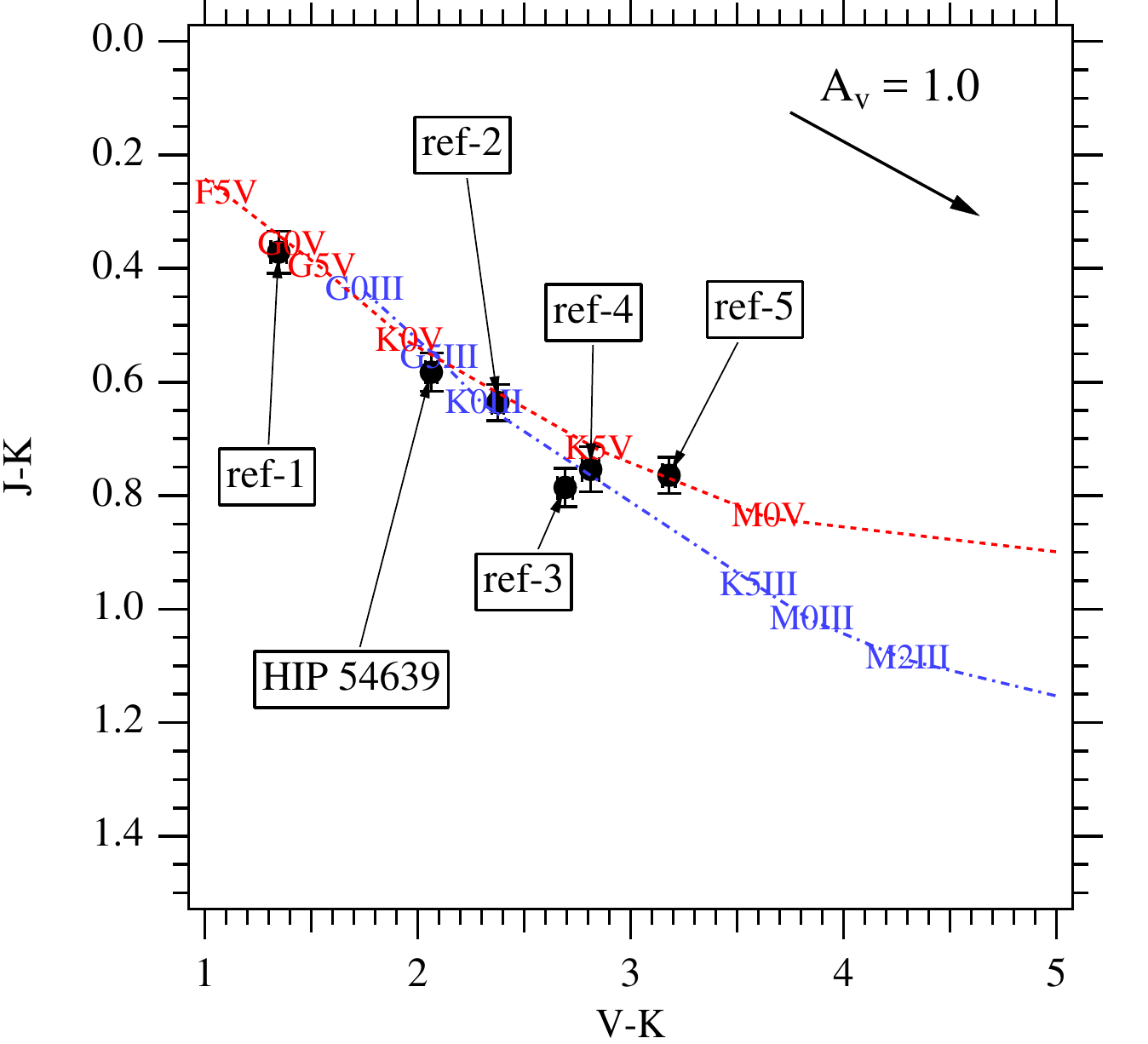}{0.33\textwidth}{}
          \fig{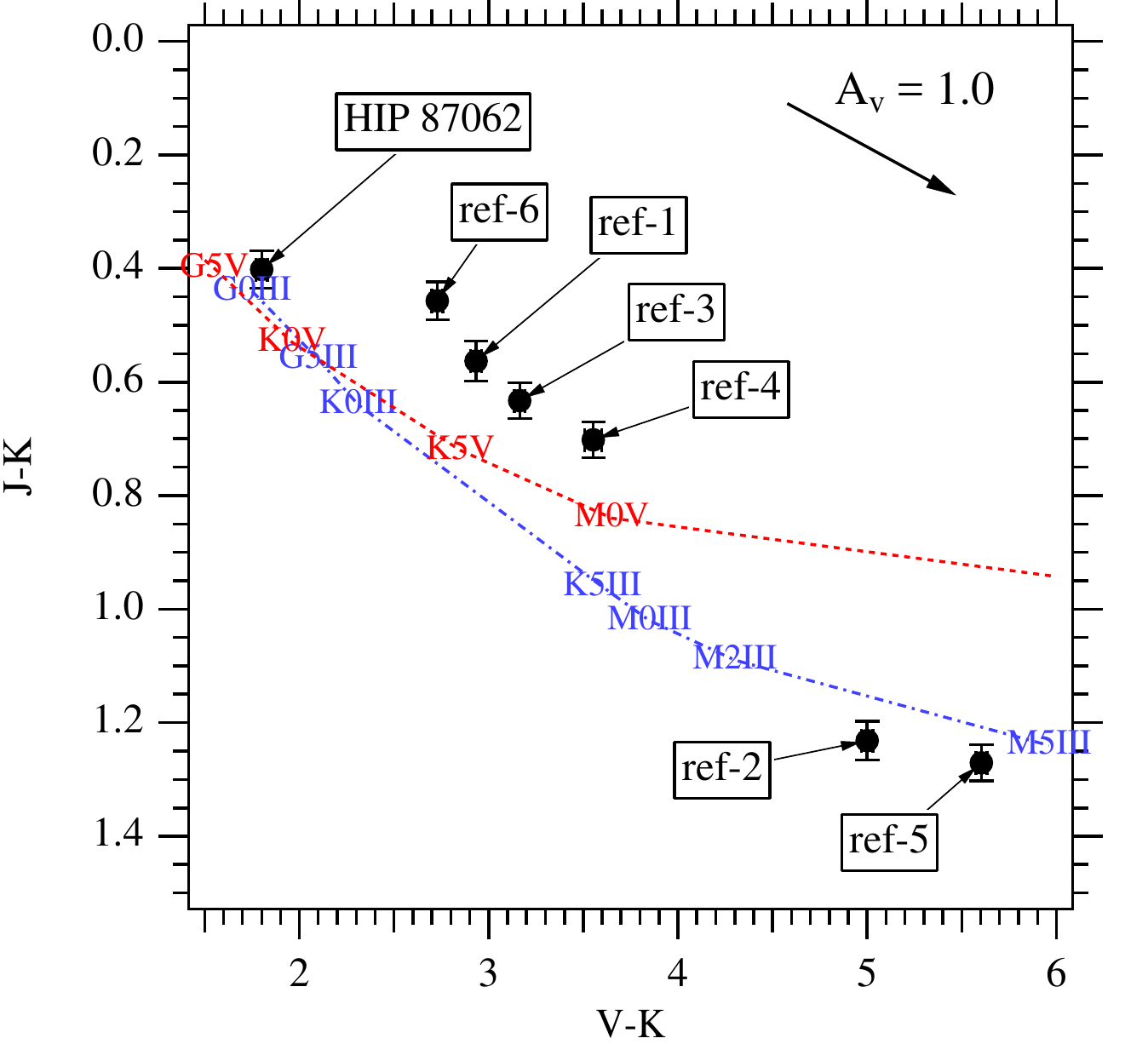}{0.33\textwidth}{}
                    }
 \gridline{\fig{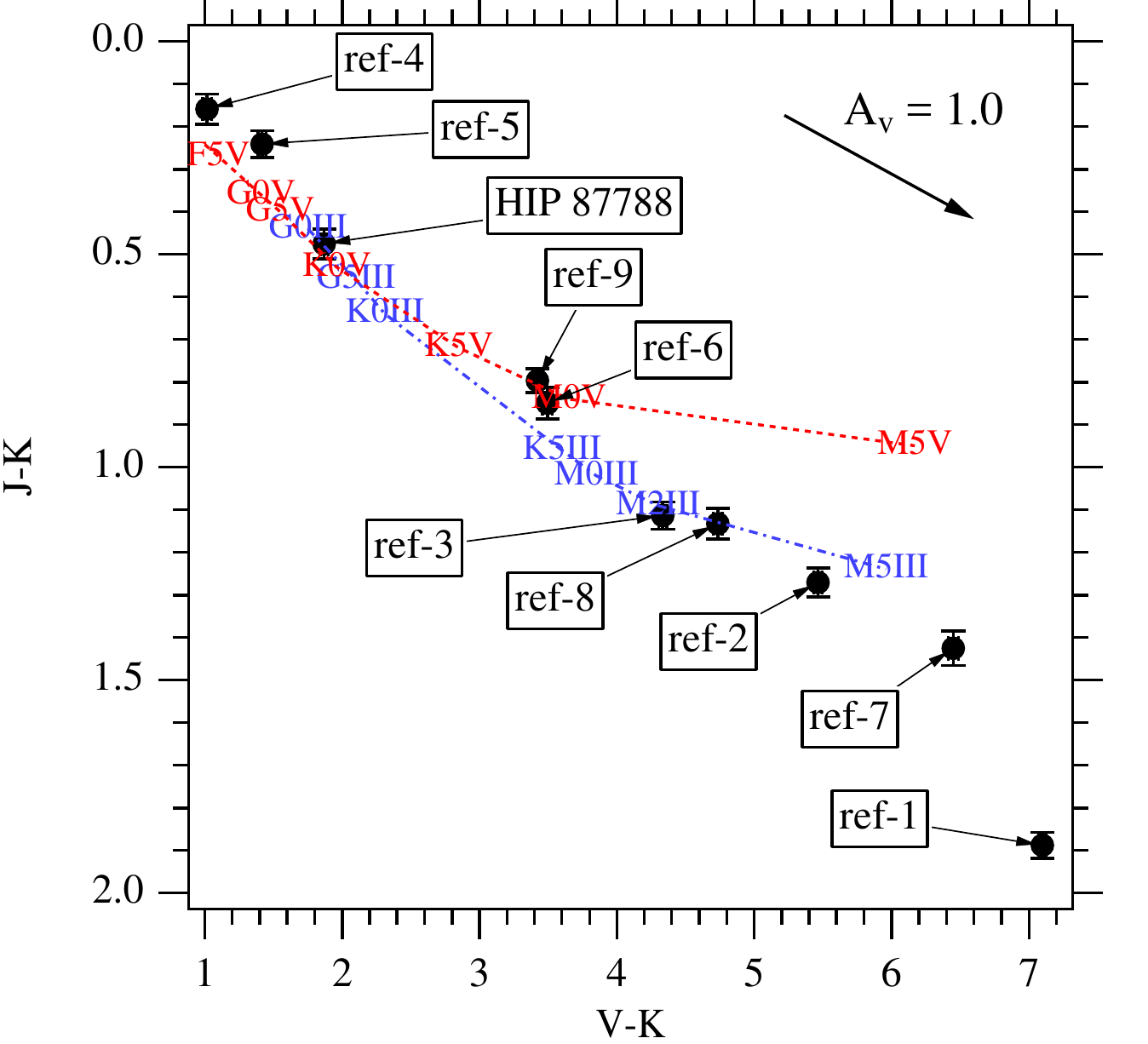}{0.33\textwidth}{}
   \fig{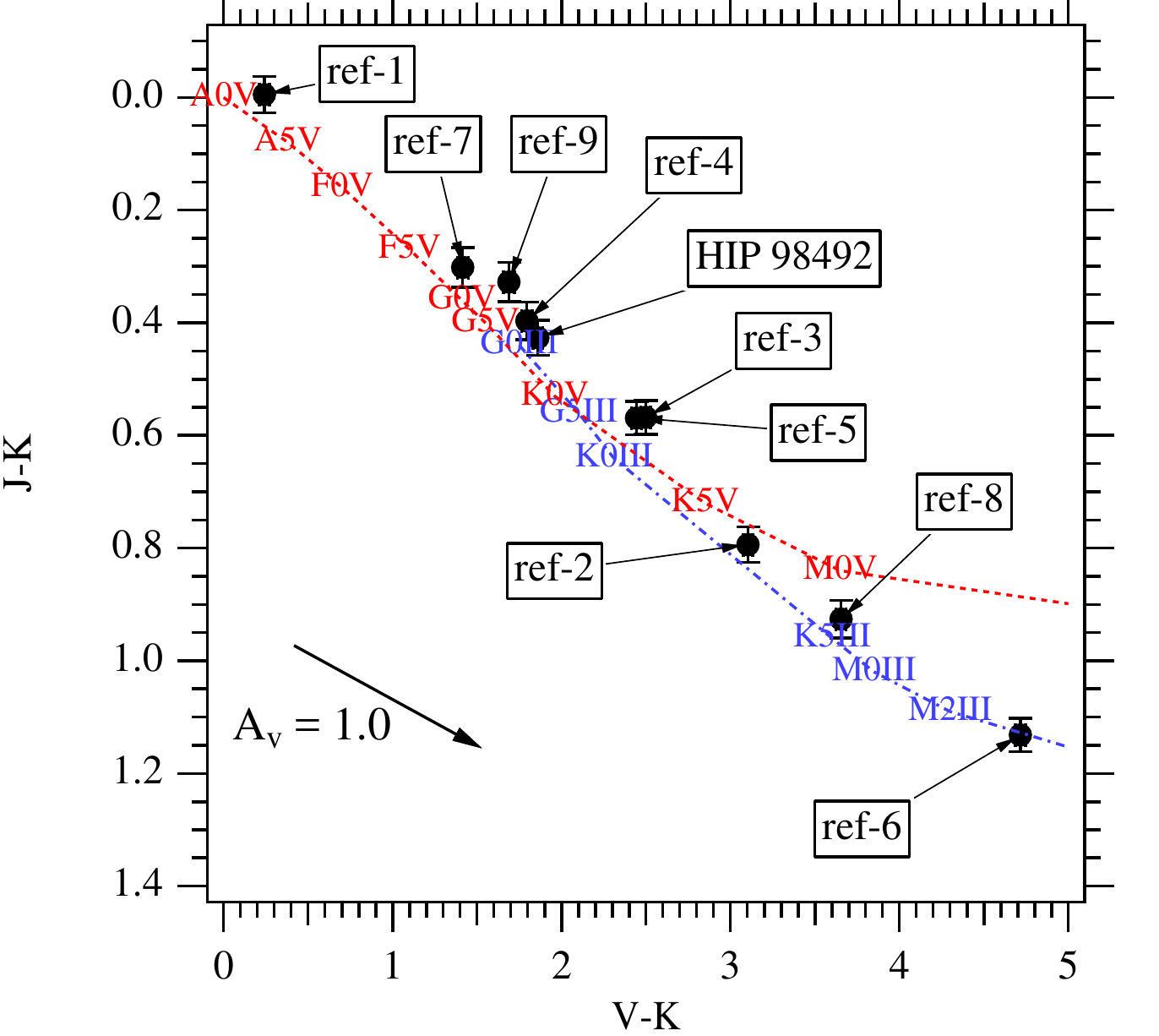}{0.33\textwidth}{}
          \fig{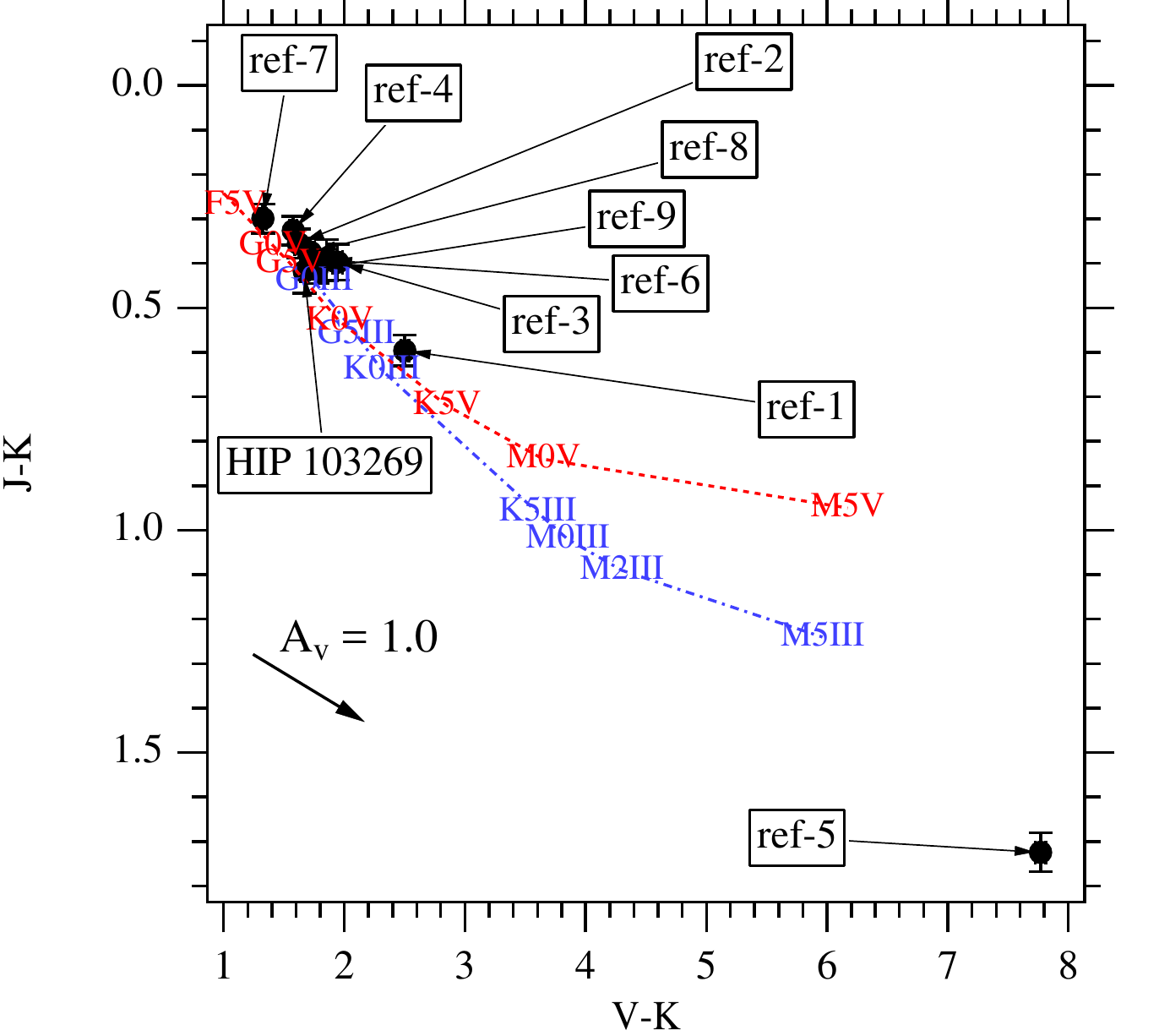}{0.33\textwidth}{}
          }
 \gridline{\fig{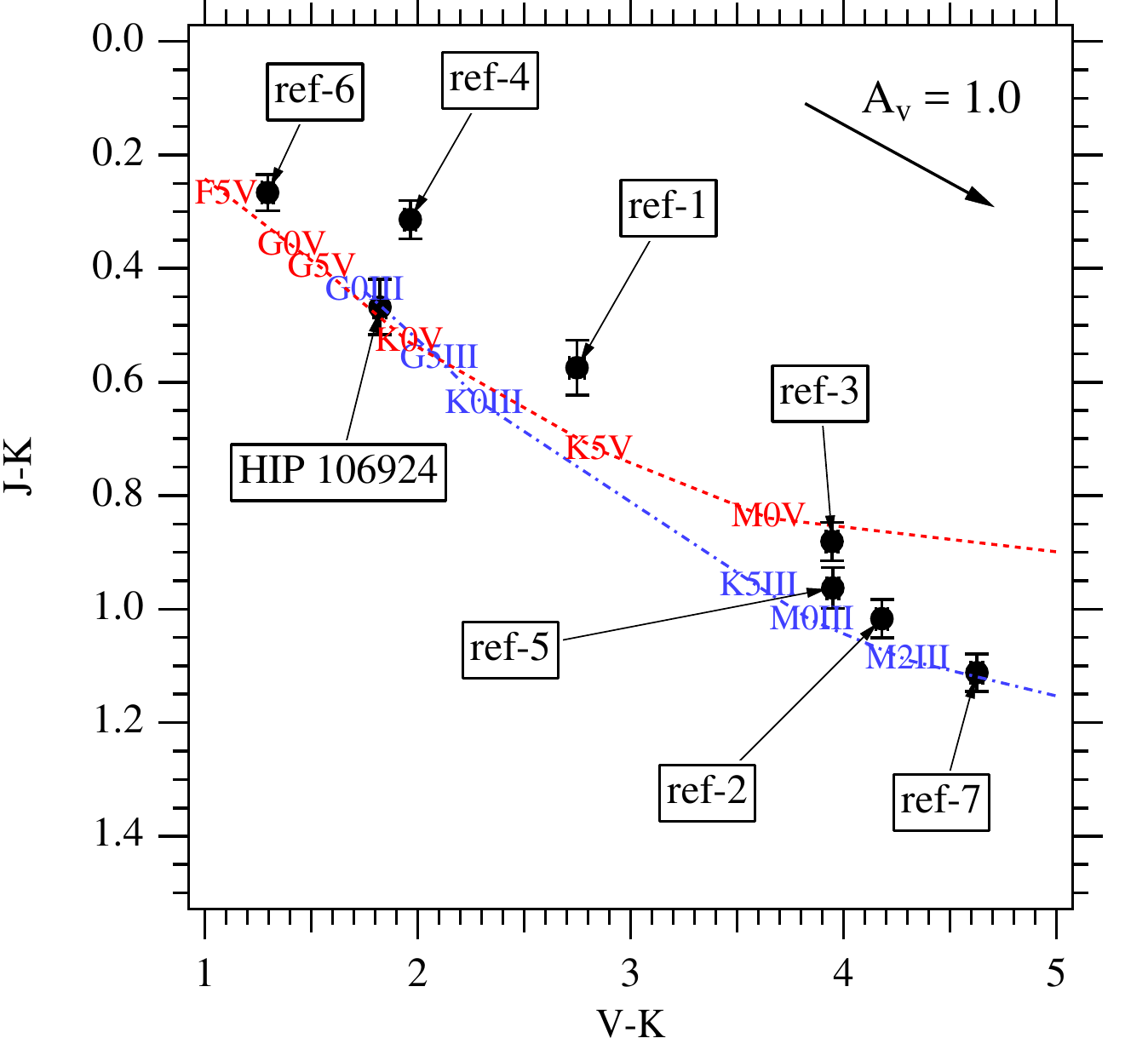}{0.33\textwidth}{}
         \fig{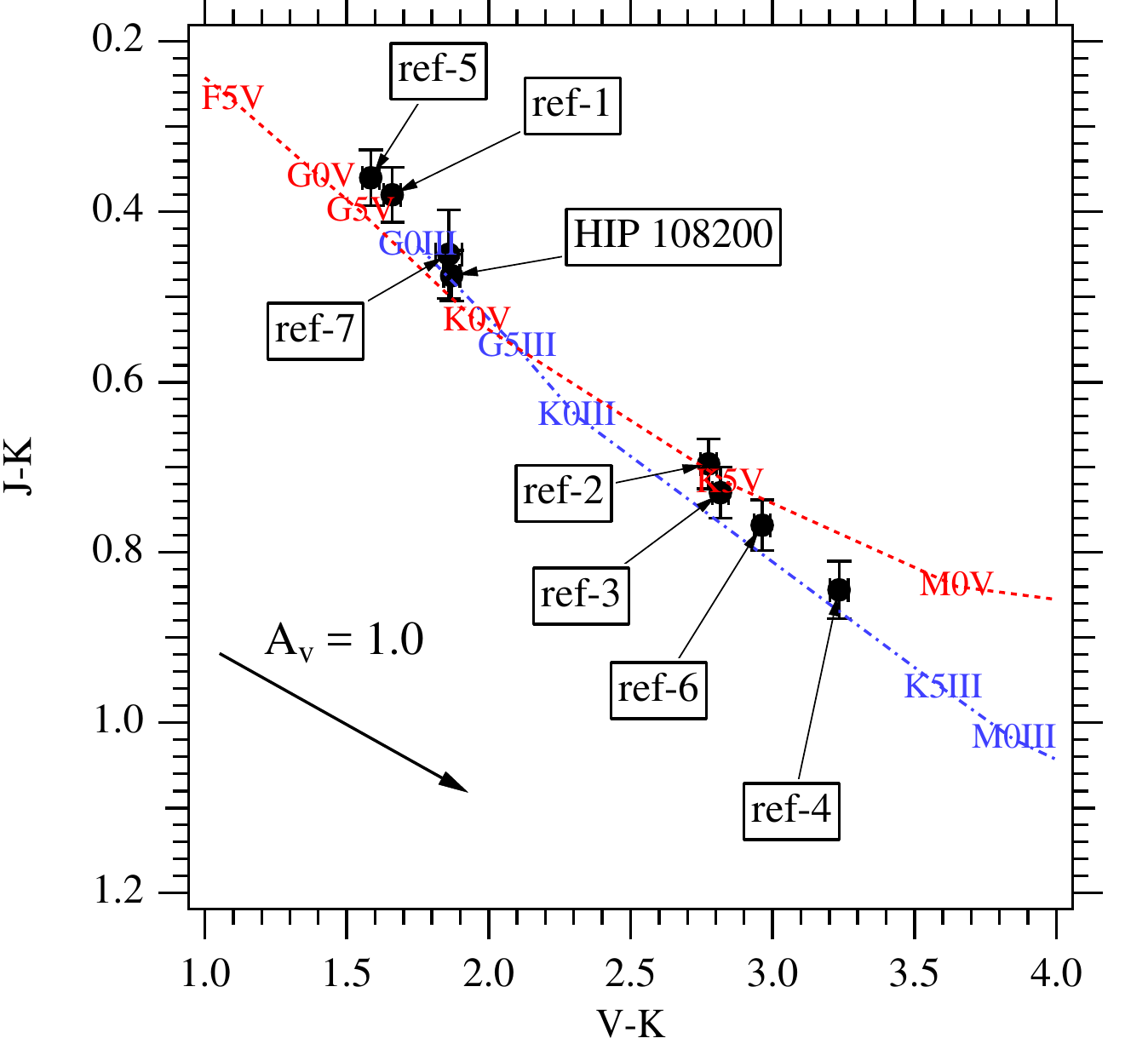}{0.33\textwidth}{}
          }
\caption{$(J-K)$ vs. $(V-K)$ color-color diagram for stars in the \hst field identified in 
Table~\ref{tab:IR}. The dashed line is the locus of  dwarf
(luminosity class V) stars of various spectral types; the dot-dashed line is for 
giants (luminosity class III).  }
\label{fig:JmKvsVmKrefs}
\end{figure}

\subsection{Spectroscopy, Luminosity Class and Reduced Proper Motion}

The derived absolute magnitudes are crucially dependent on the assumed stellar 
luminosities, a parameter impossible to obtain for all but the latest type stars 
using only  Figure~\ref{fig:JmKvsVmKrefs}.
To aid in the spectral classification of the reference stars we have compiled 
an extensive set of template spectra covering a large range of temperature 
and luminosity classes in support of our various FGS programs on both the 
APO 3.5 m, and the Blanco 4 m. We perform MK classification of each of the 
reference frame stars with respect to these templates, as well as use the 
temperature and luminosity classification characteristics listed in  \citet{yamashita78}. For the DIS spectra, our temperature
classifications are generally good to $\pm$ 1 subclass. For the lower
resolution CTIO data, however, there is more uncertainty, and we generally 
obtain spectral classifications with uncertainties of $\pm$ 2 subclasses.

To confirm the luminosity classes we  obtain  PPMXL  proper 
motions \citep{Roe10} for a 1$\degree$ square field centered on the targets shown in Figure \ref{fig:RPMvsJmK}, and then 
iteratively use the technique of reduced proper motion \citep{Yon03,Gou03}  to distinguish 
between giants and dwarfs. 

\begin{figure}
\gridline{\fig{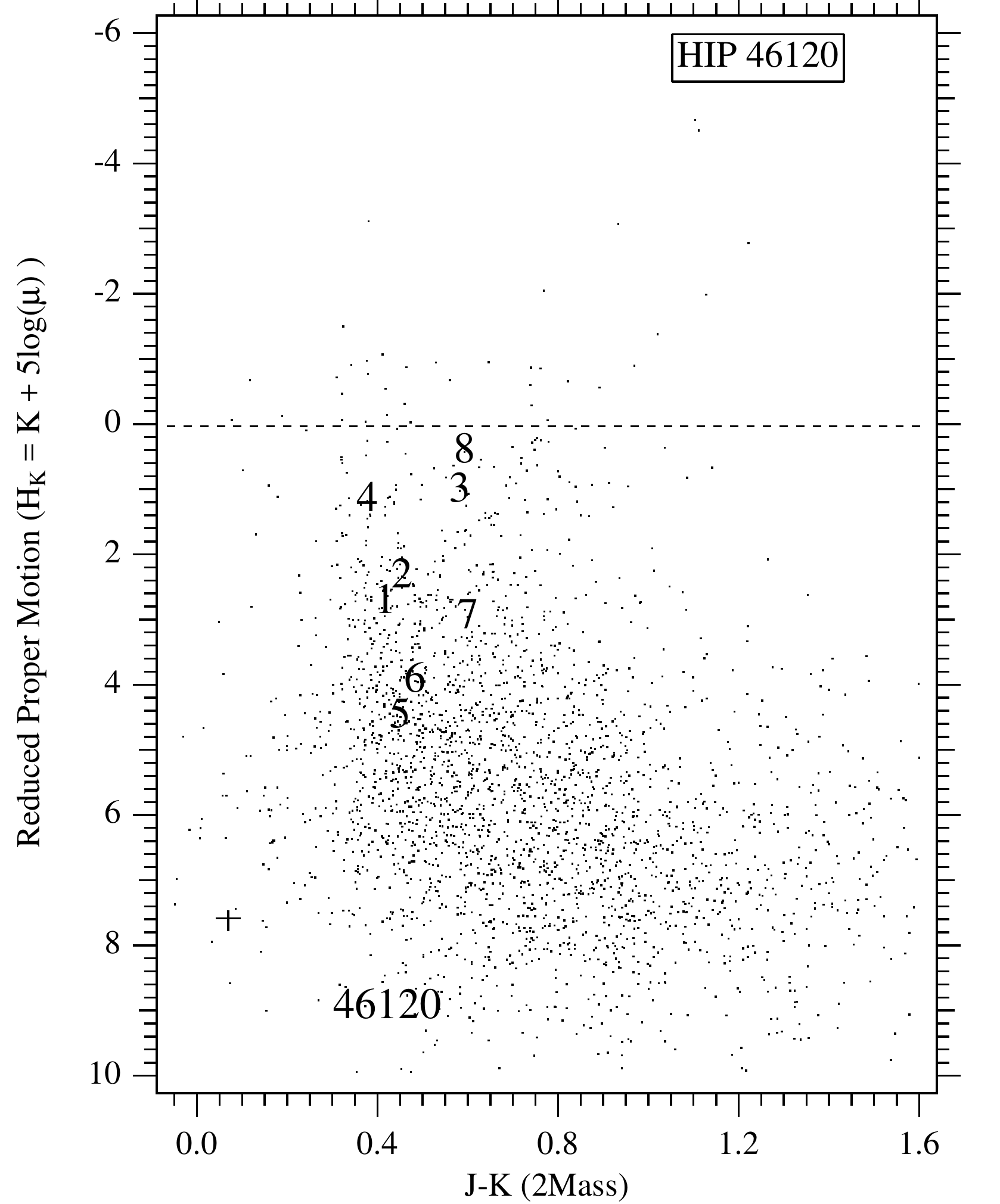}{0.3\textwidth}{}
          \fig{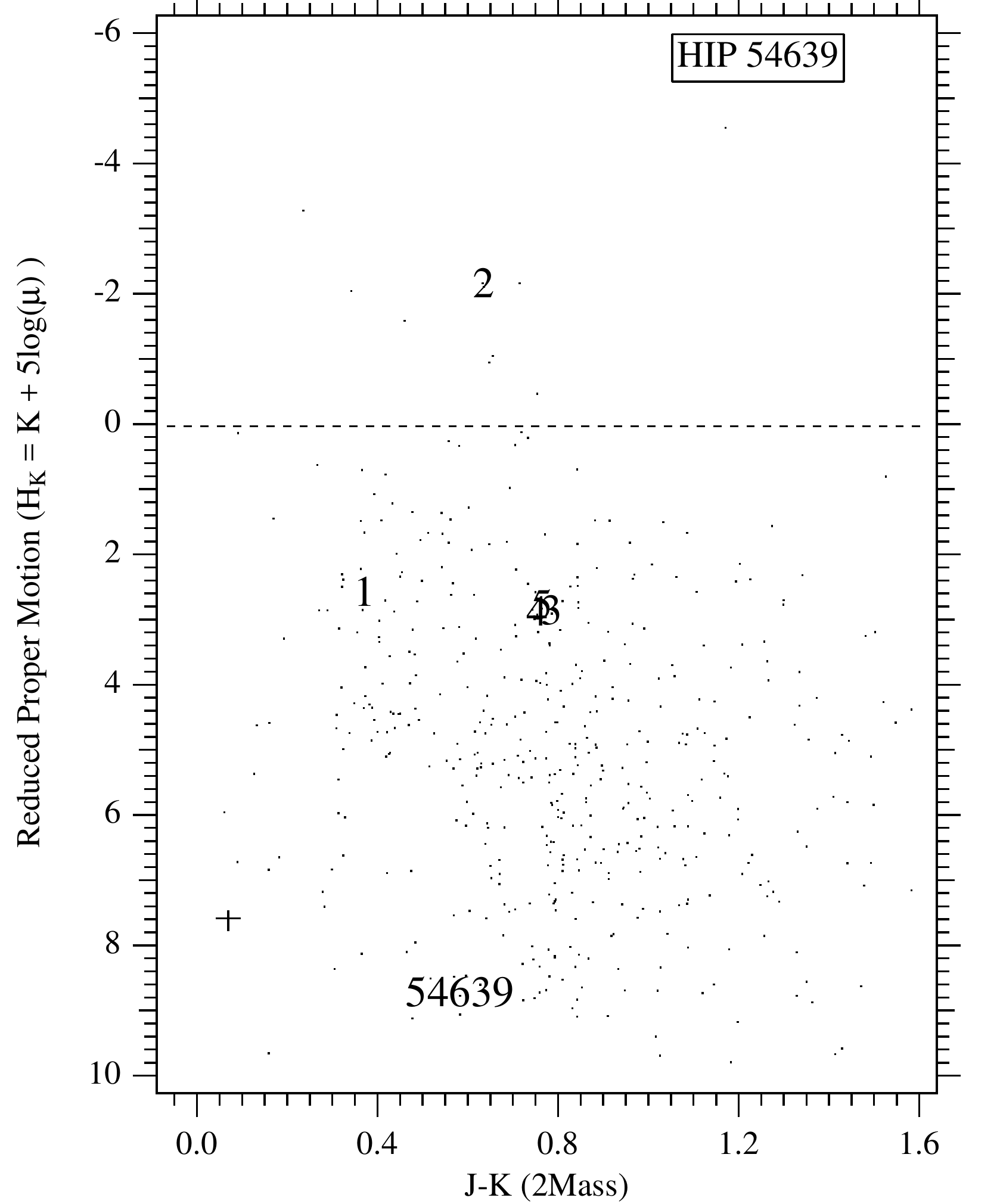}{0.3\textwidth}{}
          \fig{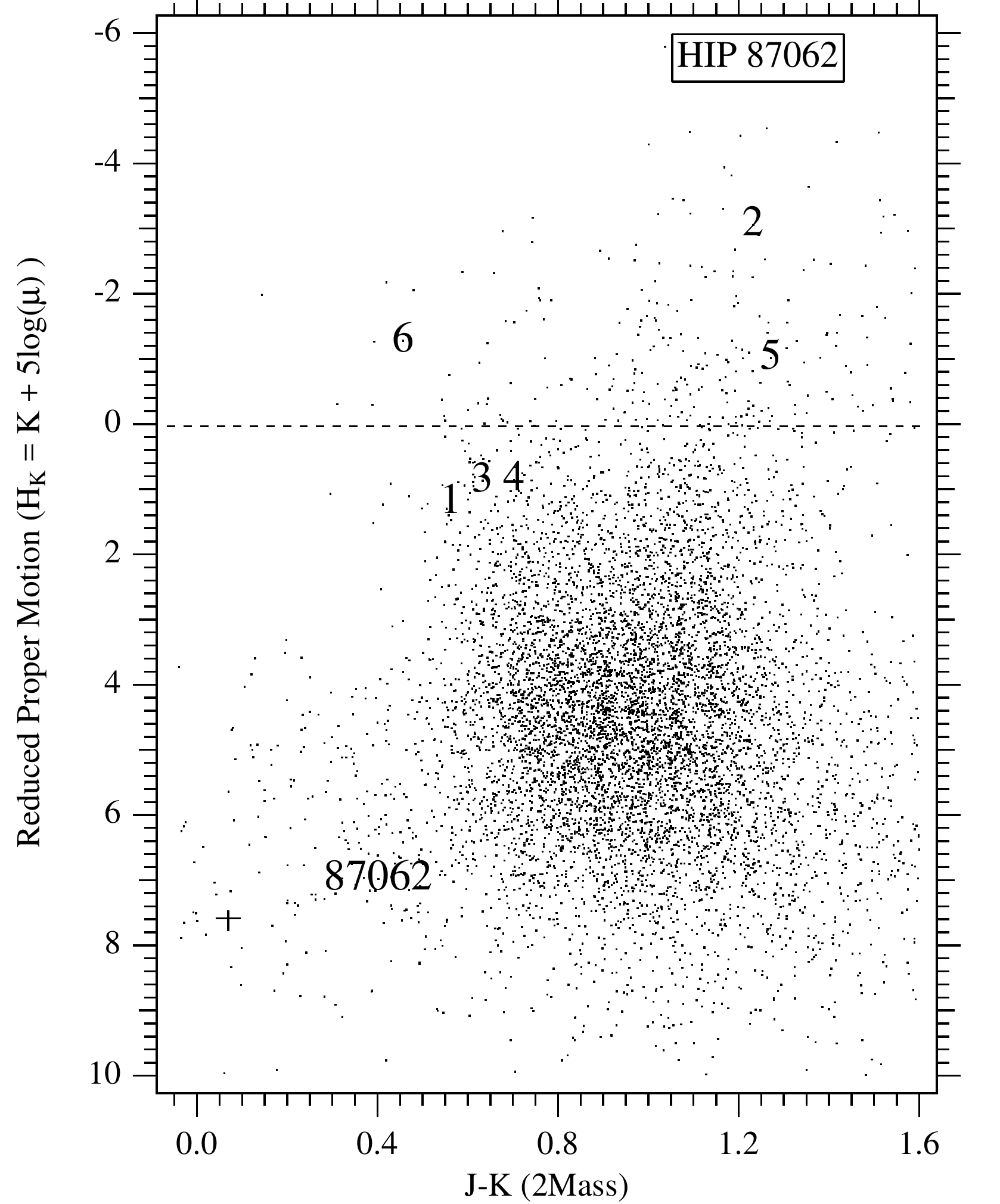}{0.3\textwidth}{}
                    }
 \gridline{\fig{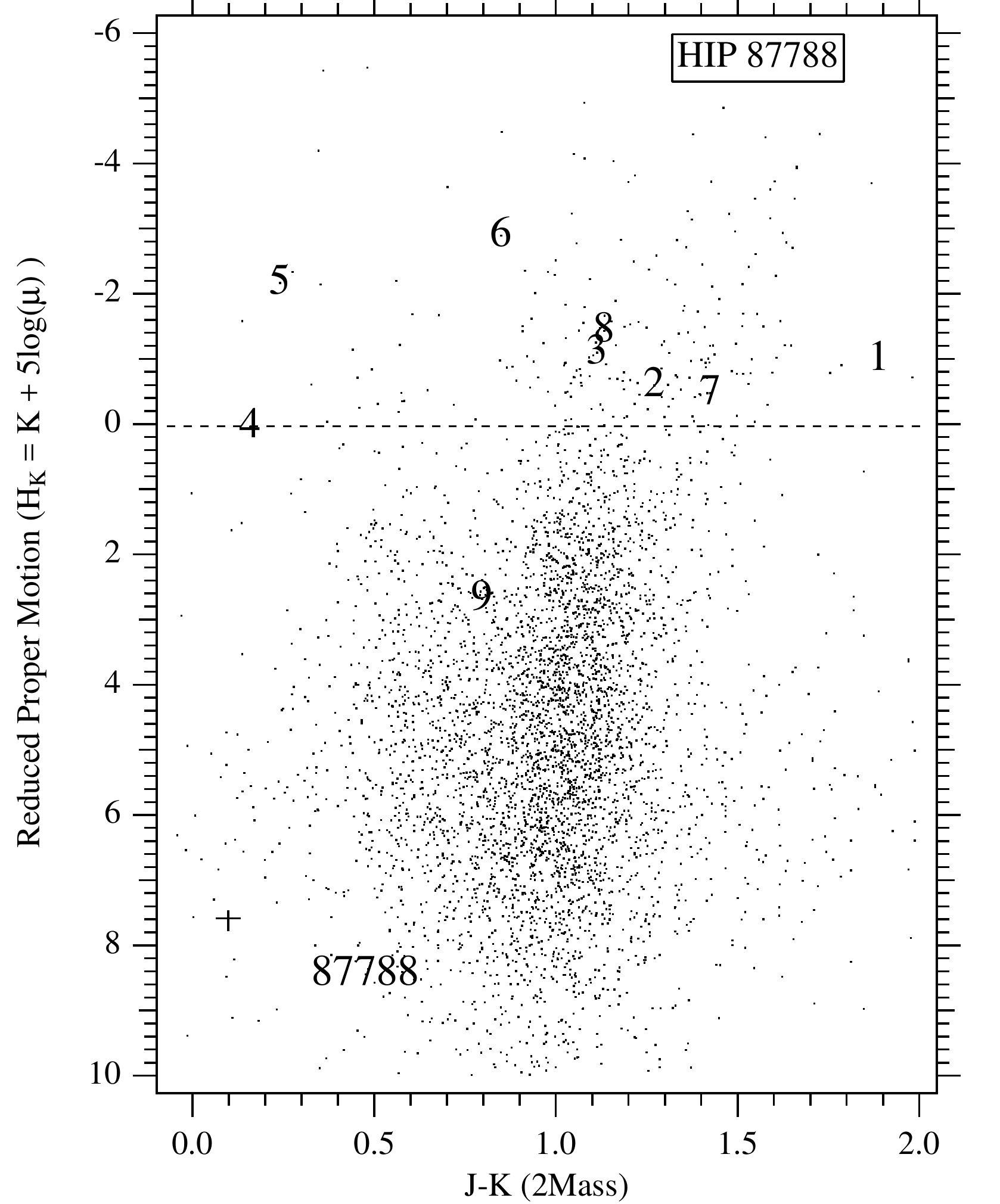}{0.3\textwidth}{}
   \fig{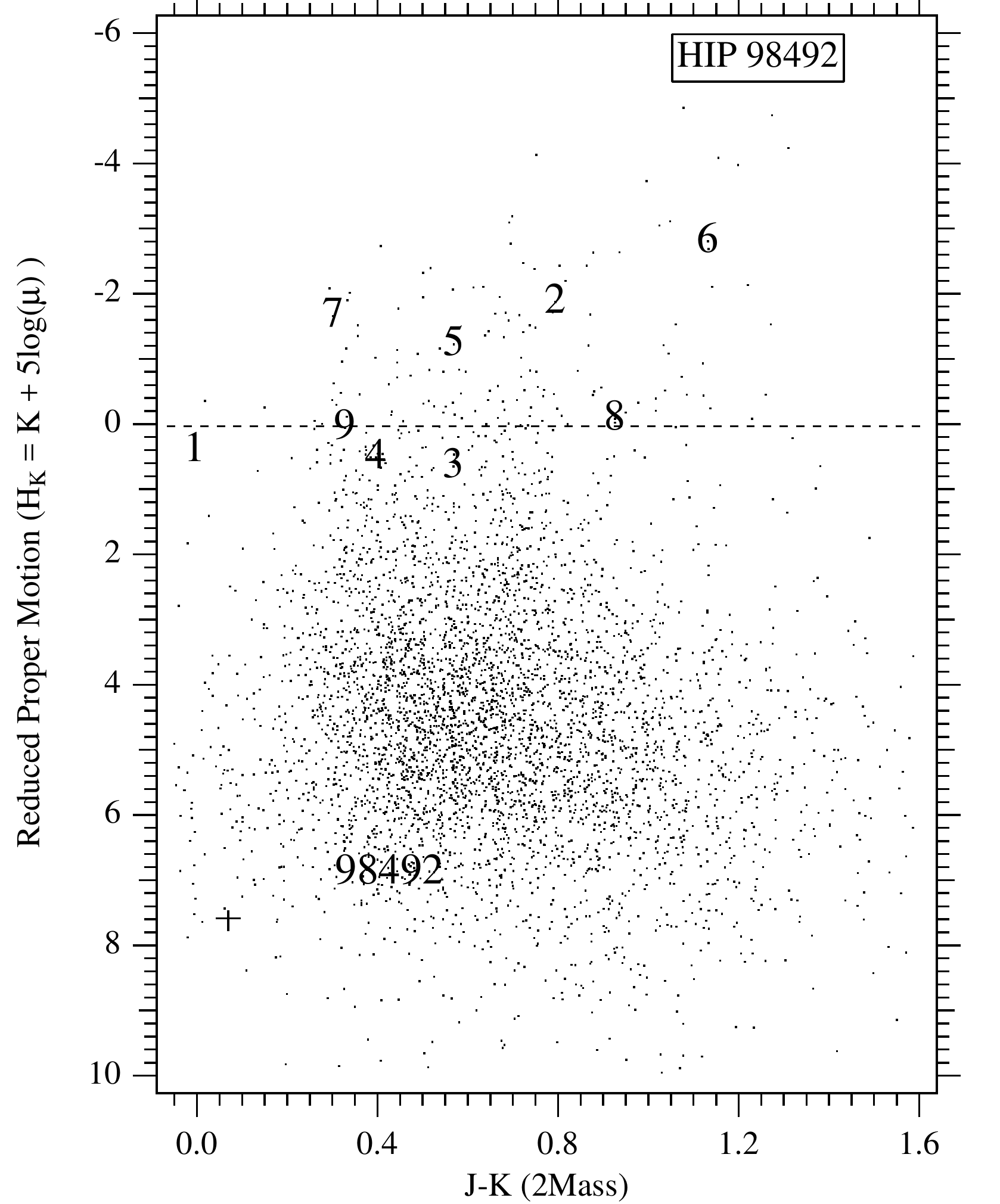}{0.3\textwidth}{}
          \fig{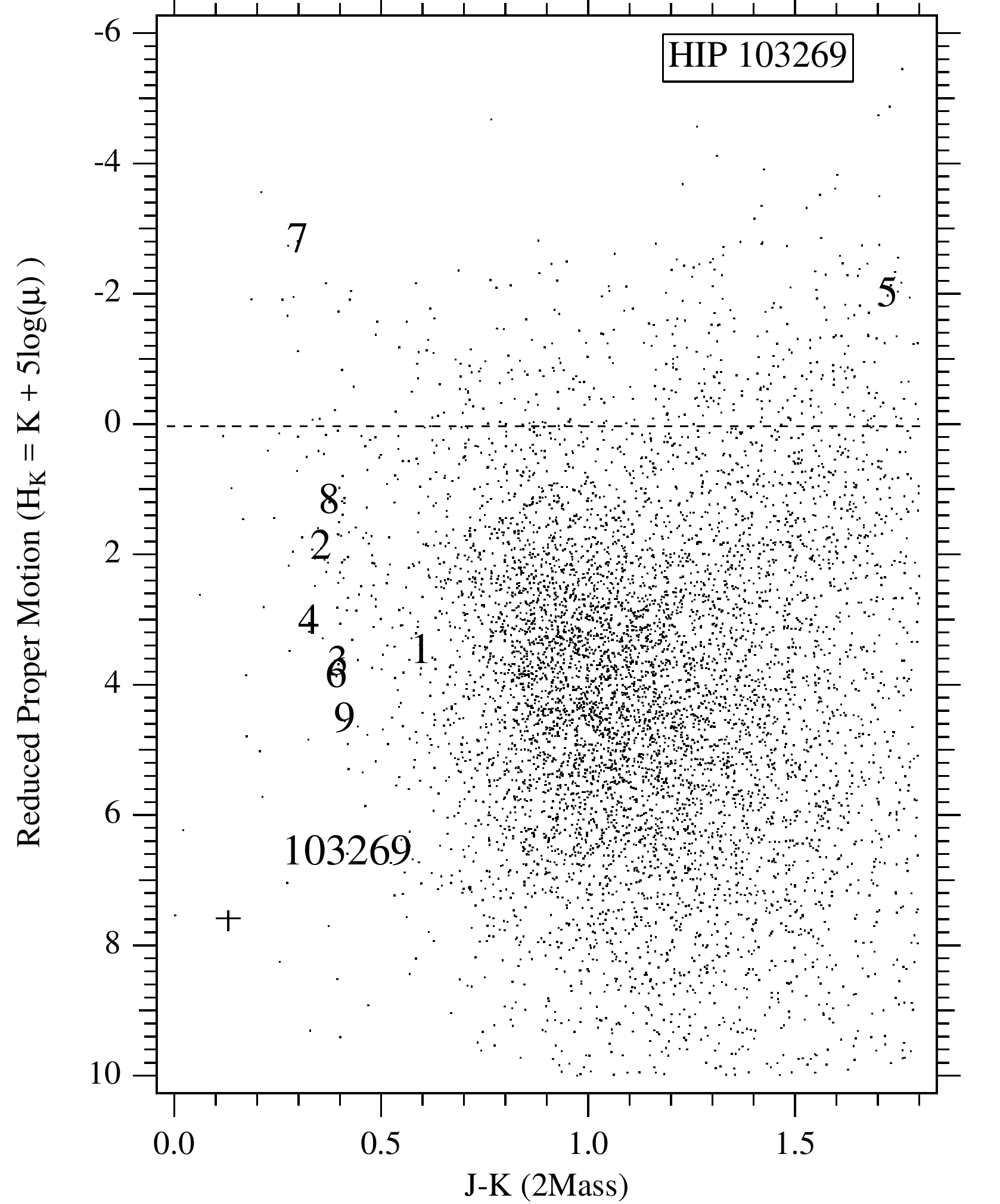}{0.3\textwidth}{}
          }
 \gridline{\fig{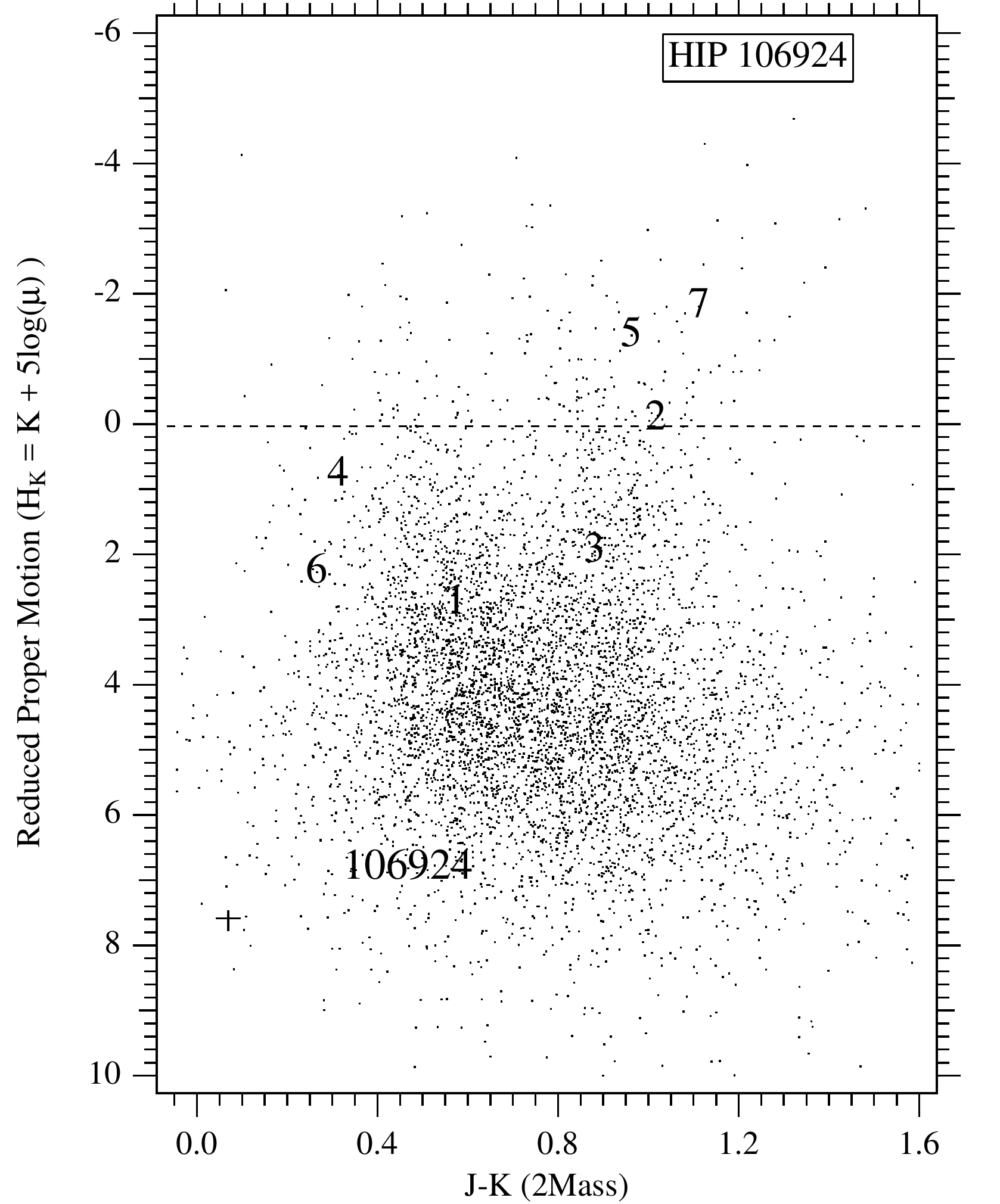}{0.3\textwidth}{}
         \fig{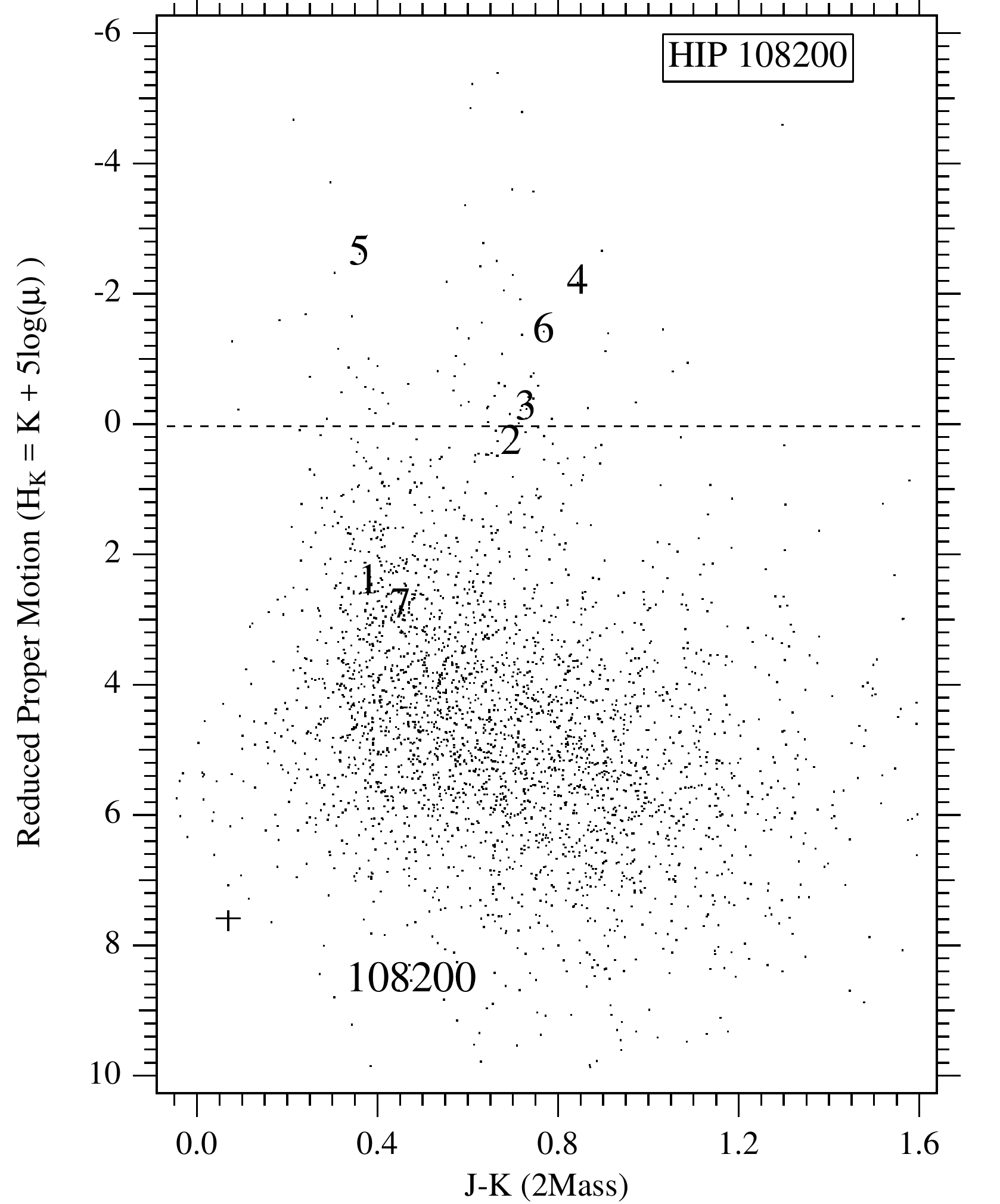}{0.3\textwidth}{}
          }
\begin{footnotesize}
\caption{Reduced proper motion diagram for 5800 stars in a 1\arcdeg ~field centered
on the metal-poor target stars. Star identifications are in Table~\ref{tab:IR}. For a given spectral type, 
giants and sub-giants have more negative $H_K$ values and are redder than dwarfs in 
$(J-K)$.  $H_K$ values are derived from `Final' proper motions in Table~\ref{tab:PM}. 
The small cross at the lower left represents a typical $(J-K)$ error of 0.04 mag and $H_K$ error of 0.17 mag. 
The horizontal dashed line is a giant-dwarf demarcation derived from a statistical analysis of the 
Tycho input catalog (Ciardi 2004, private communication). \label{fig:RPMvsJmK} }
\end{footnotesize}
\end{figure}

\subsection{Estimated Reference Frame Absolute Parallaxes}\label{sec:refpar}

With the spectral classification of the reference stars complete, we combine
the $UBVRI$ photometry we have obtained with $JHK$ photometry from 
2MASS, to derive the visual extinction to the sources using the reddening 
relationships from \citet{rieke85}.   Error bars
on the visual extinction are of order 10\%.
 Once determined, we estimate 
spectroscopic parallaxes using the absolute visual magnitude calibrations 
for main sequence stars listed in \citet{Houk97}, and for giant stars 
using \citet{cox00}.  These spectroscopic reference star parallaxes are input to our model with errors of 20\%.  The error bars on the $V$ magnitudes 
are $\pm$ 0.02 mag, and are $\pm$ 0.04 mag for the ($B - V$) colors. Table~\ref{tab:SPP} lists  all reference star absolute parallax estimates,  spectral types and luminosities. 

\clearpage

\begin{deluxetable}{lrrrrrrrrrr}
\tabletypesize{\scriptsize}
\tablecaption{ Adopted
Spectrophotometric and $\hst$ Resolved Reference Star Parallaxes  \label{tab:SPP}}
\tablehead{
\colhead{ID}& 
\colhead{$\alpha_{\rm 2000}$ } &
\colhead{$\delta_{\rm 2000}$} &
\colhead{Sp. T.\tablenotemark{a}}&
\colhead{V} &
\colhead{B-V}&
\colhead{$M_v$}&
\colhead{$A_v$}&
\colhead{D}&
\colhead{$\pi_{spec}$}&
\colhead{$\pi_{HST}$\tablenotemark{b}}\\
\colhead{}& 
\colhead{ } &
\colhead{} &\colhead{}&
\colhead{} &\colhead{}&\colhead{}&
\colhead{}&
\colhead{(pc)}&
\colhead{(mas)}&
\colhead{(mas)}
}
\startdata
\multicolumn{11}{c}{HIP 46120} \\													
\hline																	
Ref-1	&	 09:24:17.86	&	$-$80:36:17.5	&	G2V	&	14.34	&	0.74	&	4.56	&	0.37	&759	&	1.317	&	1.436	\\
Ref-2	&	 09:24:32.59	&	$-$80:34:07.1	&	G5V	&	13.96	&	0.81	&	4.93	&	0.43	&529	&	1.89	&	1.550	\\
Ref-3	&	 09:24:13.14	&	$-$80:32:58.5	&	K1V	&	14.02	&	1	&	6.13	&	0.32	&329	&	3.035	&	2.355	\\
Ref-4	&	 09:24:09.40	&	$-$80:31:59.6	&	G2V	&	12.58	&	0.75	&	4.56	&	0.27	&353	&	2.829	&	2.961	\\
Ref-5	&	 09:24:00.60	&	$-$80:31:47.9	&	G5V	&	14.76	&	0.82	&	4.93	&	0.38	&786	&	1.273	&	1.393	\\
Ref-6	&	 09:23:58.95	&	$-$80:30:45.2	&	G7V	&	13.48	&	0.89	&	5.32	&	0.44	&352	&	2.841	&	2.664	\\
Ref-7	&	 09:24:57.80	&	$-$80:29:29.1	&	K1V	&	14.91	&	1.01	&	6.13	&	0.46	&463	&	2.162	&	2.497	\\
Ref-8	&	 09:24:37.90	&	$-$80:28:24.5	&	K1V	&	13.21	&	1.03	&	6.13	&	0.43	&215	&	4.66	&	3.796	\\
\hline																	
\multicolumn{11}{c}{HIP 54639} \\													
\hline																	
Ref-1	&	 11:11:02.39	&	$+$06:24:15.7	&	F5V	&	13.87	&	0.47	&	3.34	&	0.1	&1182	&	0.847	&	0.843	\\
Ref-2	&	 11:11:19.49	&	$+$06:24:20.6	&	G9III	&	13.45	&	0.97	&	0.8	&	0.11	&3222	&	0.31	&	0.309	\\
Ref-3	&	 11:11:04.77	&	$+$06:26:39.2	&	K3V	&	14.72	&	1.09	&	6.75	&	0	&382	&	2.617	&	2.408	\\
Ref-4	&	 11:10:57.12	&	$+$06:26:37.0	&	K4V	&	14.72	&	1.06	&	7.12	&	0	&336	&	2.967	&	3.812	\\
Ref-5	&	 11:10:43.02	&	$+$06:28:16.5	&	K7V	&	13.23	&	1.24	&	8.23	&	0	&129	&	9.947	&	10.306	\\
\hline																	
\multicolumn{11}{c}{HIP 87062} \\													
\hline																	
Ref-1	&	 17:47:09.39	&	$-$08:45:06.2	&	F7V	&	13.36	&	1.07	&	3.72	&	1.89	&348	&	2.876	&	$-$\tablenotemark{c}	\\
Ref-2	&	 17:47:11.03	&	$-$08:44:37.6	&	K1III	&	13.32	&	1.88	&	0.6	&	2.65	&1009	&	0.992	&	$-$\tablenotemark{c}	\\
Ref-3	&	 17:47:16.18	&	$-$08:46:17.1	&	G0V	&	14.58	&	1.24	&	4.2	&	1.96	&490	&	2.04	&	2.034	\\
Ref-4	&	 17:47:25.99	&	$-$08:45:37.6	&	G1V	&	15.67	&	1.27	&	4.24	&	2.3	&650	&	1.541	&	1.549	\\
Ref-5	&	 17:47:33.27	&	$-$08:45:33.7	&	K2III	&	14.65	&	2.02	&	0.42	&	3.21	&1598	&	0.626	&	0.625	\\
Ref-6	&	 17:47:36.10	&	$-$08:47:58.0	&	F0V	&	13.77	&	1.05	&	2.4	&	2.19	&682	&	1.466	&	1.467	\\
\hline																	
\multicolumn{11}{c}{HIP 87788} \\													
\hline																	
Ref-1	&	 17:56:17.24	&	$-$16:25:36.1	&	Carbon*	&	14.45	&	2.35	&	-.-	&	-.- & ? & ? &	$-$\tablenotemark{c}	\\
Ref-2	&	 17:56:06.10	&	$-$16:23:38.1	&	K5III	&	14.57	&	2.13	&	-0.2	&	1.93	&3654	&	0.274	&	2.530	\\
Ref-3	&	 17:55:58.10	&	$-$16:23:24.9	&	K2III	&	10.55	&	1.63	&	0.42	&	1.89	&454	&	2.202	&	1.362	\\
Ref-4	&	 17:56:04.28	&	$-$16:24:53.1	&	A1V	&	11.98	&	0.38	&	1.01	&	1	&982	&	1.018	&	1.024	\\
Ref-5	&	 17:56:03.73	&	$-$16:26:12.4	&	A4V	&	11.04	&	0.54	&	1.57	&	1.01	&475	&	2.105	&	1.751	\\
Ref-6	&	 17:55:57.02	&	$-$16:24:41.2	&	K1III	&	11.55	&	1.39	&	0.6	&	1.05	&952	&	1.05	&	1.138	\\
Ref-7	&	 17:55:50.07	&	$-$16:23:33.2	&	M3III	&	13.49	&	2.06	&	-0.67	&	1.82	&2840	&	0.353	&	0.356	\\
Ref-8	&	 17:55:44.75	&	$-$16:25:45.0	&	K5III	&	13.75	&	1.85	&	-0.2	&	1.09	&3625	&	0.276	&	0.273	\\
Ref-9	&	 17:55:38.46	&	$-$16:23:47.5	&	G9.5V	&	13.47	&	1.3	&	5.79	&	1.63	&143	&	7.144	&$-$\tablenotemark{c}	\\
Ref-7r	&	17:55:42.35	&	$-$16:23:39.7	&	K7III	&	14.41	&	2.16	&	-0.3	&	2.26	&3067	&	0.326	&	0.325	\\
Ref-8r	&	17:55:41.08	&	$-$16:25:37.2	&	G1V	&	13.52	&	0.88	&		&		&498	&	2.007	&	2.533	\\
\hline																	
\multicolumn{11}{c}{HIP 98492} \\													
\hline
Ref-1	&	 20:00:49.97	&	$+$09:21:57.1	&	B8V	&	11.18	&	0.03	&	-0.25	&	0.45	&1561	&	0.641	&	0.561	\\
Ref-2	&	 20:00:42.85	&	$+$09:22:06.7	&	K2III	&	11.89	&	0.7	&	0.6	&	0.46	&1612	&	0.621	&	0.785	\\
Ref-3	&	 20:00:38.94	&	$+$09:23:03.4	&	K1V	&	14.19	&	1.01	&	6.14	&	0.39	&340	&	2.944	&	2.331	\\
Ref-4	&	 20:00:33.99	&	$+$09:20:44.6	&	G2V	&	13.24	&	0.74	&	4.56	&	0.29	&473	&	2.113	&	2.072	\\
Ref-5	&	 20:00:21.92	&	$+$09:22:15.0	&	*G9III	&	12.35	&	1.04	&	0.8	&	0.25	&1848	&	0.541	&	0.571	\\
Ref-6	&	 20:00:17.93	&	$+$09:22:01.5	&	M1III	&	13.32	&	1.64	&	-0.5	&	0.62	&4300	&	0.233	&	0.225	\\
Ref-7	&	 20:00:16.45	&	$+$09:19:34.6	&	F5V	&	13.61	&	0.57	&	3.35	&	0.35	&969	&	1.032	&	1.050	\\
Ref-8	&	 20:00:18.57	&	$+$09:19:30.5	&	K3III	&	13.71	&	1.44	&	0.24	&	0.6	&3688	&	0.271	&	0.270	\\
Ref-9	&	 20:00:14.28	&	$+$09:21:22.6	&	G1V	&	14	&	0.7	&	4.24	&	0.19	&786	&	1.274	&	1.308	\\							
\hline																	
\multicolumn{11}{c}{HIP 103269} \\													
\hline
Ref-1	&	20:55:27.7	&	$+$42:22:08.2	&	K1V	&	13.04	&	0.99	&	6.14	&
	0.44	&198	&	5.065	&	5.888	\\
Ref-2	&	20:55:34.9	&	$+$42:19:43.8	&	G1V	&	12.37	&	0.65	&	4.24	&	0.19	&377	&	2.654	&	1.259	\\		
Ref-3        &       20:55:31.7	&	$+$42:19:40.5	&	G2V	&	14.64	&	0.8	&	4.56	&	0.51	&823	&	1.214	&	1.354	\\							
Ref-4	&      20:55:22.3     	&	$+$42:18:44.0	&	G1V	&	13.36	&	0.67	&	4.24	&	0.13	&610	&	1.642	&	1.536	\\							
Ref-5	&     20:55:16.3	         &	$+$42:19:25.4	&	M1III	&	15.46	&	2.61	&	-0.5	&	4.14	&	2312	&	0.432	&	0.431	\\							
Ref-6	&	20:55:10.7	&	$+$42:19:39.4	&	G2V	&	14.14	&	0.82	&	4.56	&	0.5	&670	&	1.494	&	1.654	\\	
Ref-7	&     20:55:09.4	         &	$+$42:19:10.5	&	F5V	&	11.59	&	0.55	&	3.35	&	0.29	&	397	&	2.523	&	2.564	\\				
Ref-8	&      20:55:06.3	         &	$+$42:19:15.1	&	G2V	&	12.77	&	0.72	&	4.56	&	0.19	&	398	&	2.516	&	0.978	\\	
Ref-9	&	20:55:10.7	&	$+$42:17:24.4	&	G5V	&	13.49	&	0.77	&	4.93	&	0.23	&469	&	2.134	&	2.517	\\
\hline																	
\multicolumn{11}{c}{HIP 106924} \\													
\hline																	
Ref-1	&	21:39:26.9	&	$+$60:19:20.1	&	G2V	&	14.64	&	1.06	&	4.56	&	1.4	&544	&	1.838	&	1.924	\\
Ref-2	&	21:39:38.7	&	$+$60:17:29.6	&	K2III	&	13.25	&	1.64	&	0.42	&	1.64	&1738	&	0.575	&	0.573	\\
Ref-3	&	21:39:23.0	&	$+$60:17:17.7	&	K0III	&	14.73	&	1.54	&	0.7	&	1.7	&2867	&	0.349	&	0.345	\\
Ref-4	&	21:39:17.8	&	$+$60:17:21.4	&	F0V	&	14.3	&	0.76	&	2.4	&	1.41	&1261	&	0.793	&	0.795	\\
Ref-5	&	21:39:14.3	&	$+$60:16:39.5	&	K2III	&	14.63	&	1.61	&	0.42	&	1.41&	3670	&	0.273	&	0.272	\\
Ref-6	&	21:39:19.8	&	$+$60:15:34.0	&	F5V	&	11.44	&	0.56	&	3.35	&	0.21	&379	&	2.636	&	3.025	\\
Ref-7	&	21:39:03.8	&	$+$60:15:58.0	&	K2III	&	14.1	&	1.8	&	0.42	&	2.13	& 2044	&	0.489	&	0.489	\\
\hline																	
\multicolumn{11}{c}{HIP 108200} \\													
\hline																	
Ref-1	&	21:55:43.1	&	$+$32:39:08.1	&	G5V	&	13.06	&	0.69	&	4.93	&	0.05	&414	&	2.416	&	2.623	\\
Ref-2	&	21:55:39.2	&	$+$32:38:28.0	&	K0III	&	13.65	&	1.15	&	0.7	&	0.44	&3159	&	0.317	&	0.316	\\
Ref-3	&	21:55:14.0	&	$+$32:37:24.9	&	K0III	&	11.15	&	1.17	&	0.7	&	0.58	&953	&	1.05	&	0.988	\\
Ref-4	&	21:55:03.0	&	$+$32:36:27.4	&	K2III	&	12.18	&	1.3	&	0.42	&	0.59	&1729	&	0.578	&	0.589	\\
Ref-5	&	21:54:53.4	&	$+$32:35:42.7	&	G0V	&	13.18	&	0.65	&	4.2	&	0.15	&584	&	1.711	&	$-$\tablenotemark{c}	\\
Ref-6	&	21:55:05.4	&	$+$32:38:34.5	&	K2III	&	13.89	&	1.25	&	0.42	&	0.29	&4367	&	0.229	&	0.229	\\
Ref-7	&	21:55:19.1	&	$+$32:37:41.3	&	F8V	&	15.29	&	0.67	&	3.87	&	0.58	&1496	&	0.669	&	0.668	\\
\enddata
\tablenotetext{a}{Spectral types and luminosity class estimated from colors and
reduced proper motion diagram.}
\tablenotetext{b}{The $\hst$ reference star parallaxes use spectroscopic priors and are not independent.}
\tablenotetext{c}{Reference star not included in model}
\end{deluxetable}

Two objects in Table \ref{tab:SPP} warrant mention. Ref-08 for HIP46120
has very weak metal lines, and appears to have a low metallicity. The
photometry and spectral continuum are consistent with the temperature
of an early K dwarf. HIP87788 Ref-01 appears to be a J-type carbon star.

\subsection{Estimated Reference Frame Proper Motions}\label{sec:refpm}

As priors,  we test proper motion values when available from the  SPM4 \citep{Gir11}, URAT1\citep{Finch16}, or the PPMXL  \citep{Roe10} catalogs as observations with error in
the  model.    The SPM4 catalog was only available for the HIP 46120 field, while the HIP 87788 and HIP 108200 fields did not have URAT1 values available.  The catalogs used for input proper motions are listed in Table~\ref{tab:pmsource}.   The input catalog proper motions and our final $\hst$ model values are found in Table~\ref{tab:PM}.

\begin{deluxetable}{lc}
\tablewidth{0in}
\tablecaption{Astrometric Reference Star Proper Motion Input Catalogs   \label{tab:pmsource}}
\tablehead{\colhead{Target}&
\colhead{Catalog} 
}
\startdata
HIP 46120	&	SPM4	\\
HIP 54639	&	URAT	\\
HIP 87062	&	URAT	\\
HIP 87788	&	PPMXL\tablenotemark{a}	\\
HIP 98492	&	URAT	\\
HIP 103269	&	URAT	\\
HIP 106924	&	URAT	\\
HIP 108200	&	PPMXL\tablenotemark{a}	\\
\enddata
\tablenotetext{a}{URAT catalog does not contain this reference frame}  
\end{deluxetable}

\begin{deluxetable}{lrrrrrr}
\tablewidth{0in}
\tablecolumns{7} 
\tablecaption{Astrometric  Reference Star Proper Motions \label{tab:PM}}
\tablehead{\colhead{}&
\multicolumn{2}{c}{Input} &
\multicolumn{4}{c}{Final ({$\hst$})}  \\
\colhead{ID}&
\colhead{$\mu_\alpha$\tablenotemark{a} } &
\colhead{$\mu_\delta$\tablenotemark{a} }&
\colhead{$\mu_\alpha$} &
\colhead{$\sigma_{\mu_\alpha}$ } &
\colhead{$\mu_\delta$} &
\colhead{$\sigma_{\mu_\alpha}$ } 
 }
 \startdata
\hline													
\multicolumn{7}{c}{HIP 46120} \\													
\hline													
Ref-1	&	2.6	&	-2.18	&	2.190	&	0.218	&	-0.054	&	0.253	\\
Ref-2	&	0.36	&	-6.48	&	2.860	&	0.220	&	-7.039	&	0.245	\\
Ref-3	&	-9.44	&	-0.64	&	-11.918	&	0.175	&	0.740	&	0.195	\\
Ref-4	&	-4.16	&	3.2	&	-5.831	&	0.199	&	-2.828	&	0.209	\\
Ref-5	&	-15.48	&	15.76	&	-12.927	&	0.252	&	15.989	&	0.280	\\
Ref-6	&	-36.28	&	16.28	&	-35.213	&	0.258	&	19.854	&	0.249	\\
Ref-7	&	-3.94	&	-2.16	&	-4.655	&	0.319	&	-1.065	&	0.321	\\
Ref-8	&	-4.16	&	3.2	&	-4.013	&	0.211	&	3.332	&	0.209	\\
\hline													
\multicolumn{7}{c}{HIP 54639} \\													
\hline													
Ref-1	&	-11	&	2	&	-11.627	&	0.181	&	1.457	&	0.158	\\
Ref-2	&	-19.7	&	-3.5	&	-17.837	&	0.143	&	-1.330	&	0.169	\\
Ref-3	&	-1.7	&	-5.8	&	-4.135	&	0.194	&	-6.577	&	0.214	\\
Ref-4	&	-16	&	-8.3	&	-15.605	&	0.199	&	-12.139	&	0.200	\\
Ref-5	&	-3.4	&	-38.2	&	-2.083	&	0.153	&	-35.581	&	0.161	\\
\hline													
\multicolumn{7}{c}{HIP 87062} \\													
\hline													
Ref-3	&	1.4	&	-2.8	&	1.463	&	0.145	&	-2.991	&	0.163	\\
Ref-4	&	-1.5	&	-3.4	&	-1.645	&	0.323	&	-2.895	&	0.258	\\
Ref-5	&	-8.9	&	-6.9	&	-8.761	&	0.141	&	-7.232	&	0.132	\\
Ref-6	&	-0.1	&	-3.6	&	-0.162	&	0.149	&	-3.556	&	0.129	\\
\hline													
\multicolumn{7}{c}{HIP 87788} \\													
\hline													
Ref-2	&	-4.5	&	1.9	&	-4.640	&	0.233	&	0.373	&	0.217	\\
Ref-3	&	-19.3	&	-28.6	&	-20.319	&	0.109	&	-30.225	&	0.111	\\
Ref-4	&	-5.6	&	3.4	&	-4.310	&	0.138	&	2.854	&	0.149	\\
Ref-5	&	-2.5	&	3.6	&	-5.158	&	0.100	&	7.424	&	0.110	\\
Ref-6	&	-6.4	&	-0.9	&	-4.611	&	0.121	&	-0.107	&	0.107	\\
Ref-7	&	-2.1	&	-14.4	&	-3.955	&	0.627	&	-8.168	&	0.841	\\
Ref-8	&	-4	&	7.1	&	-2.892	&	1.012	&	-8.544	&	0.730	\\
Ref-101	&	-4.6	&	-8.8	&	-3.084	&	0.248	&	-4.772	&	0.216	\\
Ref-102	&	3.9	&	-0.3	&	3.352	&	0.143	&	-5.042	&	0.168	\\
\hline													
\multicolumn{7}{c}{HIP 98492} \\													
\hline													
Ref-1	&	7.8	&	7.7	&	10.411	&	0.139	&	6.070	&	0.129	\\
Ref-2	&	5.9	&	2.3	&	2.638	&	0.171	&	2.327	&	0.199	\\
Ref-3	&	-4.5	&	-3.8	&	-7.883	&	0.260	&	-3.574	&	0.243	\\
Ref-4	&	5.8	&	8.4	&	10.563	&	0.197	&	3.334	&	0.199	\\
Ref-5	&	4.4	&	7.9	&	2.180	&	0.296	&	11.309	&	0.289	\\
Ref-6	&	-2.3	&	-4.7	&	-3.765	&	0.166	&	-2.458	&	0.186	\\
Ref-7	&	3.9	&	-3	&	5.721	&	0.218	&	-1.275	&	0.266	\\
Ref-8	&	2	&	-6.3	&	6.457	&	0.161	&	-6.758	&	0.180	\\
Ref-9	&	-2.4	&	-0.7	&	-3.868	&	0.322	&	-2.408	&	0.400	\\
\hline													
\multicolumn{7}{c}{HIP 103269} \\													
\hline													
Ref-1	&	-7.6	&	-26.4	&	-9.340	&	0.203	&	-27.285	&	0.186	\\
Ref-2	&	23.5	&	4.5	&	26.538	&	0.197	&	4.983	&	0.206	\\
Ref-3	&	5	&	-2	&	3.882	&	0.334	&	-3.574	&	0.367	\\
Ref-4	&	-7.3	&	-10.8	&	-7.860	&	0.236	&	-8.644	&	0.238	\\
Ref-5	&	-1.8	&	-7.1	&	-1.910	&	0.335	&	-4.605	&	0.314	\\
Ref-6	&	-13.2	&	-7.5	&	-10.666	&	0.324	&	-8.810	&	0.331	\\
Ref-7	&	0.7	&	6.1	&	0.111	&	0.197	&	8.271	&	0.193	\\
Ref-8	&	-9.2	&	-8.3	&	-8.570	&	0.256	&	-9.039	&	0.264	\\
Ref-9	&	-27.8	&	-12.4	&	-29.631	&	0.266	&	-15.317	&	0.301	\\
\hline													
\multicolumn{7}{c}{HIP 106924} \\													
\hline													
Ref-1	&	-11	&	-20.3	&	-11.413	&	0.168	&	-18.489	&	0.176	\\
Ref-2	&	3.6	&	-8.9	&	2.351	&	0.145	&	-8.225	&	0.157	\\
Ref-3	&	-7.2	&	-6.6	&	-5.027	&	0.192	&	-8.631	&	0.228	\\
Ref-4	&	-5	&	-1.9	&	-3.750	&	0.150	&	-5.008	&	0.160	\\
Ref-5	&	-1.4	&	1	&	-1.473	&	0.175	&	-0.265	&	0.184	\\
Ref-6	&	20.3	&	12.7	&	20.795	&	0.092	&	13.156	&	0.103	\\
Ref-7	&	-6.5	&	-4.1	&	-8.683	&	0.209	&	-0.730	&	0.226	\\
\hline													
\multicolumn{7}{c}{HIP 108200} \\													
\hline													
Ref-1	&	-0.8	&	-16.1	&	-0.316	&	0.175	&	-18.600	&	0.185	\\
Ref-2	&	-2	&	-7.4	&	-2.630	&	0.199	&	-4.581	&	0.171	\\
Ref-3	&	-11.2	&	-15.8	&	-10.469	&	0.104	&	-17.051	&	0.116	\\
Ref-4	&	9.5	&	2.6	&	9.442	&	0.161	&	2.615	&	0.166	\\
Ref-6	&	-2.4	&	-2.4	&	-2.542	&	0.244	&	-1.950	&	0.261	\\
Ref-7	&	-6.6	&	-3.5	&	-7.836	&	0.476	&	-1.338	&	0.445	\\
\enddata
\tablenotetext{a}{Spectral types and luminosity class estimated from colors and
reduced proper motion diagram.}
\end{deluxetable}

\subsection{Astrometric Reference Frame Residual  Assessment}

Without the target stars, the reference frame stars are modeled many times to assess  various plate models,   spectrophotometric parallaxes, catalog proper motions, and general stability  as a reference star.    We graphed  reference frame $\it x$ and $\it y$ residuals against a number of spacecraft, instrumental, and astronomical parameters. These included $\it x$,  and $\it y$ position, radial distance from the center of the field-of-view,  V magnitude and B - V color of the reference stars, and time of observation.  We see no trends indicating systematic instrumental effects, except for  the expected small upward trend in residuals with the faintest star.   

\subsection{Astrometric Catalog Residual  Assessment}
 
 $\hst$ FGS1r has raw distortions of more than an arcsecond, but the OFAD (McArthur et al. 2002) calibration reduces these distortions to around 1 mas in the center of the pickle and below 2 mas over much of the FGS 1r field.  The goodness of fit is shown in the table of the astrometric residuals  statistics (Table~\ref{tab:refhstres}) which lists the average position errors and median average deviations of the positions 

\begin{deluxetable}{lcccc}
\tablewidth{0in}
\tablecaption{Metal-Poor Target Star Field Quality    \label{tab:refhstres}}
\tablehead{\colhead{ID}&
\colhead{Number of} &
\colhead{RMS} &
\colhead{ Xres}  &
\colhead{ Yres} \\
\colhead{}&
\colhead{Residuals} &
\colhead{(mas)} &
\colhead{  (mas)}  &
\colhead{  (mas)} 
}
\startdata
HIP 46120 & 351 &  1.370 &1.198 &1.225 \\
HIP 54639 & 322 & 1.366 &1.261 &1.252 \\
HIP 87062 & 256 & 1.133 &0.944 &1.183 \\
HIP 87788 & 267  &1.379 &1.122  &1.287 \\
HIP 98492 & 353  &1.366 &1.330  &1.172 \\
HIP103269& 324  &1.194 &1.294 &0.808 \\
HIP106924& 370  &1.144 &1.116 &0.963\\
HIP108200 &295  &0.927 &0.95 &0.907\\
\enddata
\end{deluxetable}

\subsection{Astrometric Catalog}

The astrometric catalogs from the combined modeling are shown in Table~\ref{tab:cat}
 
 \begin{deluxetable}{lrrrrrrr}
\tablewidth{0in}
\tablecaption{Astrometric Catalogs  of  Metal Poor Target and Reference Stars \label{tab:cat}}
\tablecolumns{8} 
\centering
\tablehead{\colhead{Star}&\colhead{Mag\tablenotemark{a} }&\colhead{R.A.\tablenotemark{b} }&  \colhead{Dec.\tablenotemark{b}} & \colhead{$\xi$\tablenotemark{c} } & \colhead{$\sigma_{\xi}$} &  \colhead{$\eta$\tablenotemark{c} } &  \colhead{$\sigma_{\eta}$} \\
\colhead{}&\colhead{V} & \colhead{deg} & \colhead{deg}&\colhead{arcsec}&\colhead{arcsec}&\colhead{arcsec}&\colhead{arcsec}}
\startdata
\hline															
HIP 46120	&	10.14	&	141.09068	&	-80.51926	&	-1.03559	&	0.00013	&	730.34794	&	0.00014	\\
Ref-1	&	14.42	&	141.07442	&	-80.60486	&	306.77912	&	0.00024	&	748.33586	&	0.00022	\\
Ref-2	&	14.02	&	141.13579	&	-80.56864	&	173.78383	&	0.00020	&	772.70577	&	0.00021	\\
Ref-3	&	14.09	&	141.05475	&	-80.54958	&	109.62841	&	0.00021	&	718.73084	&	0.00023	\\
Ref-4	&	12.63	&	141.03917	&	-80.53322	&	51.75303	&	0.00022	&	704.40134	&	0.00023	\\
Ref-5	&	14.81	&	141.0025	&	-80.52997	&	41.97647	&	0.00026	&	681.61466	&	0.00026	\\
Ref-6	&	13.53	&	140.99562	&	-80.51256	&	-20.36440	&	0.00030	&	671.44051	&	0.00030	\\
Ref-7	&	14.96	&	141.24083	&	-80.49142	&	-108.58955	&	0.00035	&	810.34446	&	0.00039	\\
Ref-8	&	13.28	&	141.15792	&	-80.47347	&	-168.80552	&	0.00023	&	755.47584	&	0.00022	\\
\hline															
HIP 54639	&	11.38	&	167.74858	&	6.41859	&	-0.93955	&	0.00011	&	675.41028	&	0.00011	\\
Ref-1	&	13.85	&	167.75996	&	6.40436	&	56.67655	&	0.00018	&	644.87230	&	0.00018	\\
Ref-2	&	13.47	&	167.83121	&	6.40572	&	290.49465	&	0.00018	&	749.24212	&	0.00019	\\
Ref-3	&	14.71	&	167.76988	&	6.44422	&	34.36981	&	0.00025	&	790.93086	&	0.00026	\\
Ref-4	&	14.73	&	167.738	&	6.44361	&	-70.07244	&	0.00022	&	744.85367	&	0.00023	\\
Ref-5	&	13.24	&	167.67925	&	6.47125	&	-302.26219	&	0.00019	&	755.48500	&	0.00020	\\
\hline															
HIP87062	&		&		&		&		&		&		&		\\
Ref-1	&	10.59	&	266.8672	&	-8.78091	&	199.16810	&	0.00011	&	675.88272	&	0.00010	\\
Ref-2	&	13.37	&	266.78913	&	-8.75172	&	-80.40516	&	0.00025	&	775.29420	&	0.00027	\\
Ref-3	&	13.3	&	266.79596	&	-8.74378	&	-56.82072	&	0.00019	&	804.55110	&	0.00018	\\
Ref-4	&	14.57	&	266.81742	&	-8.77142	&	21.57447	&	0.00020	&	705.92984	&	0.00019	\\
Ref-5	&	15.62	&	266.85829	&	-8.76044	&	166.15922	&	0.00045	&	748.87200	&	0.00029	\\
Ref-6	&	14.58	&	266.88862	&	-8.75936	&	273.67705	&	0.00017	&	754.89600	&	0.00016	\\
\hline															
HIP87788	&	11.32	&	268.99363	&	-16.41082	&	-0.91890	&	0.00012	&	710.27933	&	0.00014	\\
Ref-2	&	14.46	&	269.02542	&	-16.39392	&	-109.37292	&	0.00033	&	649.80905	&	0.00031	\\
Ref-3	&	10.56	&	268.99208	&	-16.39025	&	4.55778	&	0.00018	&	637.04227	&	0.00018	\\
Ref-4	&	11.89	&	269.01783	&	-16.41475	&	-84.63560	&	0.00020	&	724.71871	&	0.00021	\\
Ref-5	&	10.99	&	269.01554	&	-16.43678	&	-76.76387	&	0.00017	&	803.91410	&	0.00018	\\
Ref-6	&	11.56	&	268.98758	&	-16.41144	&	19.99242	&	0.00018	&	712.94625	&	0.00017	\\
Ref-7	&	13.48	&	268.95862	&	-16.39256	&	119.64677	&	0.00043	&	646.17077	&	0.00045	\\
Ref-8	&	13.87	&	268.93646	&	-16.42917	&	196.56365	&	0.00046	&	776.87247	&	0.00041	\\
Ref-101	&	14.41	&	268.92646	&	-16.39436	&	231.08096	&	0.00049	&	652.13340	&	0.00043	\\
Ref-102	&	13.52	&	268.92117	&	-16.427	&	248.61500	&	0.00028	&	769.72135	&	0.00033	\\
\hline															
HIP 98492	&	11.58	&	300.13994	&	9.353	&	-1.04389	&	0.00013	&	690.42745	&	0.00013	\\
Ref-1	&	11.13	&	300.20821	&	9.36586	&	245.71686	&	0.00015	&	694.40018	&	0.00015	\\
Ref-2	&	11.92	&	300.17854	&	9.36853	&	143.56687	&	0.00021	&	721.97302	&	0.00021	\\
Ref-3	&	14.19	&	300.16225	&	9.38428	&	96.06615	&	0.00032	&	787.67128	&	0.00031	\\
Ref-4	&	13.24	&	300.14162	&	9.34572	&	0.35508	&	0.00021	&	663.61701	&	0.00022	\\
Ref-5	&	12.36	&	300.09133	&	9.37083	&	-160.11894	&	0.00036	&	783.23070	&	0.00034	\\
Ref-6	&	13.3	&	300.07471	&	9.36708	&	-220.85237	&	0.00021	&	780.01338	&	0.00022	\\
Ref-7	&	13.59	&	300.06854	&	9.32628	&	-267.59091	&	0.00025	&	639.16527	&	0.00031	\\
Ref-8	&	13.69	&	300.07738	&	9.32514	&	-237.36517	&	0.00019	&	629.58129	&	0.00021	\\
Ref-9	&	14	&	300.0595	&	9.35628	&	-280.58999	&	0.00036	&	750.99604	&	0.00038	\\
\hline															
HIP 103269	&	10.3	&	313.82004	&	42.29906	&	-1.06961	&	0.00010	&	765.31094	&	0.00010	\\
Ref-1	&	13.07	&	313.86533	&	42.36894	&	-235.75461	&	0.00018	&	615.48889	&	0.00017	\\
Ref-2	&	12.39	&	313.89562	&	42.32883	&	-228.81074	&	0.00020	&	780.62742	&	0.00018	\\
Ref-3	&	14.66	&	313.88225	&	42.32792	&	-196.45390	&	0.00033	&	764.52578	&	0.00031	\\
Ref-4	&	13.4	&	313.843	&	42.31222	&	-77.67290	&	0.00021	&	757.33490	&	0.00021	\\
Ref-5	&	15.29	&	313.81775	&	42.32372	&	-42.60106	&	0.00031	&	686.87482	&	0.00031	\\
Ref-6	&	14.18	&	313.79475	&	42.32761	&	2.27333	&	0.00030	&	642.63245	&	0.00030	\\
Ref-7	&	11.63	&	313.78933	&	42.31958	&	29.19090	&	0.00016	&	659.34975	&	0.00017	\\
Ref-8	&	12.83	&	313.77604	&	42.32086	&	57.13444	&	0.00024	&	636.81119	&	0.00022	\\
Ref-9	&	13.54	&	313.79454	&	42.29011	&	74.22246	&	0.00025	&	756.71502	&	0.00025	\\
\hline															
HIP 106924	&	10.37	&	324.81503	&	60.28452	&	-0.74170	&	0.00010	&	680.29568	&	0.00009	\\
Ref-1	&	14.67	&	324.86208	&	60.32225	&	-158.70460	&	0.00019	&	701.12039	&	0.00019	\\
Ref-2	&	13.28	&	324.91125	&	60.29156	&	-93.96140	&	0.00018	&	826.50046	&	0.00015	\\
Ref-3	&	14.75	&	324.84567	&	60.28825	&	-34.83180	&	0.00024	&	724.59364	&	0.00022	\\
Ref-4	&	14.32	&	324.82433	&	60.28928	&	-23.04487	&	0.00017	&	688.31047	&	0.00016	\\
Ref-5	&	14.65	&	324.80954	&	60.27764	&	25.94452	&	0.00021	&	681.24496	&	0.00020	\\
Ref-6	&	11.49	&	324.83242	&	60.25944	&	68.67841	&	0.00010	&	745.75221	&	0.00009	\\
Ref-7	&	14.14	&	324.766	&	60.26611	&	95.59408	&	0.00022	&	627.50410	&	0.00021	\\
\hline															
HIP 108200	&	11.01	&	328.8199	&	32.64515	&	29.01044	&	0.00008	&	660.57750	&	0.00008	\\
Ref-1	&	13.08	&	328.9295	&	32.65225	&	-292.13097	&	0.00016	&	749.66380	&	0.00016	\\
Ref-2	&	13.68	&	328.91312	&	32.64111	&	-231.76706	&	0.00017	&	770.29653	&	0.00015	\\
Ref-3	&	11.21	&	328.80829	&	32.62358	&	88.71493	&	0.00010	&	721.85232	&	0.00010	\\
Ref-4	&	12.26	&	328.76229	&	32.60761	&	239.24013	&	0.00016	&	728.26817	&	0.00015	\\
Ref-6	&	13.94	&	328.77263	&	32.64292	&	166.52997	&	0.00023	&	619.34112	&	0.00022	\\
Ref-7	&	15.26	&	328.8295	&	32.62814	&	22.67684	&	0.00041	&	728.01902	&	0.00033	\\
\enddata
\tablenotetext{a}{V mag calculated from Hubble Photometry}   
 \tablenotetext{b}{Predicted coordinates for equinox J2000.0}   
 \tablenotetext{c}{Relative coordinates in the reference frame of the constrained plate .}
\end{deluxetable}

\end{document}